\documentclass{jfm}

\usepackage{amsmath}

\usepackage{graphicx}
\usepackage{placeins}
\usepackage{subfig}
\usepackage{dcolumn}
\usepackage{bm}
\usepackage{hyperref}
\usepackage[mathlines]{lineno}
\usepackage{setspace}
 
\hypersetup{
    colorlinks=true,
    linkcolor=blue,
    filecolor=blue,      
    urlcolor=blue,
    citecolor=blue
}
\usepackage{tikz}
\usepackage{soul}
\usepackage{setspace}
\usepackage{mathrsfs}

\usepackage{fancyhdr} 
\usepackage{makecell}

\newcommand{\hs}[1]{\hspace{#1 cm}}

\title{Stability and dynamics of the flow past of a bullet-shaped blunt body moving in a pipe}

\author{Paul Bonnefis, David Fabre  \corresp{\email{david.fabre@imft.fr}} \& Christophe Airiau.}
 
\affiliation{%
IMFT, Universit{\'e} de Toulouse, UMR 5502 CNRS / INPT / UPS,\\
Allée du Pr Camille Soula, 31400 Toulouse, France
}%

\begin{document}
\maketitle
\date{\today}

\begin{abstract}
The flow past a bullet-shaped blunt body moving in a pipe 
is investigated through global linear stability analysis (LSA) and direct numerical simulation (DNS).
A cartography of the bifurcation curves is provided thanks to LSA, covering the range of parameters corresponding to Reynolds number $Re = [50-110]$, confinement ratio $a/A = [ 0.01 - 0.92]$ and length-to-diameter ratio $L/d = [ 2-10]$. Results show that the first bifurcation is always a  steady bifurcation associated to a non-oscillating eigenmode with azimuthal wavenumber $m=\pm 1$ leading to a steady state with planar symmetry. For weakly confined cases ($a/A<0.6$) the second bifurcation is associated to an oscillating mode with azimuthal wavenumber $m=\pm 1$, as in the unconfined case. On the other hand, for the strongly confined case ($a/A>0.8$), on observes destabilization of non-oscillating modes with $|m| =2,3$ and a restabilization of the $m=\pm1$ eigenmodes.
The aspect ratio $L/d$ is shown to have a minor influence for weakly confined cases and almost no influence for strongly confined cases. DNS is subsequently used to characterize the nonlinear dynamics. The results confirm the steady bifurcation predicted by LSA with excellent agreement for the threshold Reynolds. For weakly confined cases, the second bifurcation is a Hopf bifurcation leading to a periodic, planar-symmetric state in qualitative accordance with LSA predictions. For more confined cases, more complex dynamics is obtained, including  a steady state with $|m|=3$ geometry and  aperiodic states. 
\end{abstract}





\section{Introduction}
The flow past blunt bodies is a problem of practical importance, with obvious engineering applications to transport. In such applications it is important to estimate and predict the lift and drag forces exerted on the body as well as to assert the influence of the geometry on these forces for a shape optimization procedure. Characterisation of wake flows 
in the transitional regime (with Reynolds numbers of order $10^2-10^3$)
is also a problem of fundamental interest where global 
stability theory and bifurcation theory have been particularly successful 
to characterise complex nonlinear dynamics. 
The most documented case corresponds to the wake of a cylindrical body placed perpendicularly to the flow \citep{benard1908formation,von1912uber,provansal1987benard}. 
This case is characterised by a Hopf bifurcation for $Re \approx 47$ giving rise to the well-known B\'{e}nard-Von Ka\`{a}rm\`{a}n vortex street. 
Secondary bifurcations occurring in the range $Re \approx 200$ and leading to three-dimensional states have also been investigated by stability analysis of the periodic solution and bifurcation theory \citep{thompson1996three}.
Among three-dimensional geometries, the cases of a sphere and of a disk have been particularly considered as canonical geometries. Linear stability analysis (LSA) \citep{natarajan1993instability,meliga2009global} provides a powerful framework allowing to tackle this class of problems. This approach predicts that the
first unstable modes is a non-oscillating mode (i.e. with purely real global
eigenvalue) characterised by azimuthal wavenumbers $m=\pm 1$.
It leads to a steady state ($SS$) solution with planar symmetry, a presence of a pair of longitudinal vortices and finally a non-zero lift force exerted on the body. 
The LSA study also predicts the onset of a secondary eigenmode which is oscillating (i.e. a pair of complex conjugated eigenvalues) and also associated to an azimuthal  wavenumber $m=1$. 
Comparisons  with direct numerical simulations  and application of normal form theory \citep{fabre2008bifurcations,auguste2010bifurcations} and weakly nonlinear analysis (WNLA) \citep{meliga2009global} showed that this secondary mode is responsible for the onset of an oscillating state which is either reflection-symmetry preserving (RSP) for spheres and thick disks or reflection-symmetry breaking (RSB) for thin disks.
Effect of motion of the body has also been considered. 
First, the effect of imposed rotation on the wake of a sphere has been  analysed. 
In the case the axis of rotation is aligned with the flow, rotation breaks the symmetry between $m=+1$ and $m=-1$ modes and modifies the bifurcation scenario leading to the onset of quasiperiodic states \citep{pier2013periodic}.
In the case the axis is transverse, weak rotation stabilises the RSP mode but strong rotation gives rise to a new oscillating mode with a smaller frequency
\citep{citro2016linear,fabre2017flow}.
Secondly,  \cite{tchoufag2014global} have demonstrated the influence of wake dynamics on the motion of bodies in free movement submitted to 
buoyancy force. In that case, the destabilisation of the base flow field may result in a path deviation of the buoyancy-driven disk or sphere
leading to a variety of states including zig-zag paths, steady-oblique paths, etc.

Another canonical blunt body geometry which was selected by a number of studies in the literature is the bullet-shaped body, consisting in a half-ellipsoidal nose glued to a cylindrical blunt rear. It has the advantage to have a shape closer to real industrial applications, trains for instance. Experiments performed by \cite{brucker2001spatio} revealed a stabilizing effect of the presence of the ellipsoidal nose, in comparison with the flow past disks. 
An extensive study presented by \cite{bohorquez2011stability} uses three approaches, DNS, LSA and experiments. This study reveals that the bifurcation sequences and wake patterns are globaly similar to the case of a sphere, and that increasing the length of the body generally delays the bifurcations towards larger Reynold numbers.
A base-bleed flow control has also been tested and its stabilizing effect was demonstrated. The sequence of bifurcations occurring in the wake has been examined by \cite{bury2012transitions} using DNS, from the laminar axisymmetric wake to the onset of chaotic behavior. In \cite{jimenez2014global}, the effect of spinning of this blunt body around its axis of symmetry is shown to have stabilizing effect, promoting the second most amplified mode and widening the range of existence of a stable axisymmetric wake.

The present study considers the effect of confinement on wake dynamics past the bullet-shaped blunt body (cf fig. \ref{geometryBullet}).
Many industrial issues raise in the case of an object traveling in a confined environment. A good example is a high-speed train passing through a tunnel, how it enters the tunnel and how the tunnel influences the aerodynamics of the train \citep{mok2001numerical,kwon2003numerical,baron2001alleviation}.
The issue encountered relies more on the pressure wave created by the train nose and its interaction with the tunnel than the wake itself, but the drag is still of interest. In another study, \cite{choi2014effects} investigates the optimisation of the nose shape of the high-speed Korean subway and the tunnel cross-sectional area influence  on the total drag. Of course, with velocities  of several hundred kilometers per hour, the Reynolds numbers are of order $10^8$ and characterisation of nonlinear dynamics in the transitional range may be irrelevant.
The situation changes considering new technologies in train transportation such as evacuated 
tube transportation system where a capsule travels at high velocity in a near vacuum network of pipes. Numerous studies describes the limitations and opportunities arising in such configurations \citep{opgenoord2018aerodynamic,oh2019numerical,braun2017aerodynamic} and highlight differences in aerodynamics compared to standard trains. The expected operating pressure for such system is in the range $1 - 100$~Pa,
leading to Reynolds numbers in the range $10^3 - 10^5$. 
Hence, characterisation of dynamics in the transitional range using a combination of LSA, bifurcation theory, and DNS, may be relevant in this range.
The study of non-axisymmetric bifurcation giving rise to a lift force may be of practical interest in the operation of such devices. 
Such applications also operate in the transonic regime, so that for an 
accurate modelling compressibility and rarefied gas effects should also be taken into account. However, as a first approach towards these problems, it might be interesting to stick to an incompressible flow and target 
the effect of confinement regardless of 
additional effects.

Our current investigations  on a slender axisymmetric blunt-based body moving in a tube is inspired by such industrial applications. 
In order to pave the way to such complicated problems, 
the study has been limited to incompressible flows and to a Reynolds number $Re$ lower than $1500$.

The paper is organized as follows.
The configuration, the governing equations  and the Linear Stability Analysis equations and resolution methods are presented in section II. 
Section III is devoted to the characterisation of instability properties  thanks to LSA. A parametric study of the linearly unstable modes is obtained as function of the  confinement ratio, of the length-to diameter ratio and of the Reynolds numbers.
Section IV is dedicated to direct numerical simulations (DNS) and to comparisons with the results of the LSA analysis. DNS is used to confirm the predictions of LSA regarding the first bifurcation threshold and to explore the nonlinear behavior arising away from this threshold.
The paper ends by some concluding remarks. Two appendixes have been added, the first one on the analytical solution of the annular Couette-Poiseuille flow and the last one on the mesh convergence study.



\section{Methodology }

\subsection{Configuration and parameters}

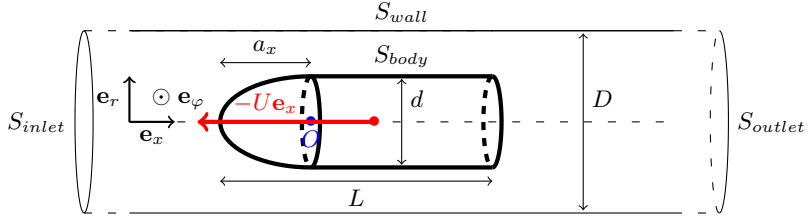
\begin{figure}
\centering
\begin{tikzpicture}[scale = 1.2]
\draw (-2,1) -- (4,1);
\draw (-2,-1) -- (4,-1);
\draw (-2,0) -- (4,0) [loosely dashed];
\draw (-2,1) -- (4,1);
\draw (-2.5,1) -- (2,1) [loosely dashed];
\draw (4,1) -- (4.5,1) [loosely dashed];
\draw (-2.5,-1) -- (-2,-1) [loosely dashed];
\draw (4,-1) -- (4.5,-1) [loosely dashed];

\draw [domain=pi:-pi, samples=40] 
plot ( {0.1*cos(deg(\x))-2.5}, {1*sin(deg(\x))} );
\draw [loosely dashed,  domain=-0.5*pi:-pi*1.5, samples=40]
plot ( {0.1*cos(deg(\x))+4.5}, {1*sin(deg(\x))} );
\draw [  domain=pi*0.5:-pi*0.5, samples=40]
plot ( {0.1*cos(deg(\x))+4.5}, {1*sin(deg(\x))} );

\draw [ultra thick,  domain=-pi*0.5:-pi*1.5, samples=40] 
 plot ( {cos(deg(\x))}, {0.5*sin(deg(\x))} );
\draw (0,0.5) -- (2,0.5) [ultra thick];
\draw (0,-0.5) -- (2,-0.5) [ultra thick];
\draw [ultra thick,  domain=-pi*0.5:pi*0.5, samples=40] 
plot ( {0.1*cos(deg(\x))+2}, {0.5*sin(deg(\x))} );
\draw [ultra thick, dashed,  domain=pi*0.5:pi*1.5, samples=40] 
plot ( {0.1*cos(deg(\x))+2}, {0.5*sin(deg(\x))} );
\draw [ultra thick,  domain=-pi*0.5:pi*0.5, samples=40] 
plot ( {0.1*cos(deg(\x))}, {0.5*sin(deg(\x))} );
\draw [ultra thick, dashed,  domain=pi*0.5:pi*1.5, samples=40]
plot ( {0.1*cos(deg(\x))}, {0.5*sin(deg(\x))} );

\draw [<-> ] (1,-0.47) -- (1,0.47)  ;
\draw  (1,0.25)  node [right]{$d$};
\draw [<-> ] (3,-0.97) -- (3,0.97)  ;
\draw  (3,0.25)  node [right]{$D$};
\draw [<-> ] (-1,-0.65) -- (2,-0.65)  ;
\draw  (0.5,-0.65)  node [below]{$L$};
\draw [<-> ] (-1,0.65) -- (0,0.65)  ;
\draw  (-0.5,0.65)  node [above]{$a_x$};
\begin{scope}[xshift=-2cm]
\draw [->](0,0) -- (0,0.5)[thick];
\draw [->](0,0) -- (0.5,0)[thick];
\draw (0,0.25) node [left]{$\textbf{e}_r$};
\draw (0.25,0) node [below]{$\textbf{e}_x$};
\draw (0.15,0.25) node [right]{$\odot\;\textbf{e}_\varphi$};
\end{scope}

\draw (-2.6,0) node[left] {$S_{inlet}$};
\draw (4.6,0) node[right] {$S_{outlet}$};
\draw (1,0.5) node[above] {$S_{body}$};
\draw (1,1) node[above] {$S_{wall}$};

\draw (0,0) node {$\textcolor{blue}\bullet$};
\draw (0,0) node[below] {$\textcolor{blue}O$};
\draw [<-](-1.25,0) -- (0.7,0)[ultra thick,red];
\draw (-0.5,0) node[above] {$\textcolor{red} {-U\textbf{e}_x}$};
\draw (0.7,0) node {$\textcolor{red}\bullet$};

\end{tikzpicture}
\caption{Geometry of the axisymmetric blunt body in a pipe.\label{geometryBullet} }
\end{figure}

The geometry of the bullet-shaped blunt body moving in a tube and a the main geometrical parameters are shown in Figure \ref{geometryBullet}.
 The body consists of a half-ellipsoid nose glued to a cylindrical rear.
The diameter of the cylinder is referred as  $d$. 
The ellipsoid of revolution is
defined by its major axis with $a_x=2a_y=d$ and its  minor axis  which fits with the cylindrical section by imposing $a_y=a_z=d/2$.


%
The diameter of the pipe is noted $D$, so that the effect of confinement will be
defined by either a diameter ratio $\xi =  d/D$ or an area ratio $a/A =  \xi^2$ with $a=\pi d^2/4$ the frontal area of the body and $A =\pi D^2/4$ the area of the tube.

The origin of the frame is taken at the junction between half-spheroidal and cylindrical parts, so that the body spans from $x=-d$ (nose) to $x=L-d$ (base).

The object moves with a velocity $U$ in the direction $-\bf{e}_x$ and the wall of the pipe does not move. Assuming the flow is incompressible and isothermal, 
the nondimensional parameters of this problem are the Reynolds number $Re= \rho U d / \mu$, the radius aspect ratio $\xi=d/D$, and the length aspect ratio $L/d$. $\rho$  and $\mu$ are respectively the constant density and dynamic viscosity of the fluid.
In most cases this parameter will be set to $L/D=2$, except in sec. \ref{sec:LsurD} where the effect of this parameter will be investigated.

The study will be conducted in the frame of reference associated to the body. The boundaries of the computational domain are given in the figure. It is limited 
by respectively an inlet section $S_{inlet}$ and an outlet section $S_{outlet}$.
In this frame, the body is fixed, and placed within an incoming flow $U$ of direction $+~\bf{e}_x$ and the  tube wall also moves at the same velocity with respect to the body. 
Hence the dimensionless incompressible Navier-Stokes equations  and associated boundary conditions are :
\begin{subequations}
\begin{align}
 \partial_t\textbf{u}  =\mathcal{NS}([\textbf{u},~p])&\equiv 
 - \textbf{u}\cdot\nabla\textbf{u} -\nabla p + \frac{2}{Re}\nabla \cdot {\bf D}(\textbf{u}) \\
\nabla\cdot\textbf{u}&=0 \\
\textbf{u}|_{S_{body}}&=\textbf{0} \\
\textbf{u}|_{S_{inlet}\cup S_{wall} }&=\textbf{e}_x \\
{\left[ -p~{\bf e}_x +\frac{2}{Re}D(\textbf{u}) \cdot {\bf e}_x  \right]}_{S_{outlet}} &=\bf{0}
\end{align}
\label{NS}
\end{subequations}
where \textbf{u} is the relative velocity, and ${\bf D}({\bf u})=(\nabla {\bf u} + \nabla^T {\bf u}) / 2$ is the rate-of-strain tensor. 
The divergence formulation for the viscous terms related to the choice of the Finite Element Method (FEM) to solve the equations. 

This equation is non-dimensionalized by the body velocity $U$, the fluid density $\rho$ and the body diameter $d$. 
The last boundary condition  is written as a no-stress condition  on the outlet section, which is a convenient choice for  outlet condition with FEM approach.

\subsection{Global linear stability analysis }
 
 \subsubsection{Equations}
 
The global linear stability approach is performed in the line of the now classical approach described for instance in \cite{sipp2007global,fabre2018stabfem}.
Within the LSA framework, the velocity and pressure are decomposed into  base flow and a small perturbation as follows:
\begin{equation}
\textbf{u}(r,\varphi,x,t)=\textbf{u}_{b}(r,x)+\epsilon~\hat{\textbf{u}}(r,x)e^{i m\varphi+\lambda t},\qquad
p(r,\varphi,x,t)=p_{b}(r,x)+\epsilon~\hat{p}(r,x)e^{i m\varphi+\lambda t}
\label{decompositonUP}
\end{equation}

Here $[{\bf u}_b;~p_p]$ is the so-called "base flow", namely the solution of the axisymmmetric,  time-independent version of equations \eqref{NS}, namely:
\begin{equation}
\mathcal{NS}([\textbf{u}_b,~p_b]) = {\bf 0} ; \quad \nabla \cdot \textbf{u}_b = 0.      
\label{NS0}
\end{equation}

In \eqref{decompositonUP}, a small-amplitude perturbation of the base-flow is assumed
in the form of an eigenmode $\left[ \hat{\textbf{u}},~\hat{p} \right]$ associated to an eigenvalue 
$\lambda=\lambda_r + i\lambda_i$.  The real part of the eigenvalue is the growth rate. A positive value indicated here an amplification. The imaginary part is a nondimensional frequency (time oscillation), which is most conveniently represented by 
 the Strouhal number $St = \lambda_i/ (2 \pi)$ thanks to the  nondimensionalization choices.
  Fourier decomposition in the azimuthal direction is possible with the 
  axisymmmetric invariance and  an azimuthal wavenumber $m\in\mathbb{Z}$ can be added
  in the exponential wave-like part of the perturbation.
  Introducing the decomposition \eqref{decompositonUP} into the Navier-Stokes equations 
and linearizing leads to an  eigenvalue problem written as:

\begin{subequations}
\begin{align}
\lambda~\hat{\textbf{u}}&=\mathcal{LNS}_{\textbf{u}_b}([\hat{\textbf{u}},~\hat{p}]) \\
\nabla_m\cdot\hat{\textbf{u}}&=0 \\
\hat{\textbf{u}}|_{S_{inlet}\cup S_{wall}\cup S_{body} }&=\textbf{0}  
\end{align}
\label{NS1}
\end{subequations}


Where $\mathcal{LNS}^m$ is the Linearized Navier-Stokes operator defined as

\begin{equation}
\label{eq:NS0}
   \mathcal{LNS}^m_{\textbf{u}_b}([\hat{\textbf{u}},~\hat{p}]) =- \textbf{u}_{b}\cdot\nabla_m\hat{\textbf{u}} - \hat{\textbf{u}}\cdot\nabla_m\textbf{u}_{b}  -\nabla_m~\hat{p} + \frac{2}{Re}\nabla_m\cdot D_m(\hat{\textbf{u}})
\end{equation}

Here quantities $\nabla_m$ and ${\bf D}_m$ are the gradient and rate-of-strain operators with $\partial_\varphi(.)$ replaced by $i \times m(.)$.


\subsubsection{Resolution methods}

The resolution methods  employed here are essentially similar to the one used in recent papers such as \cite{tchoufag2014global,fabre2019acoustic} considering stability analysis of axisymmetric incompressible flows. 

 Thanks to axisymmetry, the base-flow velocity is searched in the cylindrical frame $[{\bf e}_x,{\bf e}_r,{\bf e}_\varphi]$, as  
 ${\bf u}_b = [u_{b,x}(x,r),~u_{b,r}(x,r),~0]$ so that only two components of velocity are kept. On the other hand, for eigenmodes three components of velocity are needed, i.e.  ${\hat u} = [\hat u_{x}(x,r),~ \hat u_{r}(x,r), ~\hat u_{\phi}(x,r)]$.
 Within this assumption it is enough to consider a two-dimensional numerical domain ($\Omega$) corresponding to a meridian plane $(x,r)$. 

For both base-flow equations and linear stability equations, a Finite-Element method is used. For this sake, the equations are first turned into a weak form by introducing test functions $\textbf{v}$ and $q$ and a scalar product $\langle\varphi_1,\varphi_2\rangle=\int_\Omega\bar{\varphi_1}\cdot\varphi_2\,\mathrm{d}\Omega$. 
For instance, the weak form of the base-flow equations \ref{eq:NS0} are written as 
\begin{equation}
\forall ({\bf v},q), \quad 
\langle\textbf{v},~\mathcal{NS}([\hat{\textbf{u}_b},~\hat{p}])\rangle
+\langle q,~\nabla_0\cdot\hat{\textbf{u}_b}\rangle = 0.
\label{eq:weakNS0}
\end{equation}
An integration by part of the viscous terms is afterward performed and their derivation order is reduced. Dirichlet boundary conditions are incorporated by penalization while the stress-free outlet condition is directly satisfied thanks to integration by parts. 
The nonlinear problem is then solved using Newton iteration.
The developed form of the base-flow equations in cylindrical coordinates and details about the Newton method can be found, for instance, in  \cite{tchoufag2014global,fabre2019acoustic}.

Similarly, the leads stability problem  leads to the weak form

\begin{equation}
\forall ({\bf v},q), \qquad
\lambda~\langle\textbf{v},~\hat{\textbf{u}}\rangle=
\langle\textbf{v},~\mathcal{LNS}_{\textbf{u}_b}^m([\hat{\textbf{u}},~\hat{p}])\rangle
+\langle q,~\nabla_m\cdot\hat{\textbf{u}}\rangle
\label{eq:weakNS1}
\end{equation}

which after discretization leads to a 
generalized eigenvalue problem :

\begin{equation}
    \lambda~\mathrm{B}\hat{X}=\mathrm{A}\hat{X}
\end{equation}

A shift-and-invert method is applied to obtain a collection of eigenvalues (typically 10) located closest to a "shift" value taken as a guess of the searched eigenvalues. 
The weak form of the stability equations, details about the integration by parts and the construction of matrices $A$ and $B$ can again be found in \cite{tchoufag2014global,fabre2019acoustic}.

\subsubsection{Numerical implementation}

%

All numerical operations (generation and adaptation of a mesh; building of matricial operators, Newton iteration for base-flow problem and Shift-invert method for eigenvalue problem) are handled thanks to the Finite-Element software \textsc{FreeFem++} \citep{hecht}.

As for mesh generation, in order improve local accuracy, and adaptative mesh strategy as described in \cite{fabre2018stabfem} is adopted.
First a triangular mesh is built using the well-known Delaunay–Voronoi algorithm, and a preliminary base flow and some modes are computed.
Then, an adaptation procedure is performed with a criteria  based on 
these preliminary base flow and mode fields.
The procedure is repeated two to three times. 
This step is carried out for each set of parameters and each computed mode. Eventually, it gives converged results with very light meshes. the FEM discretization is built with the classical Taylor-Hood elements  for all computations.

The whole computational chain, including mesh generation, adaptation, loop over parameters, and post-processing, is monitored in the Matlab environement thanks to the \textsc{StabFem} suite \citep{fabre2018stabfem} which is a set of \textsc{Matlab/Octave} drivers/wrappers specifically designed to perform such studies.
%
A sample script reproducing a selection of results from the present study is available through the website of the \textsc{stabfem} project\footnote{\url{https://stabfem.gitlab.io/StabFem}}.
\\


\subsection{Direct numerical simulations}

To validate and extent the results of the stability analysis,
some full  direct  numerical simulations are performed with
the open source computational fluid dynamics software package, OpenFOAM\textregistered.
Time varying solutions of the equation \eqref{NS} are computed with 
its incompressible finite volume  solver, \textit{pimpleFOAM} built with a second-order spatial derivative schemes and  an Euler temporal scheme. A fixed Courant number set to $Co=0.5$  ensures the stability of these schemes. The meshes are built with the cfMesh software provided with OpenFOAM. A typical mesh and a mesh convergence study are presented in the appendix \ref{appendixB}.
Results and comparison with LSA is discussed in section IV.



\section{Linear Stability Analysis: Results}

With the objective of building an exhaustive cartography of instability properties, four parameters will be varied. The two first ones are geometrical parameters, namely the aspect ratio $L/d$ and the confinement parameter $a/A$.
The third input parameter is the Reynolds number
and the fourth one is the azimutal wavenumber $m$.
Table \ref{tab:parameters} indicates the range of parameters defined
for this study, and the concerned sections.
Regarding the azimuthal wavenumber, it is known for open flows past blunt bodies  that the most unstable modes are found for
$m=\pm 1$ \citep{natarajan1993instability,auguste2010bifurcations,jimenez2014global}. The justifies that our study will primarily focus on this value, and postpone other values of $m$ to Sec. III-D.


\begin{table}
    \setstretch{1.3}
    \centering
    \begin{tabular}{c|cccc}
    Section \hs{0.6} & \hs{0.6} $L/d$ \hs{0.6} &    \hs{0.6} $a/A$  \hs{0.6}  &  $Re$  \hs{0.6} &  $|m|$   \\ \hline \hline
    III - A   &  2     &  0.01, 0.75, 0.81  & 320, 200 - 1200 & 1 \\
    III - B   &  2     &  0.01 - 0.92       &  \hs{0.6} 320 - 1130, 110 - 400 & 1\\
    III - C & 2 - 10 &  0.01 - 0.75       & 110 - 140 - 200 & 1 \\
    III - D & 2 &  0.6 - 0.92       & 80 - 1100  & 2-3 \\ \hline
    \end{tabular}
    \caption{Ranges of parameters investigated and corresponding sections of the paper.}
    \label{tab:parameters}
    \setstretch{1}
\end{table}

\subsection{$m=\pm 1$ modes for sample values of the section aspect ratio $a/A$}

In this section and the next one we set  the length-to-diameter aspect ratio $L/d=2$ and 
we focus on the effect of the confinement ratio $a/A$.

\subsubsection{Weakly confined flow}

\begin{figure}
\centering
\includegraphics[scale = .8]{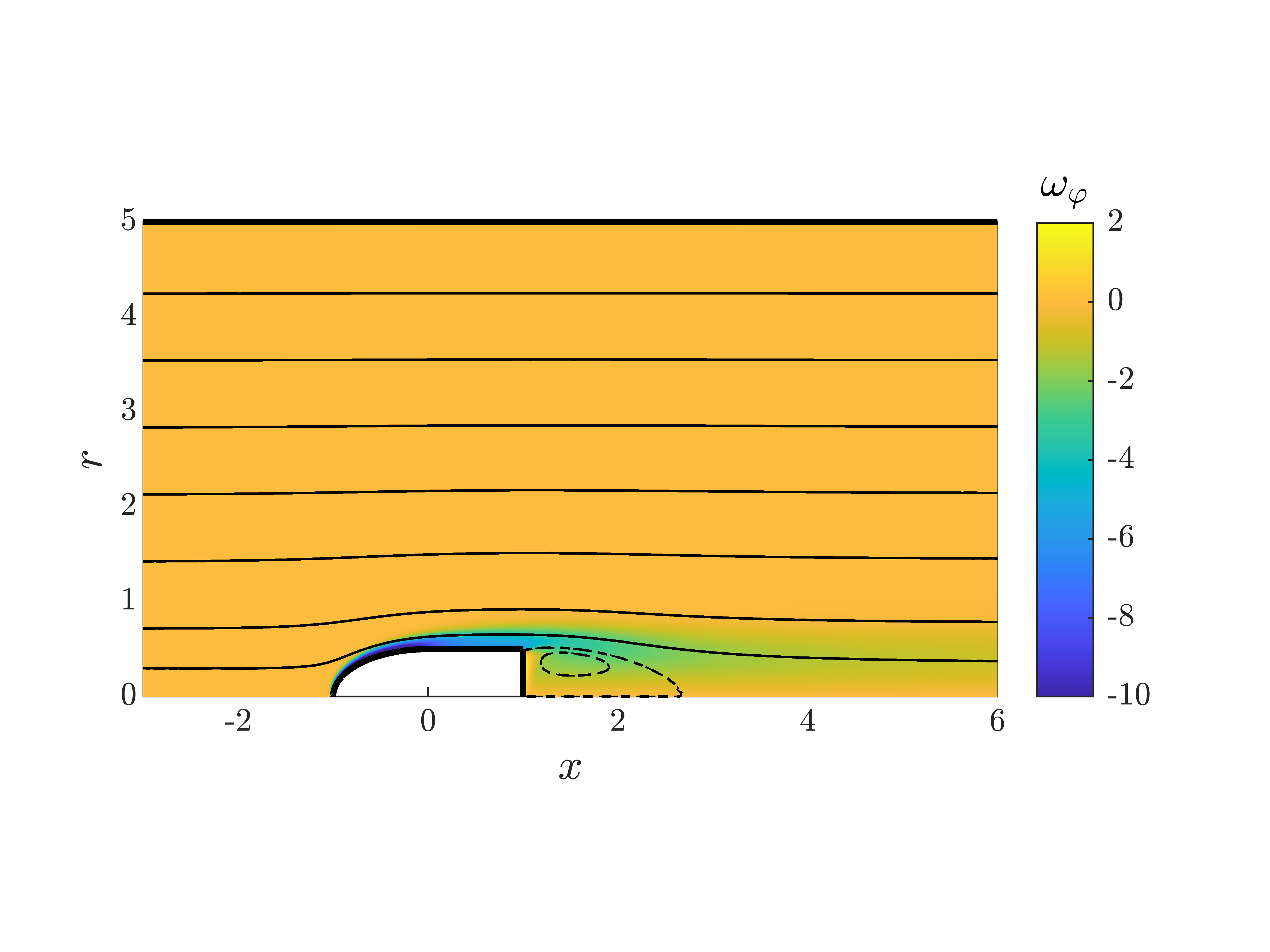}
\caption{Azimuthal vorticity ($\omega_\varphi=\nabla \times \textbf{u}_b \cdot \textbf{e}_\varphi \equiv \partial_x u_r - \partial_r u_x $) 
and streamlines of the base flow for $Re=320$, $L/d=2$, $a/A=0.01$. }
\label{fig:Baseflow_asurA0.1}
\end{figure}

\begin{figure}
\centering
\includegraphics[scale = .8]{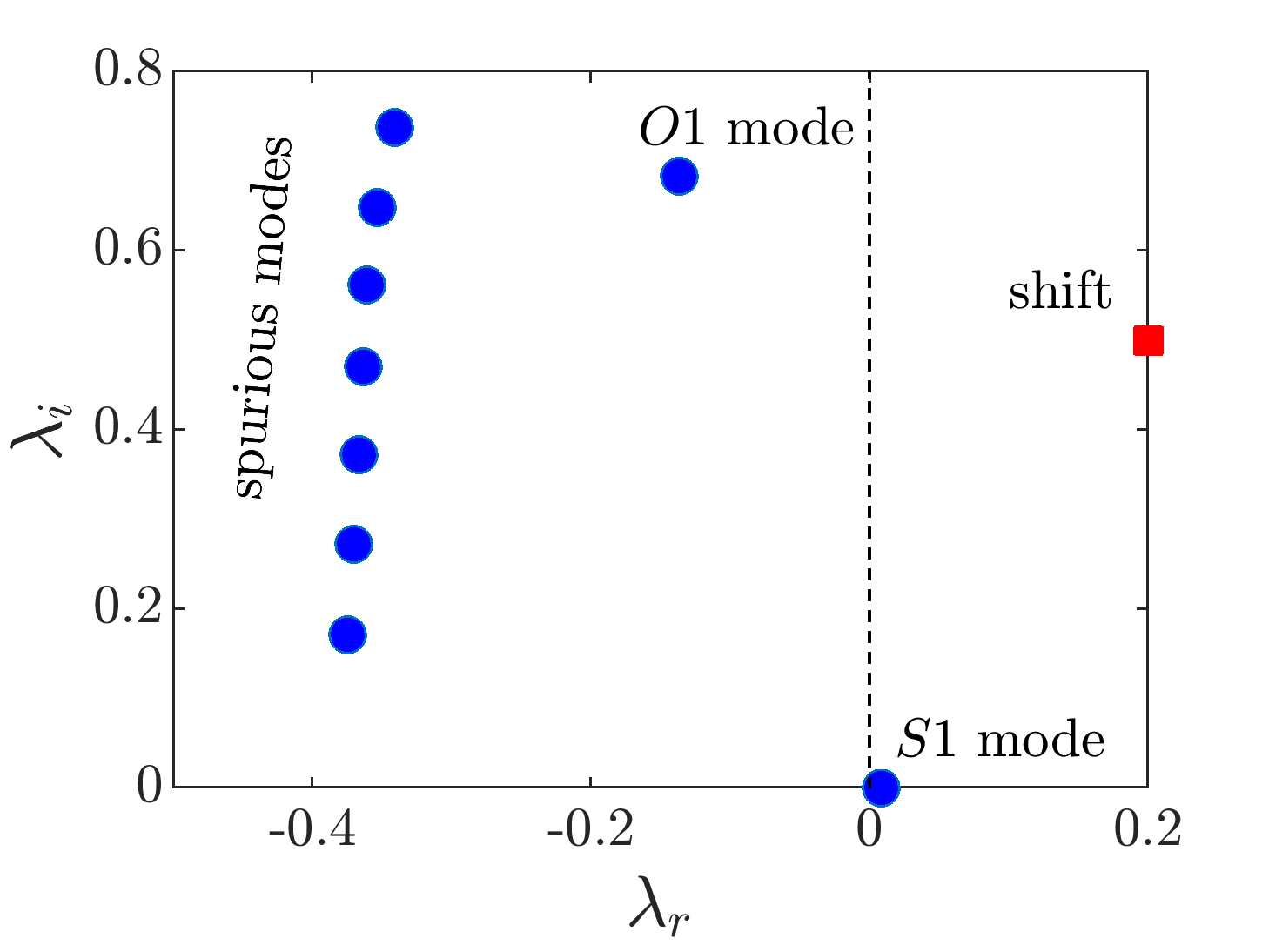}
\caption{Example of a spectrum found with the shift and invert algorithm  ($Re=320$, $L/d$=2, $a/A$=0. 01).
The 10 eigenvalues closest to the "shift" value indicated in red are computed. }
\label{fig:spectrum_asurA0.01_Re320}
\end{figure}

In this section  a weakly confined case is defined with the section aspect ratio $a/A = 0.01$ (or $d/D = 0.1$).
Figure \ref{fig:Baseflow_asurA0.1} represents the base flow around the blunt body for a Reynolds number $Re=320$.
The axisymmetric base flow field exhibits a standing eddy which has approximately the same length as the body itself.
The boundary layer present on the body surface is made visible through generation of negative azimuthal vorticity.
Overall, this structure seems to be very similar to the one found  
in \cite{jimenez2014global} for the same object and conditions in an unconfined flow.

For the same base flow, a part of spectrum found using LSA approach is shown on 
 figure \ref{fig:spectrum_asurA0.01_Re320}.
 It reveals two physical modes, the first one called $S1$ is non-oscillating (often referred as stationary)  and unstable ($\lambda_r>0$).
 The other one called $01$ is oscillating and damped.
 The other modes quasi-aligned  are some spurious modes of non-physical nature and come only from the numerical discretization.
 Similar results are observed by \cite{jimenez2014global} for a non-spinning object in unconfined space.
  In their study, the onset of the first instability (the $S1$ mode) was detected at a critical Reynolds number $Re_{c,S1}=325.21$
 whereas its value is  $Re_{c,S1}=312.21$ in the present study, leading to a less than a four percents difference.
 It can be concluded that the confinement produces a small influence over the onset of the first instability  in this case.

\begin{figure}
\begin{center}
\subfloat[]{\includegraphics[scale = .8]{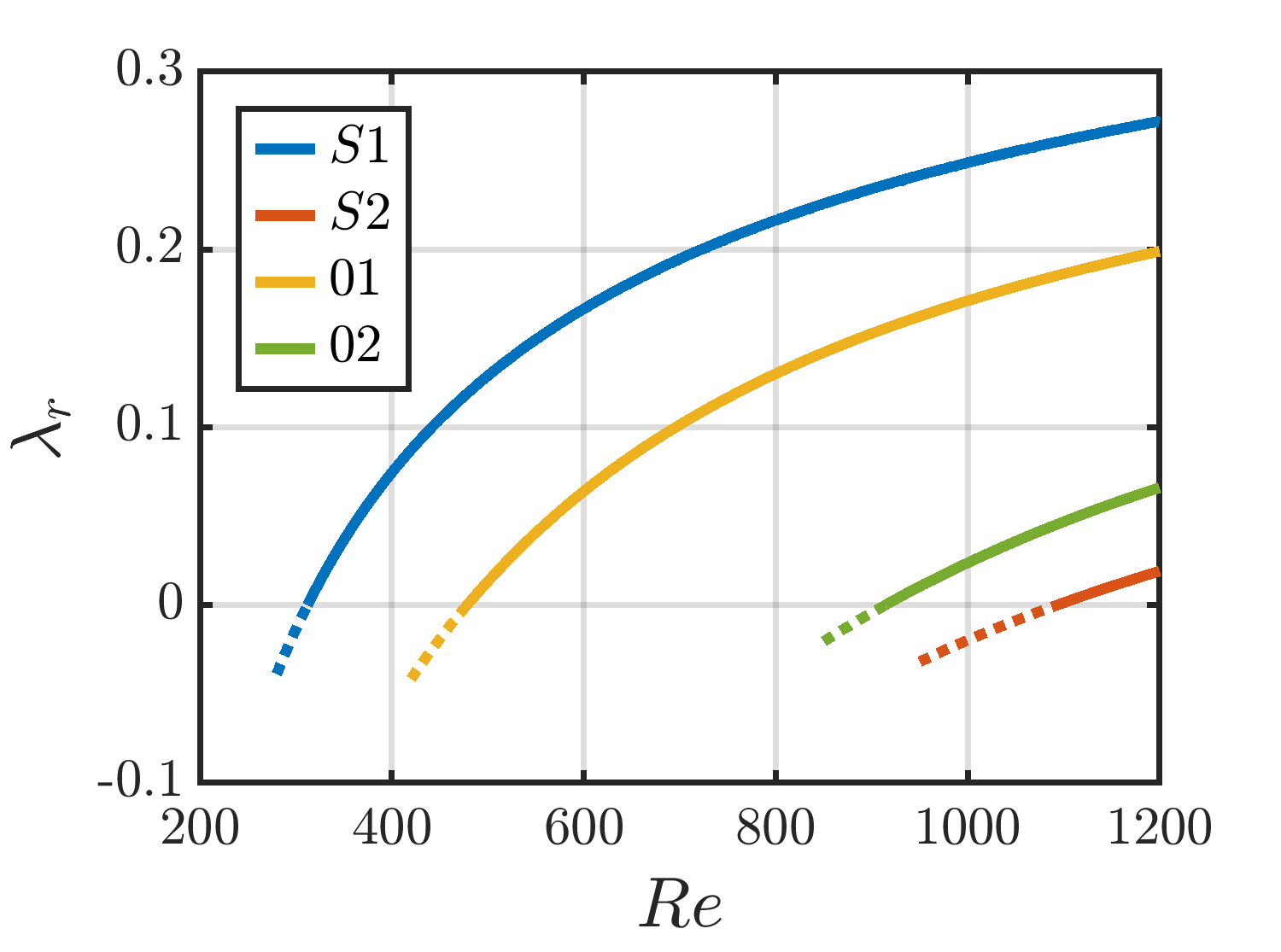}} 
\subfloat[]{\includegraphics[scale = .8]{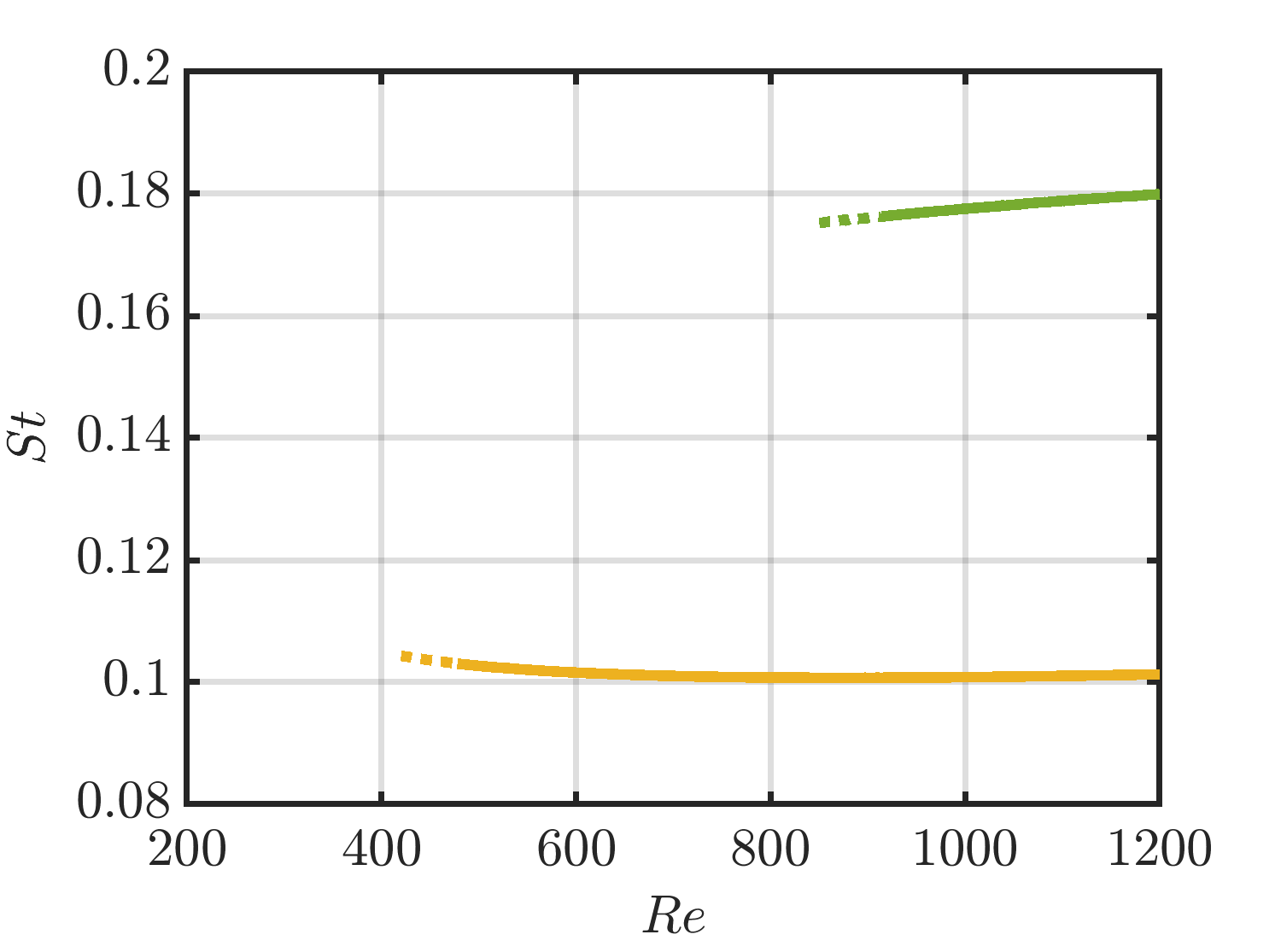}}
\end{center}
\caption{ amplification rate $(a)$ and Strouhal number \textit{(b)} as a function of the Reynolds number for the first unstable modes, $a/A=0.01$, $L/d=2$. }
\label{fig:asurA0.1_lambda}
\end{figure}

Figure \ref{fig:asurA0.1_lambda} displays the four most amplified eigenvalues as function of  $Re$, again for $a/A=0.01$. The first unstable mode appearing is non-oscillating 
and remains the most amplified  mode over the whole range of $Re$ studied.
The second most amplified mode, $O1$, becomes unstable at $Re_{c,O1}=478.26$
and $St_c=0.103$. This  Strouhal value is very close to the one found by \cite{bohorquez2011stability} who reported $St=0.102$. 
But the latter authors found a somehow larger value of the critical Reynolds number,
namely $Re_{c,O1} = 518$.

In addition to the effect of confinement,
this gap between critical Reynolds number may be explained by the fact that the computational domain defined for the stability analysis included 
only the cylindrical rear of the body and excluded the nose  in \cite{bohorquez2011stability}. 
The critical Reynolds is also notably higher than the one given by the reference case of a thin disk.
The geometry of the nose of the blunt body changes the amount of vorticity produced at its surface, as pointed out by \cite{brucker2001spatio}, and it is known that this vorticity production is responsible for triggering the instabilities \citep{magnaudet2007wake}. 
Having a profiled nose diminishes such production of vorticity and pushes back the onset of the $O1$ mode from $Re_{c,O1}=125.3$ (for a thin disk in an open flow \cite{meliga2009unsteadiness}) to $Re_{c,O1}=478.26$ in the present  case. 


Up to here, the main scenario revealed in the present configuration by the linear stability analysis is  a first non-oscillating mode $S1$ amplification followed by an oscillating one $O1$.
It is the same encountered for all axisymmetric bodies considered in literature \citep{natarajan1993instability, meliga2009unsteadiness,tchoufag2014global}. 
When pushing the Reynolds number towards higher values, two additional modes are found, an oscillating one and a non-oscillating one termed $O2$ and $S2$.
These higher modes arise at much larger Reynolds numbers,
in the range $Re\approx 1000$.
They are less likely to be observed experimentally or numerically because in such regimes the mean flow is already very far from the axisymmetric base-flow analyzes with the 
linear stability theory.
Nevertheless, when the confinement effect will be increased, 
these higher modes will turn to be relevant to obtain a consistent picture of the bifurcation scenario. 
Hence they will be kept in the analysis and their critical Reynolds $Re_{c,S2}$ and 
$Re_{c,O2}$ will be tracked.




\begin{figure}
\centering
\begin{tabular}{c l}
\multicolumn{2}{l}{$S1$, $Re=320$}\\
\includegraphics[scale = .8,trim= 5 30 25 35  ,clip]{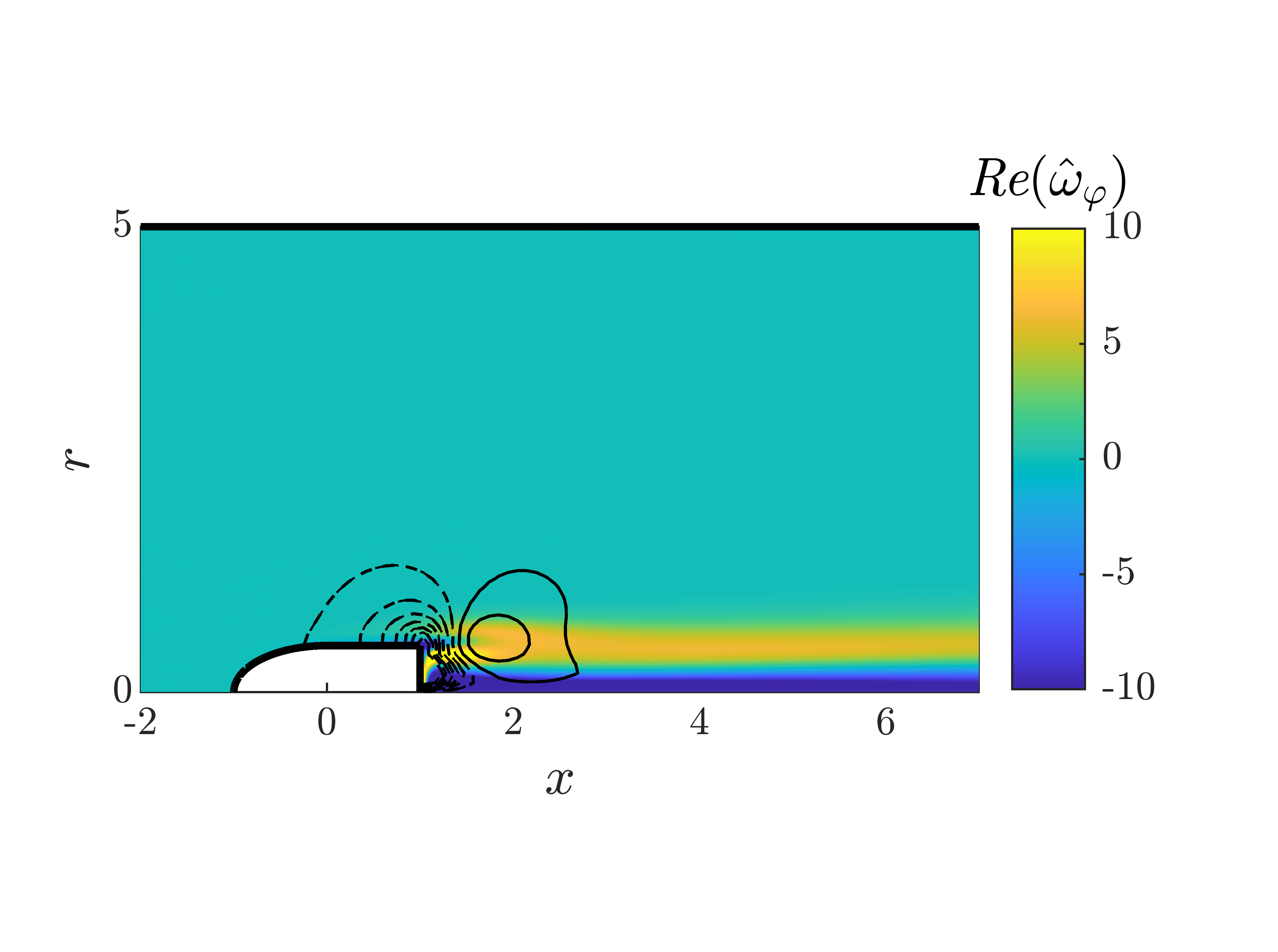} &
    \hspace{-0.5em}\includegraphics[scale = .8,trim=5 5 0 0 ,clip]{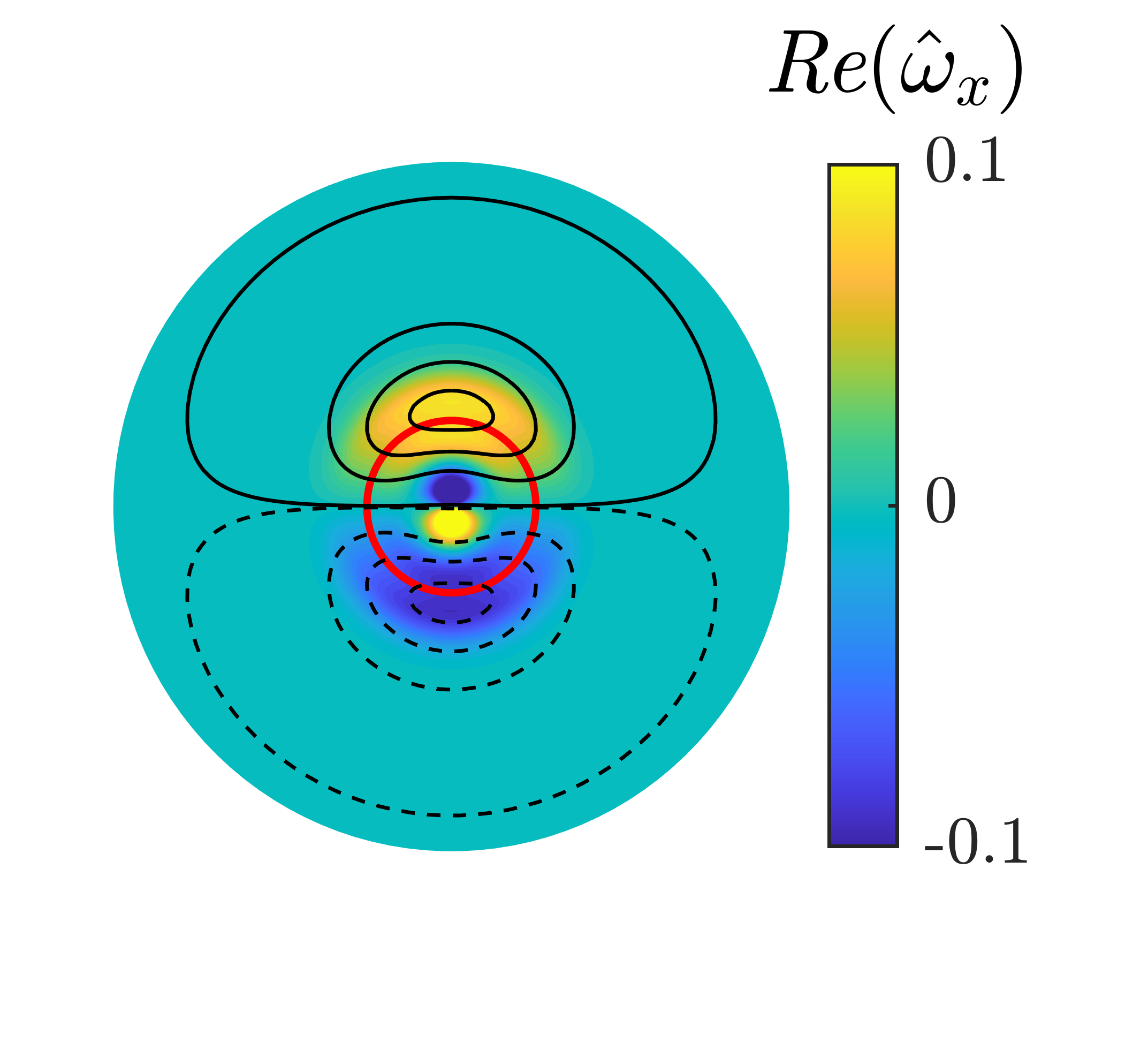}  \\
\multicolumn{2}{l}{$S2$, $Re=1130$}\\
\includegraphics[scale = .8,trim= 5 30 25 35  ,clip]{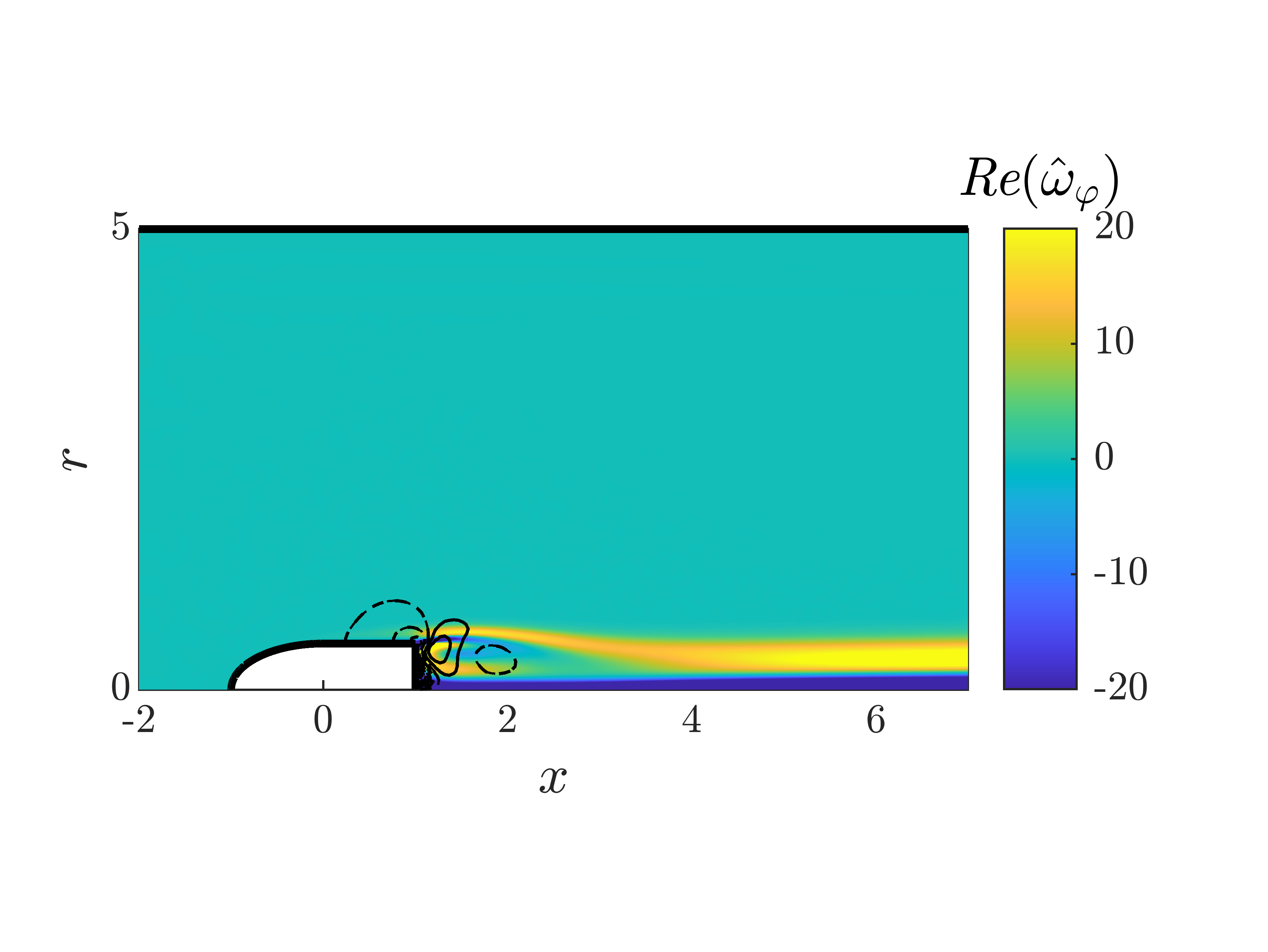} &
\hspace{-0.5em}\includegraphics[scale = .8,trim=5 5 0 0 ,clip]{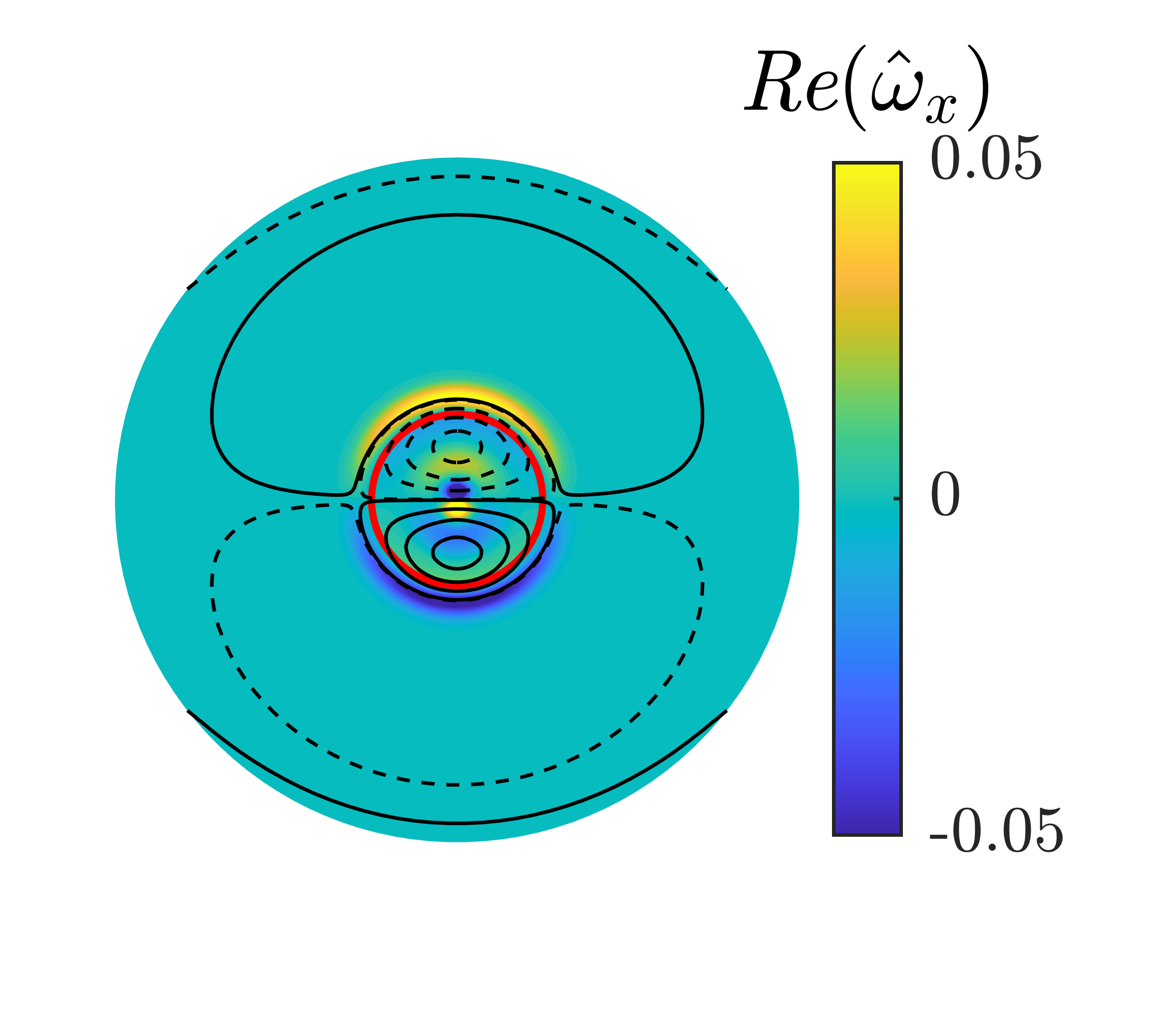}  \\
\multicolumn{2}{l}{$O1$, $Re=500$}\\
\includegraphics[scale = .8,trim= 5 30 25 59  ,clip]{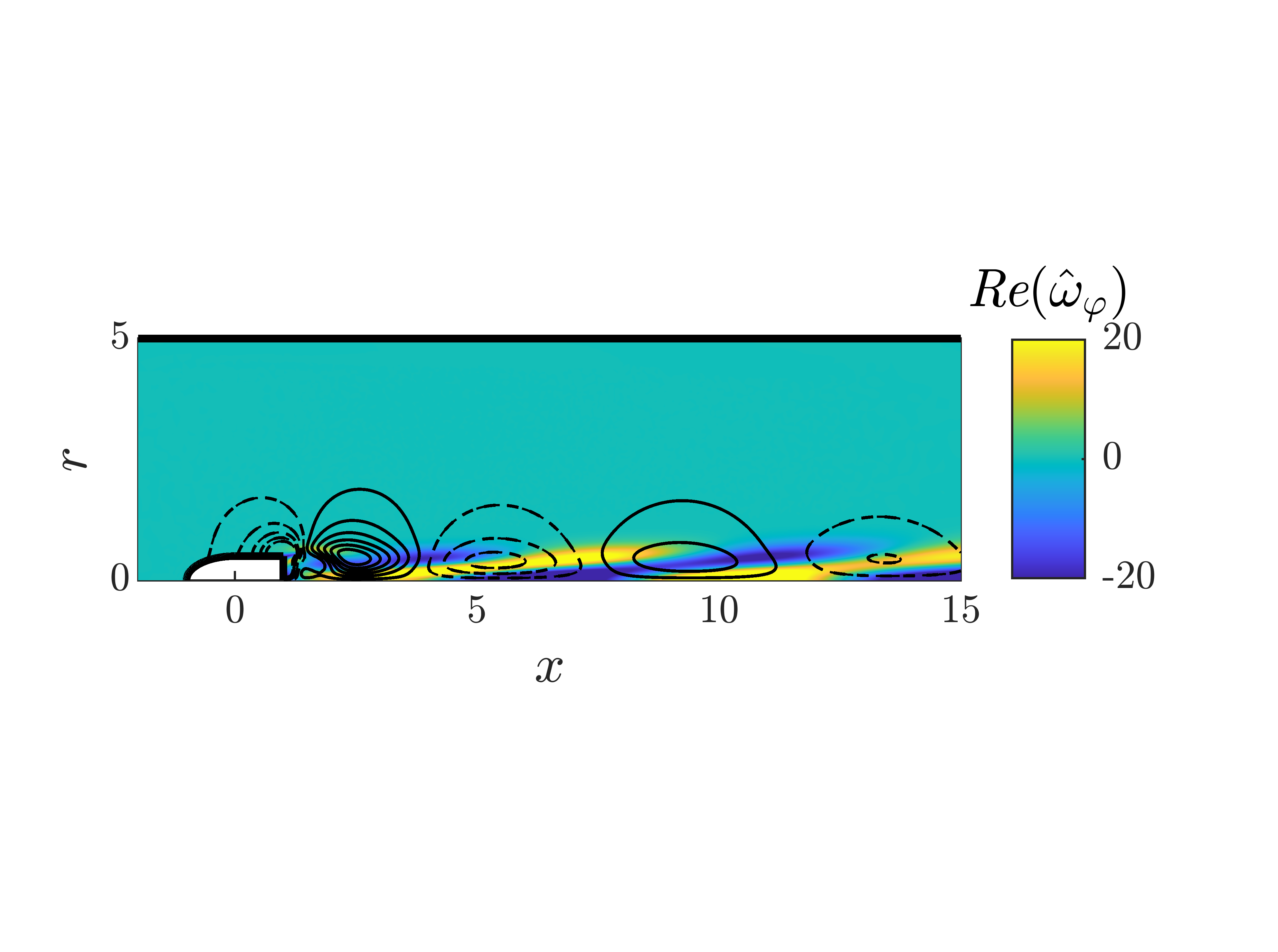} &
    \hspace{-0.5em}\includegraphics[scale = .8,trim=5 5 0 0 ,clip]{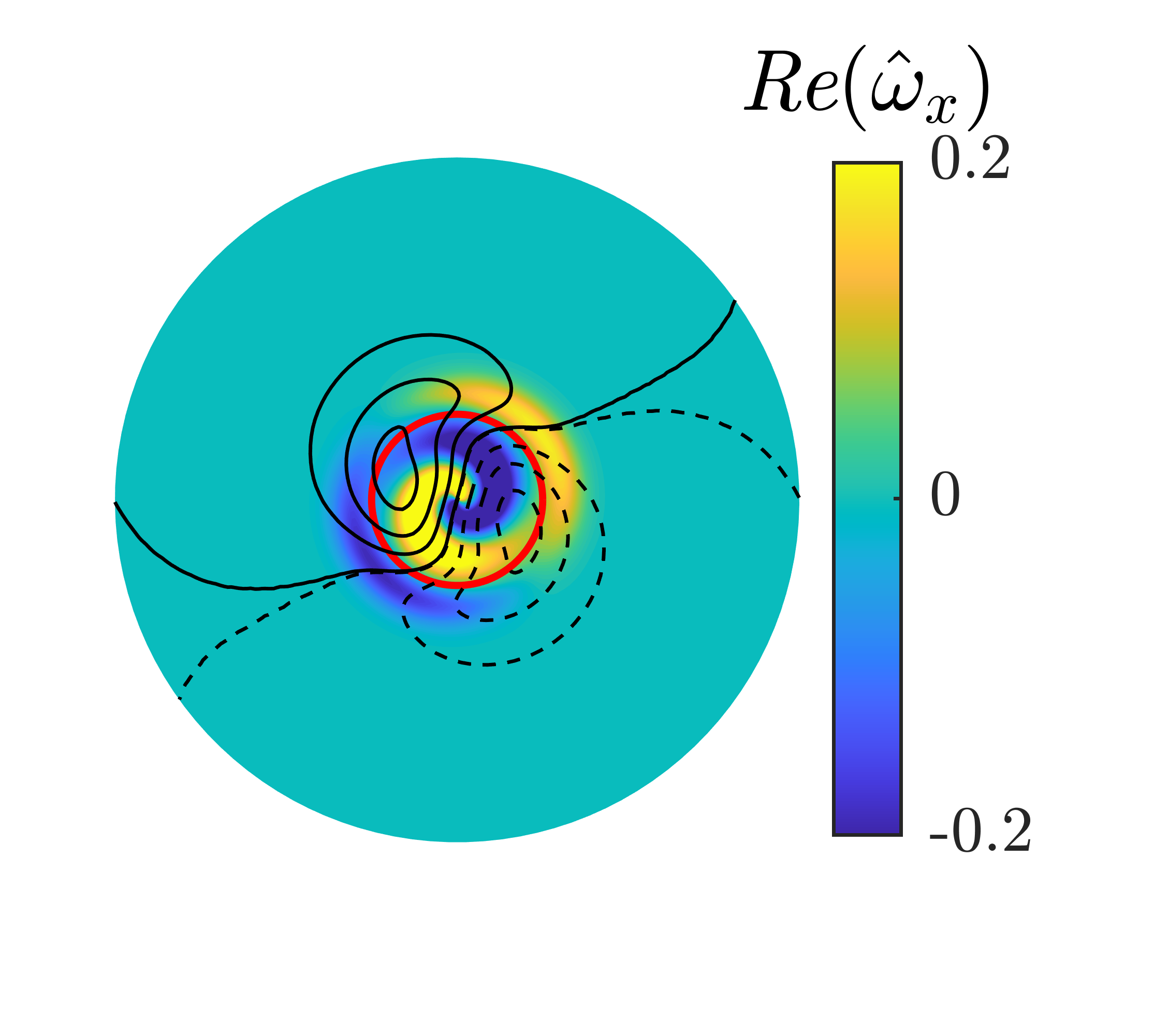}  \\
\multicolumn{2}{l}{$O2$, $Re=950$}\\
\includegraphics[scale = .8,trim= 5 30 25 59 ,clip]{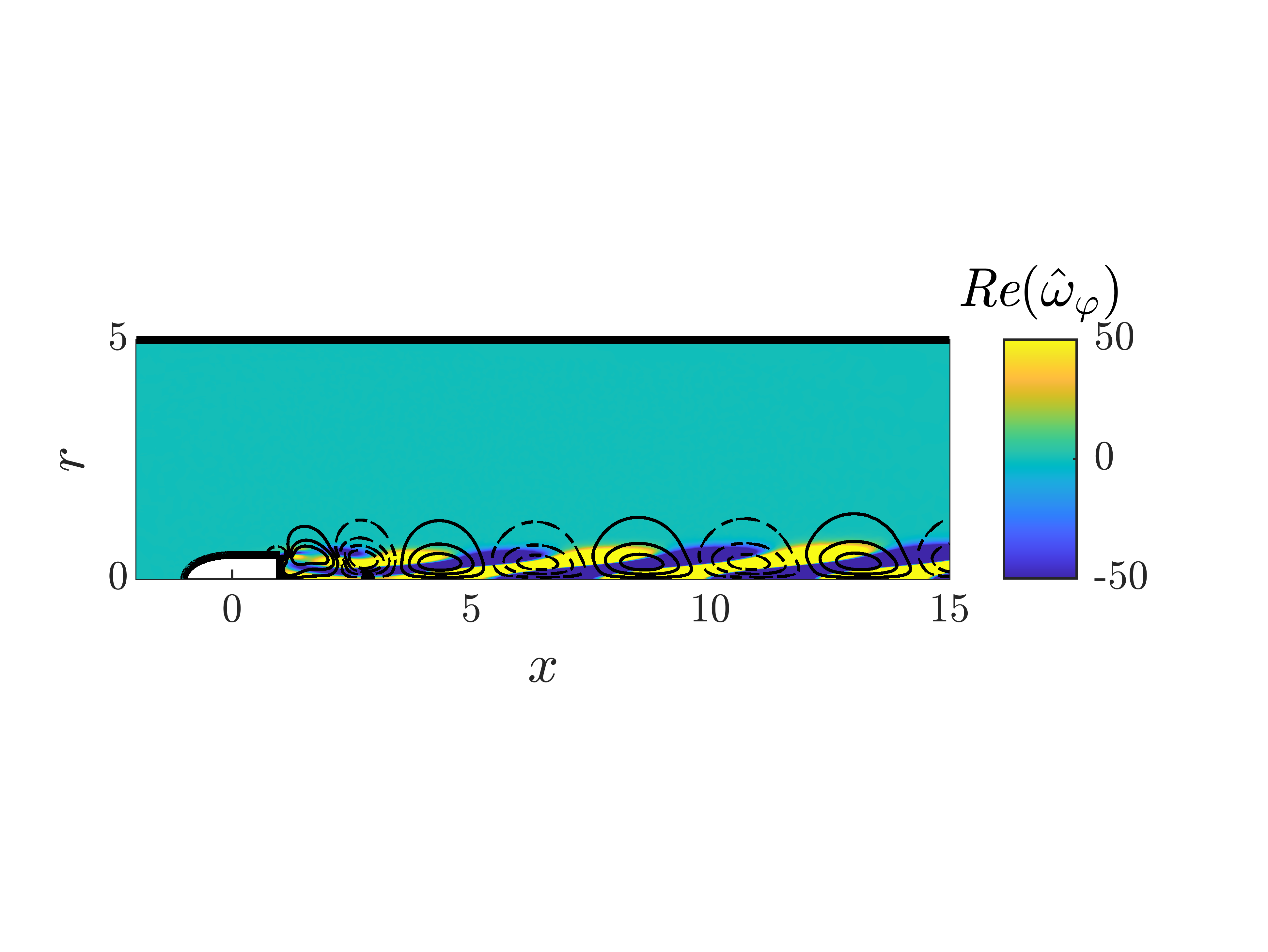} &
\hspace{-0.5em}\includegraphics[scale = .8,trim=5 5 0 0 ,clip]{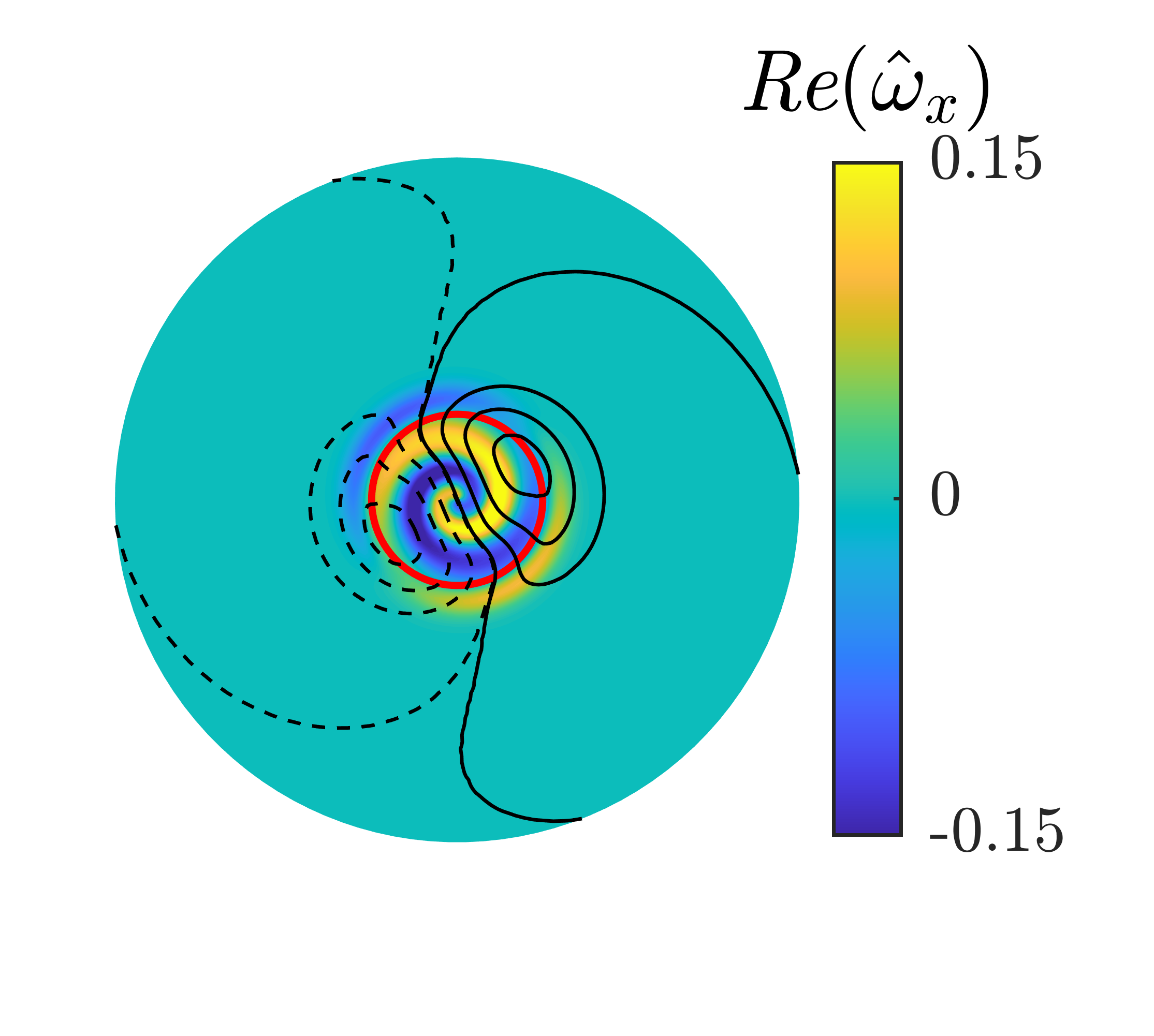}  
\end{tabular}
\caption{ Eigenmodes found for $a/A=0.01$, real parts of the vorticity with iso-levels of pressure. Slices are defined by $x=2$ and $r\le 2$.}
\label{fig:modes_asurA0.01}
\end{figure}

Figure \ref{fig:modes_asurA0.01} illustrates the structures of those modes. 
The plots display both the azimuthal velocity (colors) and pressure (lines) levels in a meridional $(x,r)$-plane (left plots) and the axial vorticity levels in a transverse $(y,z)$-plane corresponding to a cut at location $x=2$ (right plots).
The first unstable mode, $S1$, has a rather simple vorticity structure.
In the meridional plane, the azimuthal vorticity of the eigenmode is positive
in the region of the shear layer. Recalling that the vorticity of the base flow shear layer is negative (see \ref{fig:Baseflow_asurA0.1}),
the effect of the eigenmode is to decrease the net vortical intensity in the shear layer  in this region. 
Owing to the antisymmetry of $m=\pm 1$ modes, the azimuthal vorticity of the eigenmode is negative in the opposite side, meaning that the shear layer is enhanced. The axial vorticity, on the other hand, reveals a pair of counter-rotating streamwise vortices, as already noticed for disk and spheres \citep{meliga2009unsteadiness,natarajan1993instability} and similar blunt bodies in open flow \citep{jimenez2014global}. 
The $S2$ mode shows a similar structure with 
a small additional pair of counter-rotating vortices in the vicinity of the blunt body rear, and a more complex pressure field than the $S1$ mode. 

Considering now the oscillating modes $O1$ and $O2$ displayed in the bottom part of figure \ref{fig:modes_asurA0.01}. One  can observe in an azimutal plane an alternance of positive and negative streamwise vorticity which is the signature of unsteady vortex shedding.
A shorter spatial wake length scale can be seen for the $O2$ mode, 
related to its  higher shedding frequency (\textit{i.e.} larger $St$). 
Note that when observed in a transverse plane, these mode display a characteristic spiral structure. This does not imply that if these modes are present in a nonlinear solution, a spiral structure will necessarily be observed. Indeed, it is known that due to the degeneracy associated to mirror symmetry, $m=+1$ and $m=-1$ eigenmodes are mirror-images of each other and can lead to two kind of nonlinear states \citep{fabre2008bifurcations} : $(i)$ "rotating waves" corresponding to a pure $m=+1$ (or $m=-1$) eigenmodes with a spiral structure, and "standing waves" corresponding to superposition of $m=1$ and $m=-1$ eigenmodes characterized by a symmetry plane. 

\subsubsection{Confined flow with $a/A=0.75$}


\begin{figure}
\centering
\includegraphics[scale = .8, trim= 0 142 0 120 ,clip]{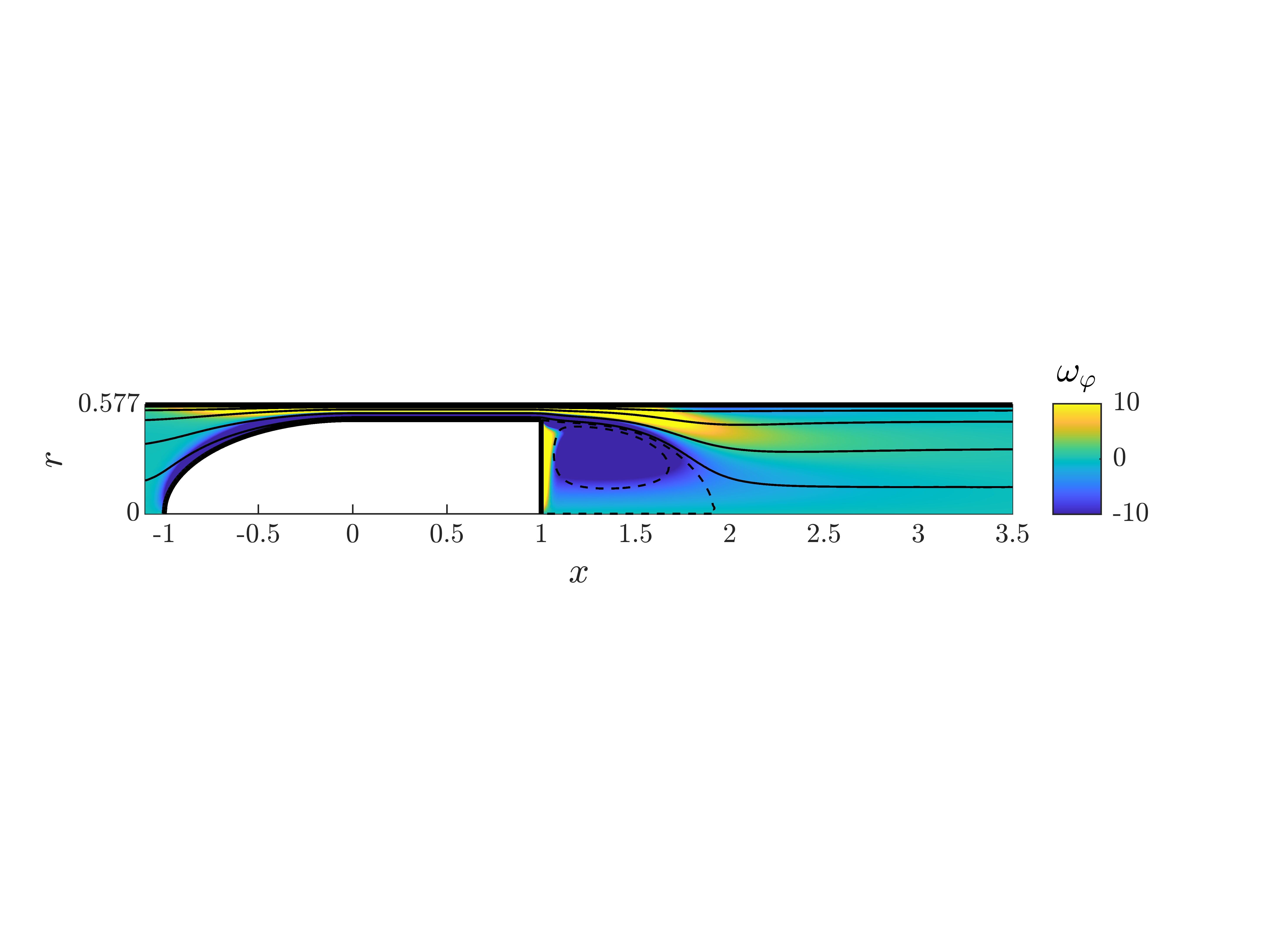}
\vspace{1em}
\\
\includegraphics[scale = .8, trim= 0 120 0 130. ,clip]{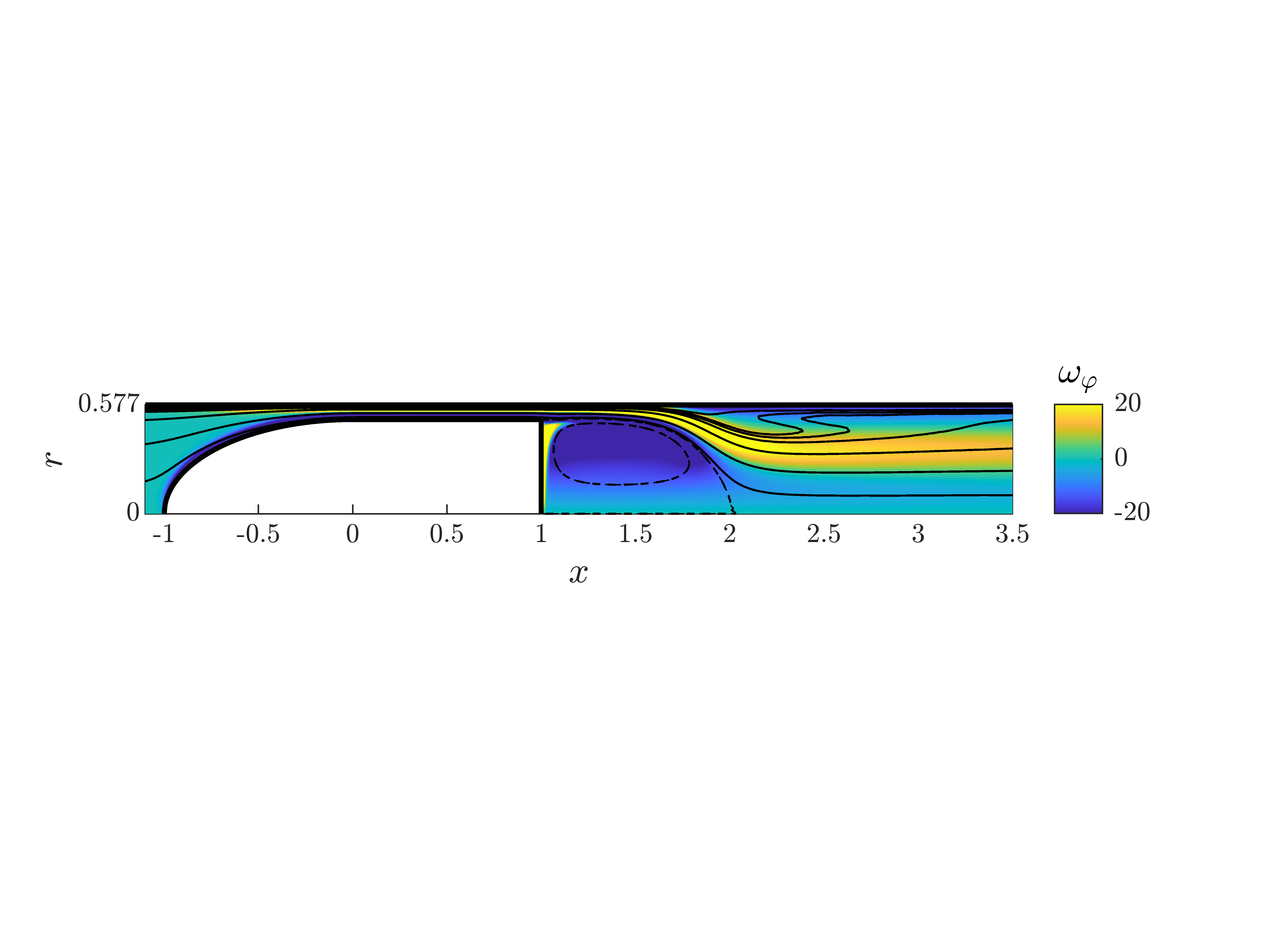}
\caption{azimuthal vorticity and streamlines of the base flow in the moving frame attached to the body for $L/d=2$ and $a/A=0.75$. Top, $Re=110$, bottom $Re=320$  }
\label{fig:Baseflow_asurA_0.75}
\end{figure}

\begin{figure}[t]
\centering
\includegraphics[scale =.5, clip]{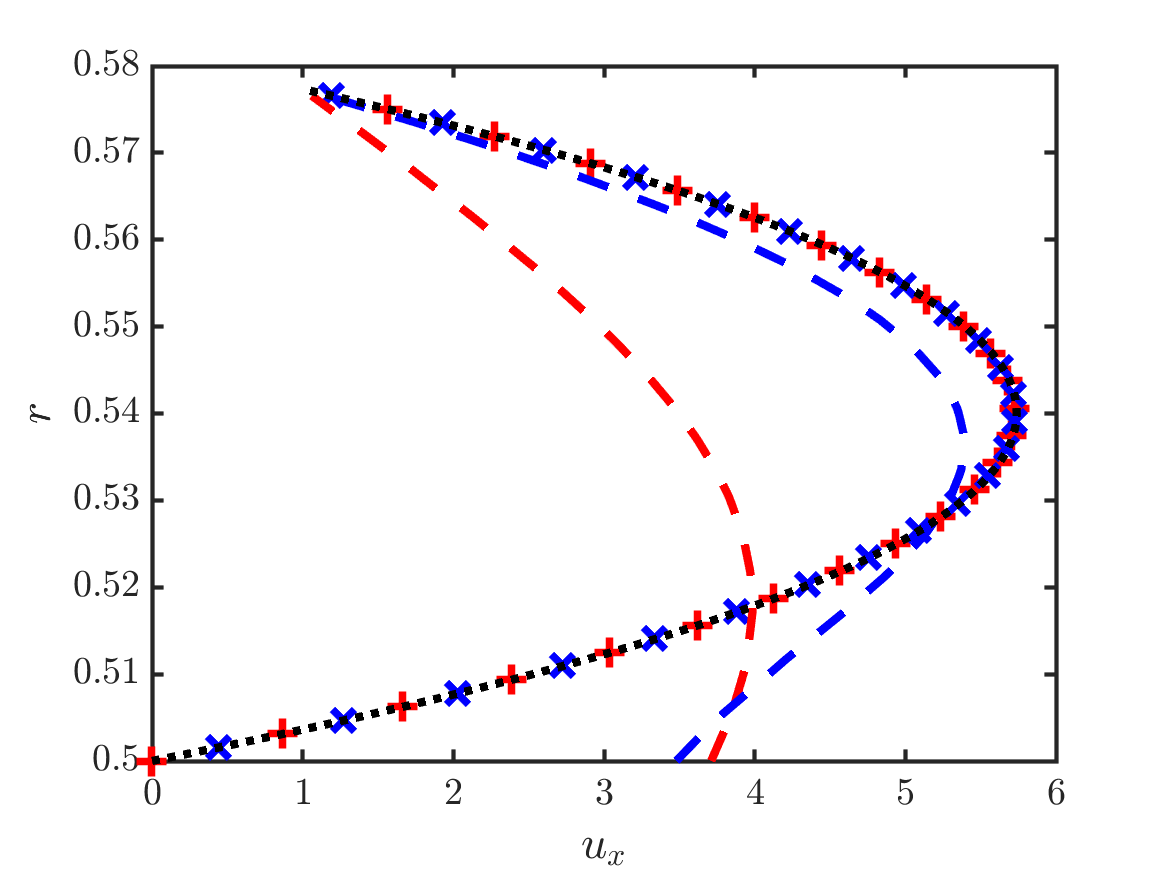}
\caption{ Axial velocity $u_x$ as function to $r$ for $Re=110$ (red) and $Re = 320$ (blue) at locations $x=0.75$ (symbols) and $x=1.25$ (dashed lines); Comparison with annular Poiseuille solution (black dotted line). }
\label{fig:CouettePoiseuille}
\end{figure}

Let us now consider a more confined flow with a section ratio of  $a/A = 0.75$ or a diameter ratio of $d/D = 0.87$.
The structure of the base flow and the influence of the tube wall are displayed in
Figure \ref{fig:Baseflow_asurA_0.75}. 
Compared to the unconfined of weakly confined flow, 
the recirculation length is shorter as the confinement becomes stronger.
For Reyndols number $Re=320$, the flow changes direction close to the pipe wall and goes downstream but without setting a closed recirculation zone attached to the wall, even for large values $Re>320$. The presence of separation  in this area is accompanied with a production of negative vorticity.
This structure reveals the presence of a confined wall jet.
Within the small gap between the body and pipe walls the flow 
can be seen as parallel ad,  one may expect the flow to be well approximated by a  parallel-flow solution called "annular Couette-Poiseuille flow".
This classical solution is reproduced in appendix \ref{appendixA}. 
The theoretical analytic nondimensional velocity profile is compared in figure \ref{fig:CouettePoiseuille} to 
the actual axial flow profile extracted from the base-flows represented in figure \ref{fig:Baseflow_asurA_0.75}. 
At location $x= 0.75$ (in the rear part of the afterbody), the observed velocity profile is 
indistinguishable from the theoretical solution, both for Reynolds numbers $Re=110$ and $320$.

The velocity profile at location $x=1.25$, slightly behind the body is also plotted in the figure. The curves show that the velocity profile turns into an annular jet, affected by some diffusion, especially for $Re =110$. 



\begin{figure}
\begin{center}
\subfloat[]{\includegraphics[scale = .8]{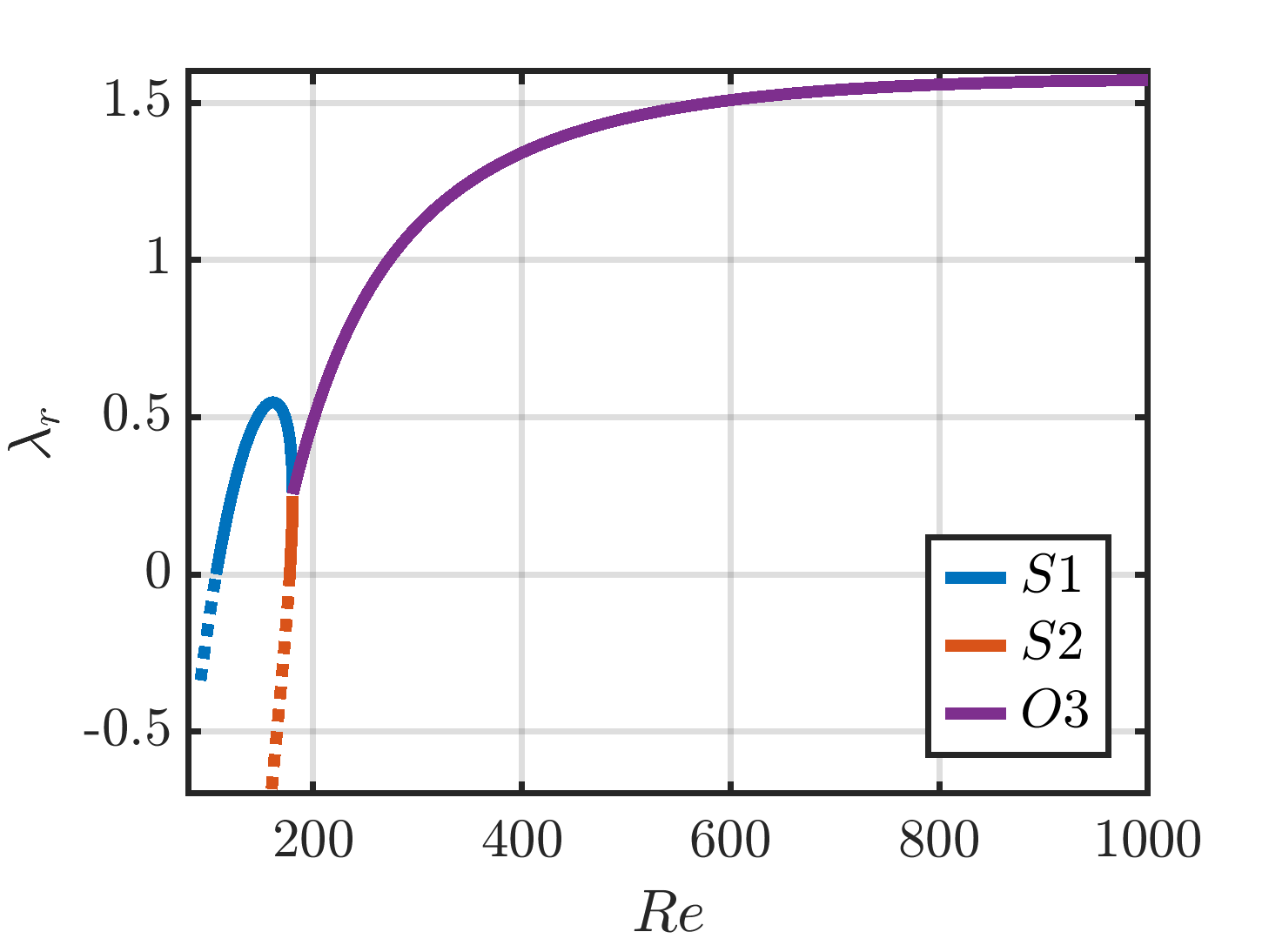}} 
\subfloat[]{\includegraphics[scale = .8]{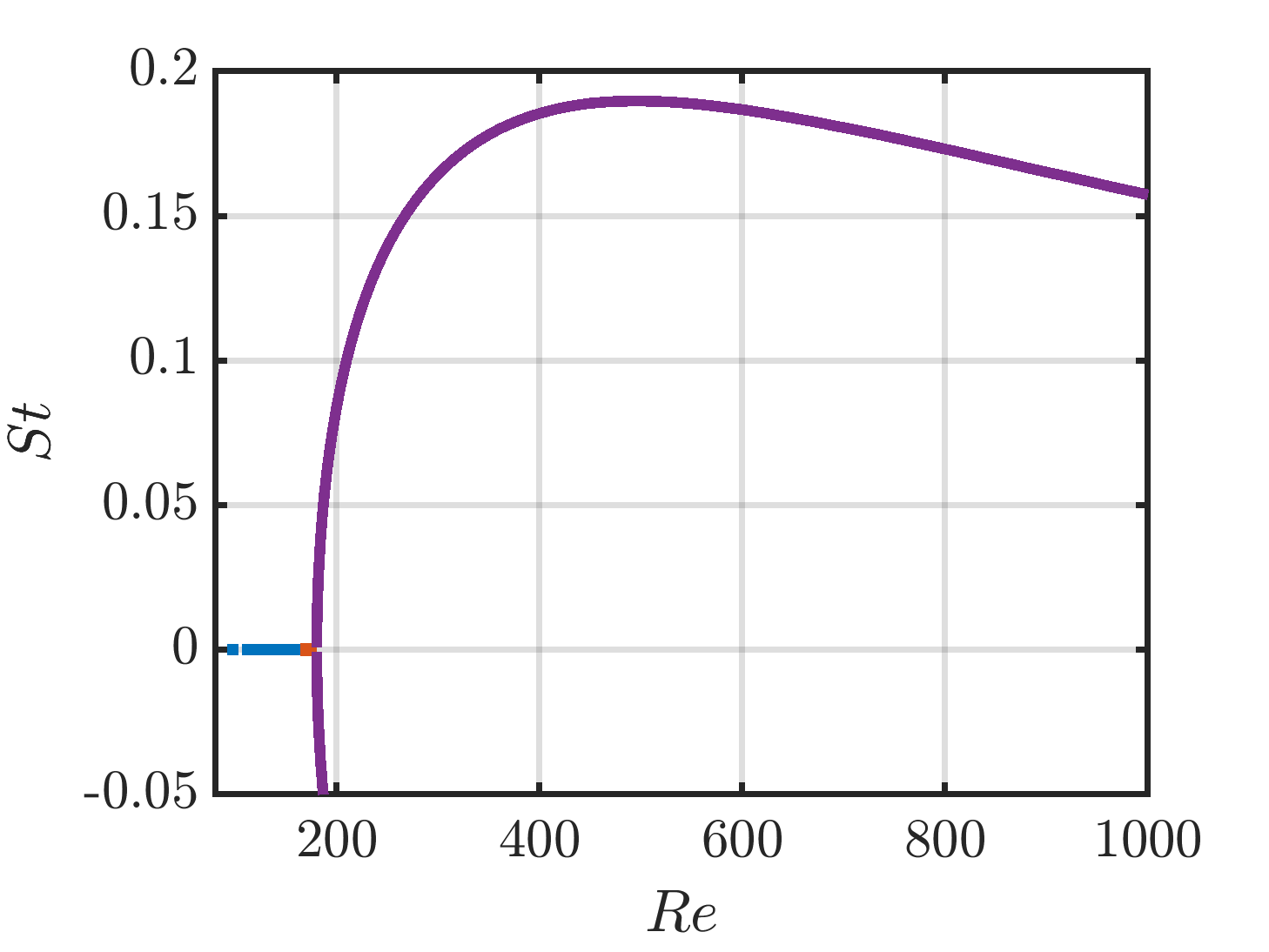}}
\end{center}
\caption{ amplification rate $\lambda_r$ $(a)$ and Strouhal number \textit{(b)} versus  the Reynolds number for the first unstable modes, $a/A=0.75$. }
\label{fig:asurA0.75_lambda}
\end{figure}






    

\begin{figure}
\centering
\begin{tabular}{c l}
\multicolumn{2}{c}{$S1$, $Re=110$}\\
\includegraphics[scale = .8,trim=0 106 0 120
,clip]{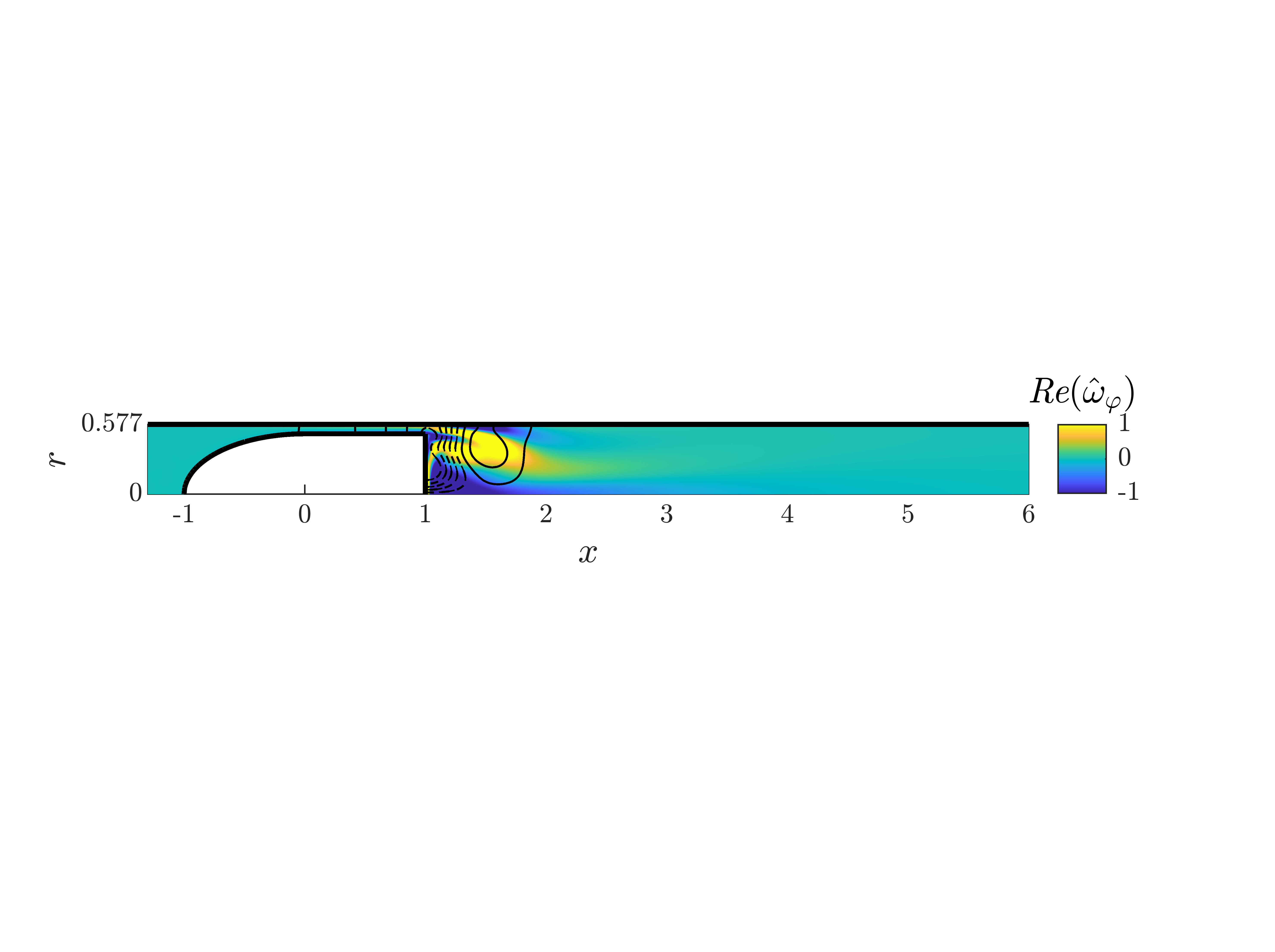} &
    \hspace{-4.em}\includegraphics[scale = .8,trim=10 5 0 0 ,clip]{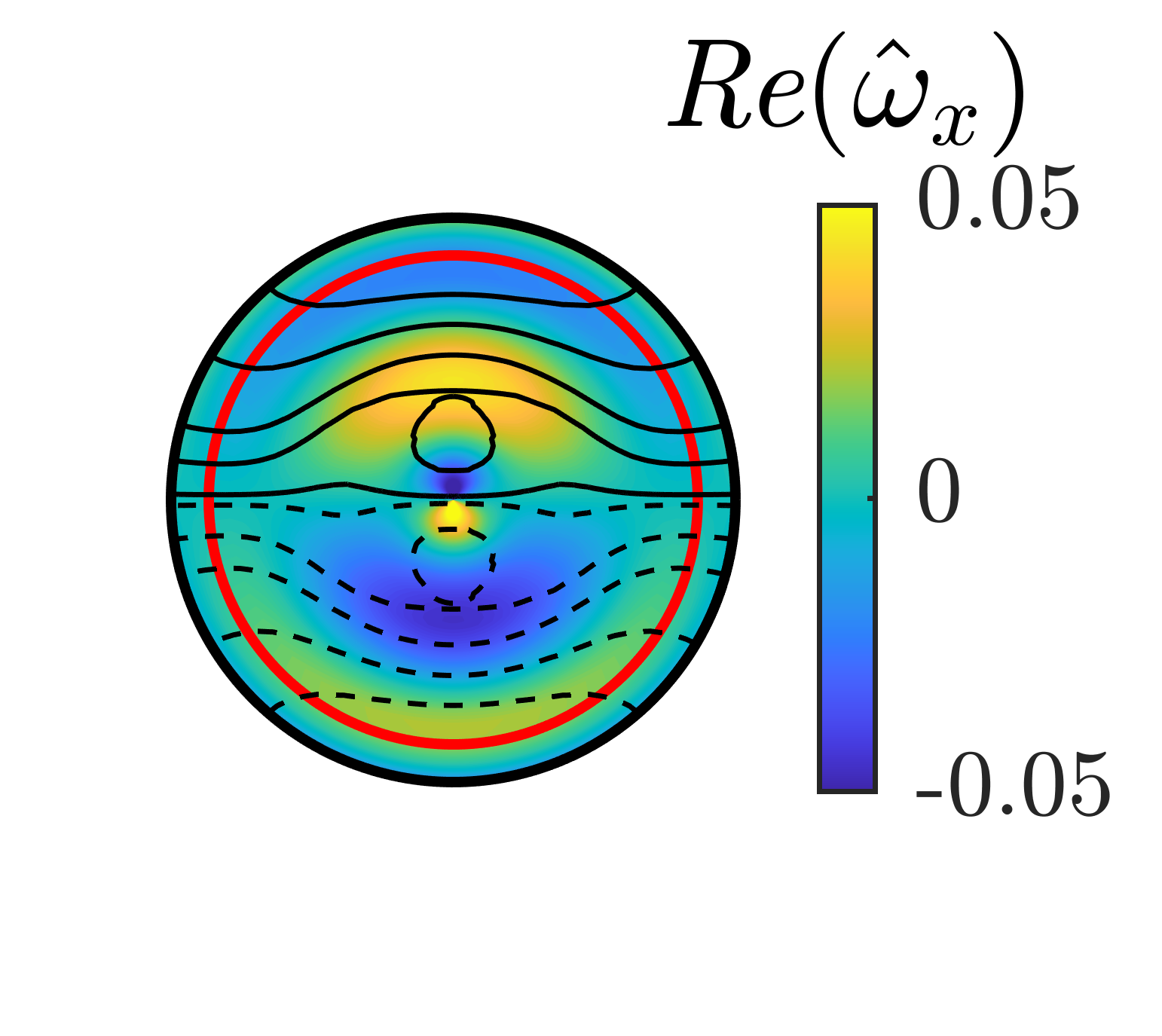}  \\
\multicolumn{2}{c}{$S2$, $Re=117$}\\
\includegraphics[scale = .8,trim=0 106 0 120 ,clip]{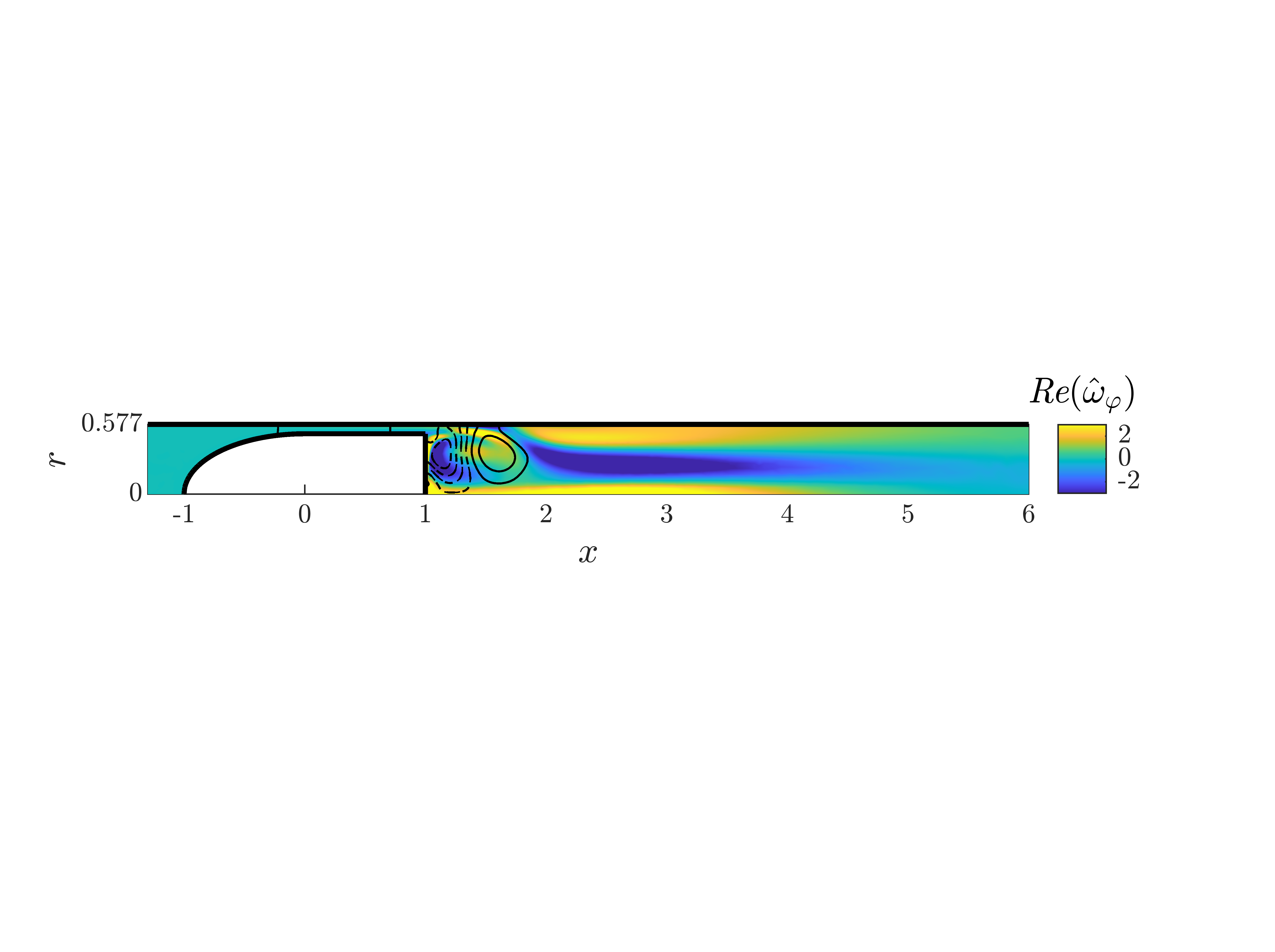} &
\hspace{-4.em}\includegraphics[scale = .8,trim=10 5 0 0 ,clip]{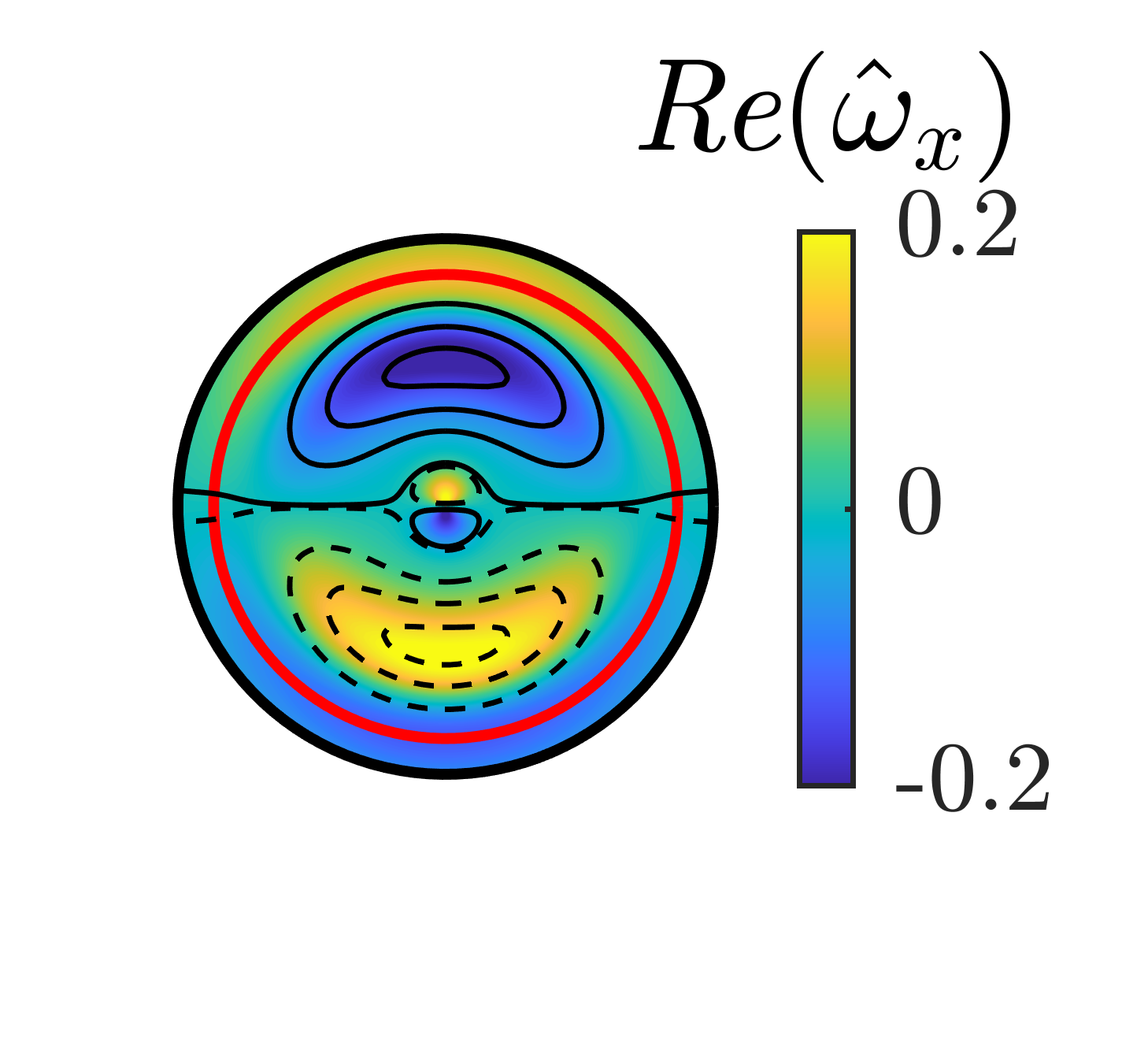}  \\
\multicolumn{2}{c}{$O3$, $Re=200$}\\
\includegraphics[scale = .8,trim=0 106 0 120 ,clip]{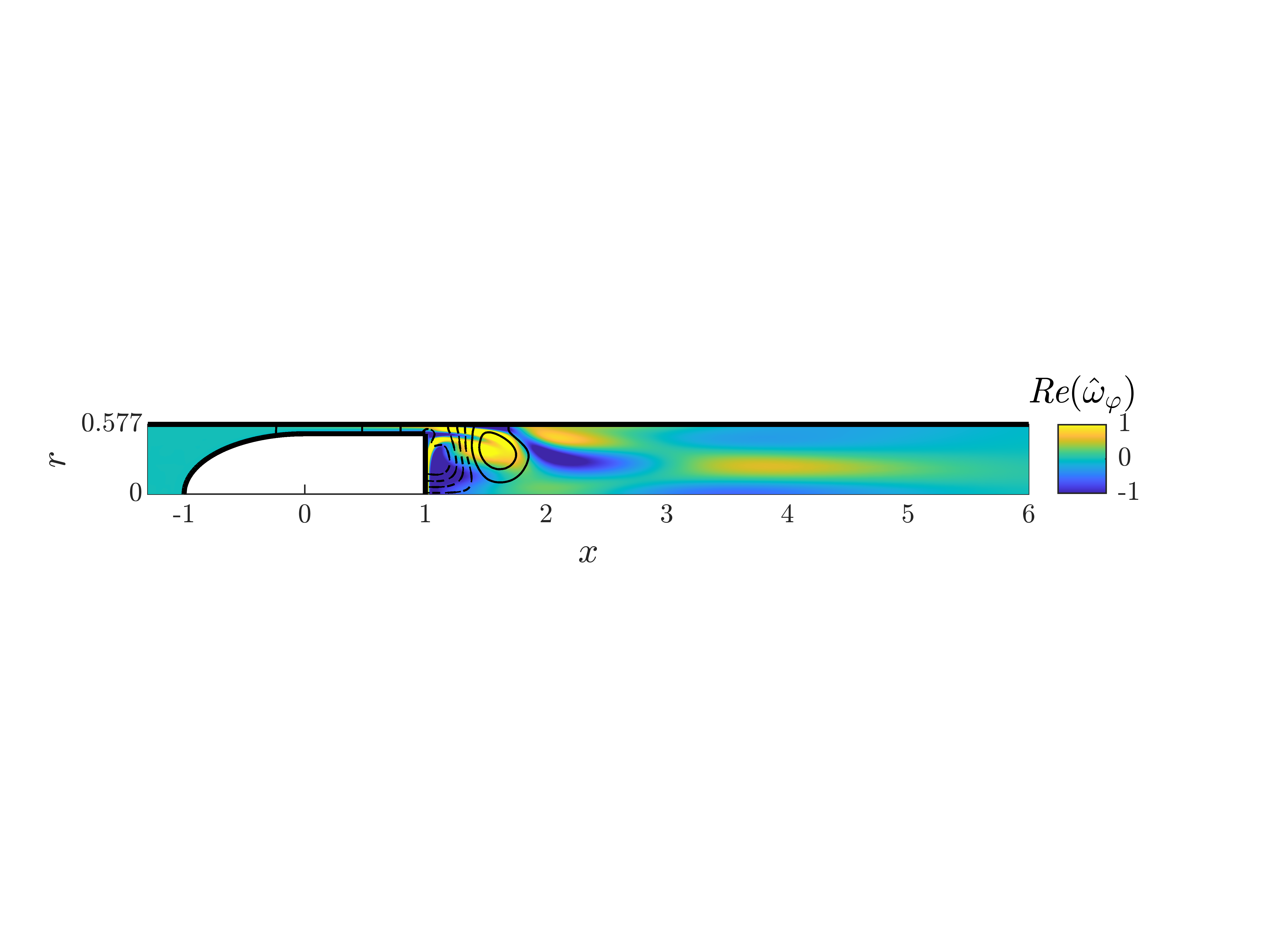} &
\hspace{-4.em}\includegraphics[scale = .8,trim=10 5 0 0 ,clip]{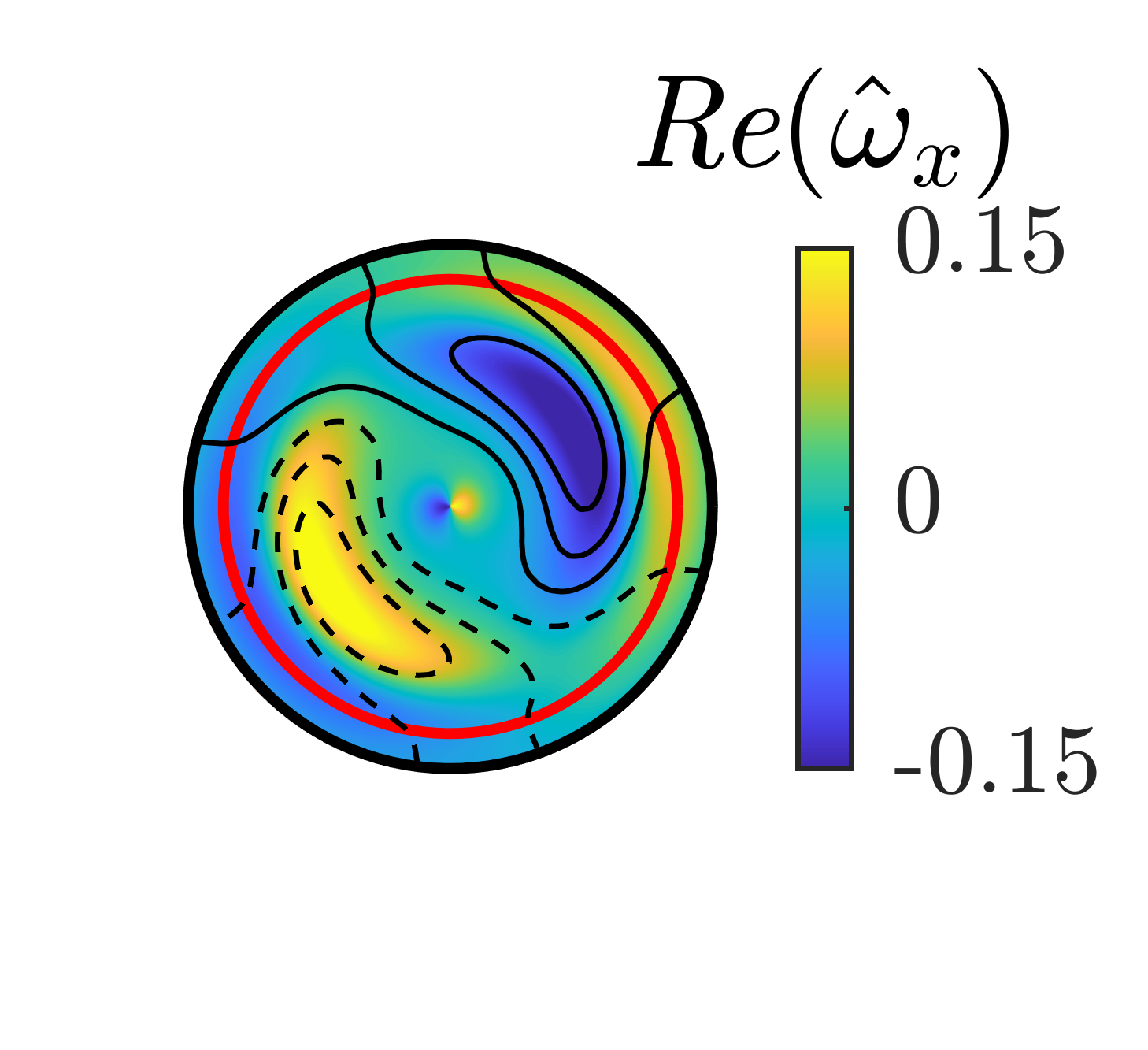}  \\    
\multicolumn{2}{c}{$O3$, $Re=400$}\\
\includegraphics[scale = .8,trim=0 106 0 120 ,clip]{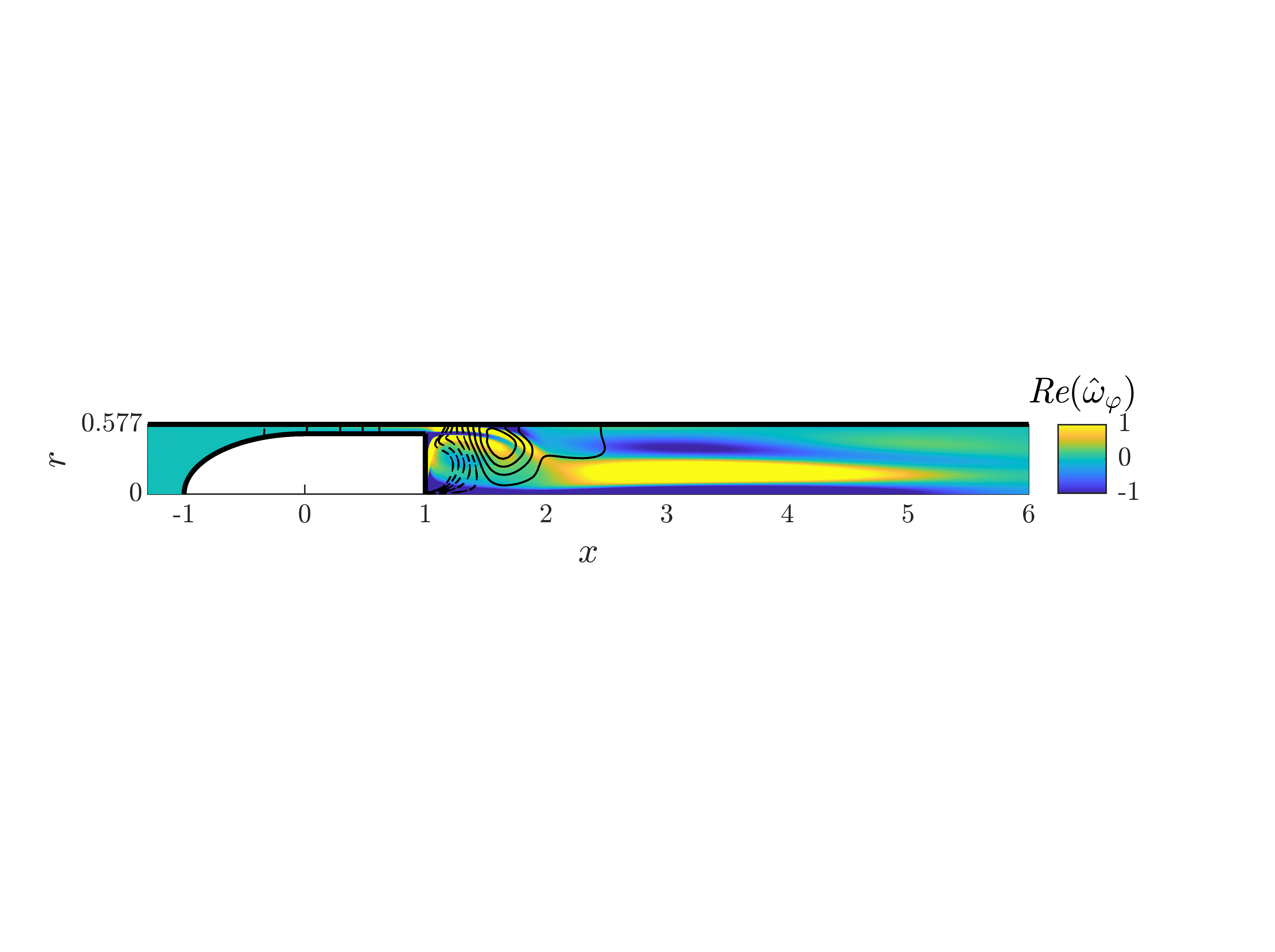} &
\hspace{-4.em}\includegraphics[scale = .8,trim=10 5 0 0 ,clip]{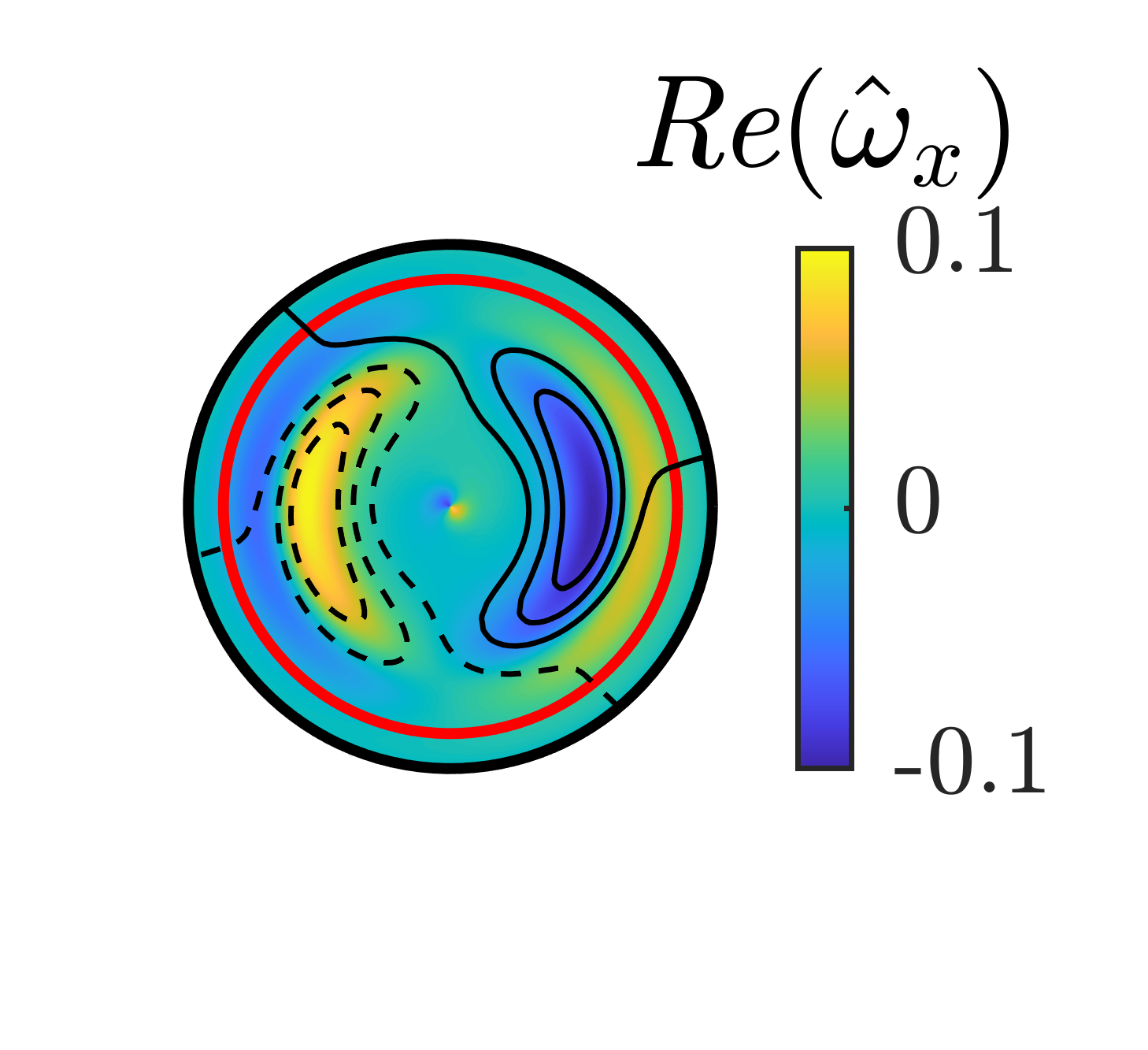}  
\end{tabular}
\caption{ Eigenmodes found for $a/A=0.75$, real parts of the streamwise vorticity with iso-levels of pressure. }
    \label{fig:modes_asurA_0.75_slice}
\end{figure}

The eigenvalues curves given by the amplification rate and the Strouhal number versus the Reynolds number and  computed by the linear stability theory are displayed  in  figure \ref{fig:asurA0.75_lambda}, for a the base flow solved with the section aspect ratio $a/A=0.75$.

Three branches are found as the Reynolds number varies corresponding to two non-oscillating (called again $S1$ and $S2$) and one oscillating mode.
The latter is of a distinct nature comparing to the modes $O1$ and $O2$ previously encountered. It is characterised by a Strouhal number in a lower range, and it is thus called $O3$.
The amplification curves of the $S1$ mode follows an inverted parabola:
as $Re$ value increases, the amplification of the $S1$ mode raises, reaches a maximum and decreases.
The decreasing $S1$ branch meets the rising $S2$ branch and both branches collide at the Reynolds number $Re=180.4$. Above this value, the collision gives rise to a pair of complex conjugate eigenvalues corresponding to the $O3$ oscillating mode. The symmetry of the problem entails that this oscillatory branch is twofold, for each eigenvalue $\lambda$ found, $\bar{\lambda}$ is also an eigenvalue. 
The Strouhal number of the $O3$ raises strongly after the collision of $S1$ and $S2$ from $St=0$ to $St=0.1897$ at $Re=490$ and then slightly decreases. 


Figure \ref{fig:modes_asurA_0.75_slice} displays the vortical structure and some iso-pressure contours of the unstable eigenmodes for
different values of the Reynolds number.
The $S1$ and $S2$ mode exhibit the same behaviours found for low confinement configuration, with a negative pressure zone at the rear of the blunt body followed by a positive pressure one. 
Nevertheless, these pressure contours are distorted by the proximity of the pipe wall as the extrema get closer to axis of symmetry. 
The vorticity of these two modes goes through the same changes and is much important close to to body.

The $S1$ mode is more active in the recirculation zone whereas for the $S2$ mode 
the azimuthal vorticity is higher in the region where the streamlines of the base flow expand,
around $x=2.5$, suggesting a different instability mechanism. At last, the structure of
the $O3$ mode seems to be a mix of the $S1$ and $S2$ mode. 
The pattern of the vorticity 
and the pressure is very similar to the $S1$ mode in the recirculation zone. 
The downstream region ($x>2$) is similar to the same region of the $S2$ mode 
but  the temporal mode oscillation, implyies alternate production of vorticity and 
its change of sign. 
For $Re=400$, the $O3$ mode is similar, the influence of a larger recirculation zone can be noticed. Alternate values of vorticity in the streamwise direction are still present but they pushed downstream, outside the scope of the plot. Stronger vorticity is also observable because of an important contraction of the base flow due to its local reversal.

\subsubsection{Strongly confined flow with  $a/A=0.81$ }

\begin{figure}[t]
\centering
\includegraphics[scale = .8, trim= 0 136 0 120 ,clip]{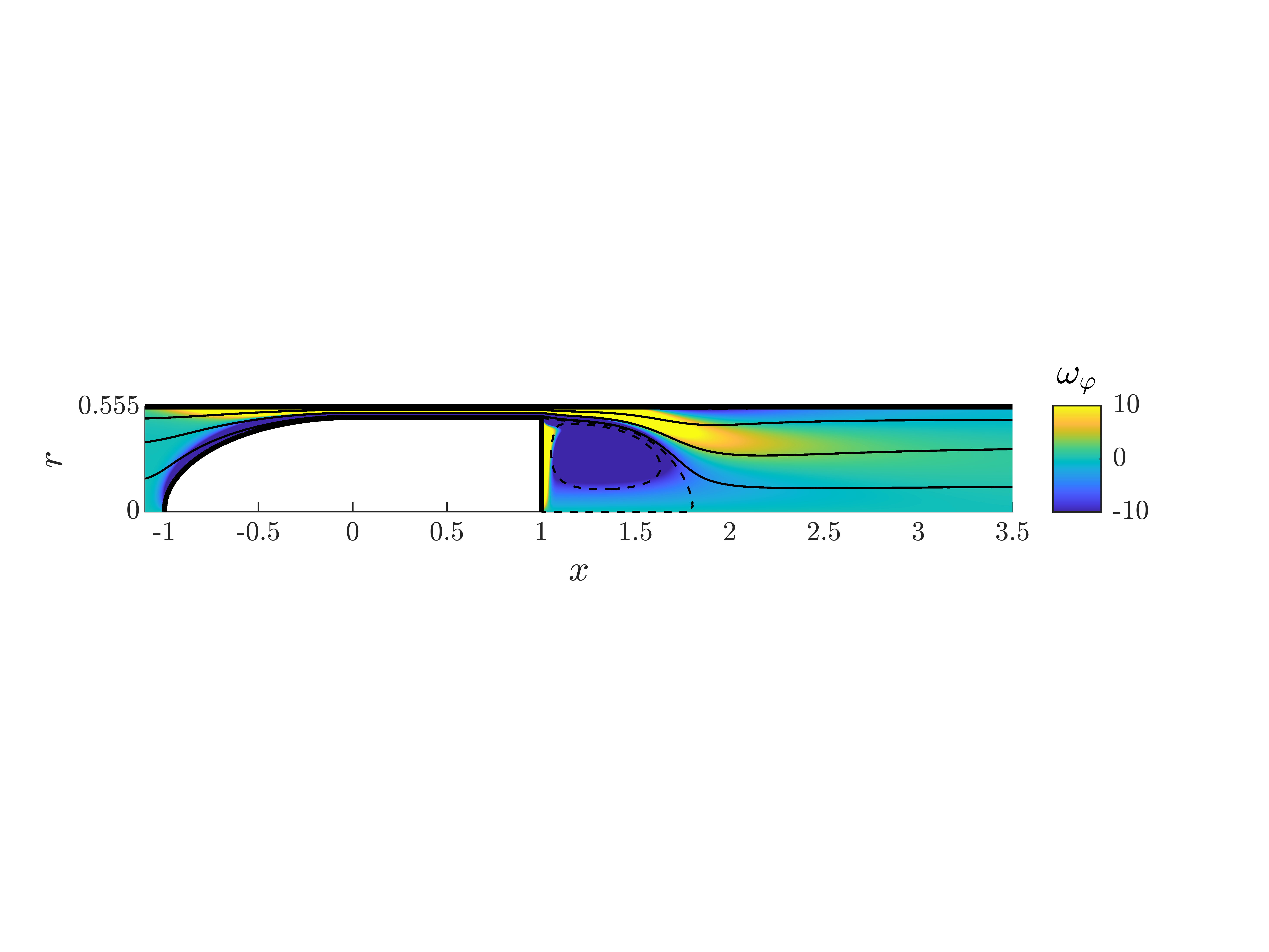}
\vspace{1em}

\caption{Azimuthal vorticity and streamlines of the base flow in the moving frame attached to the body for $L/d=2$ and $a/A=0.81$, $Re=110$}
\label{fig:Baseflow_asurA_0.81}
\end{figure}

\begin{figure}
\subfloat[]{\includegraphics[scale = .8]{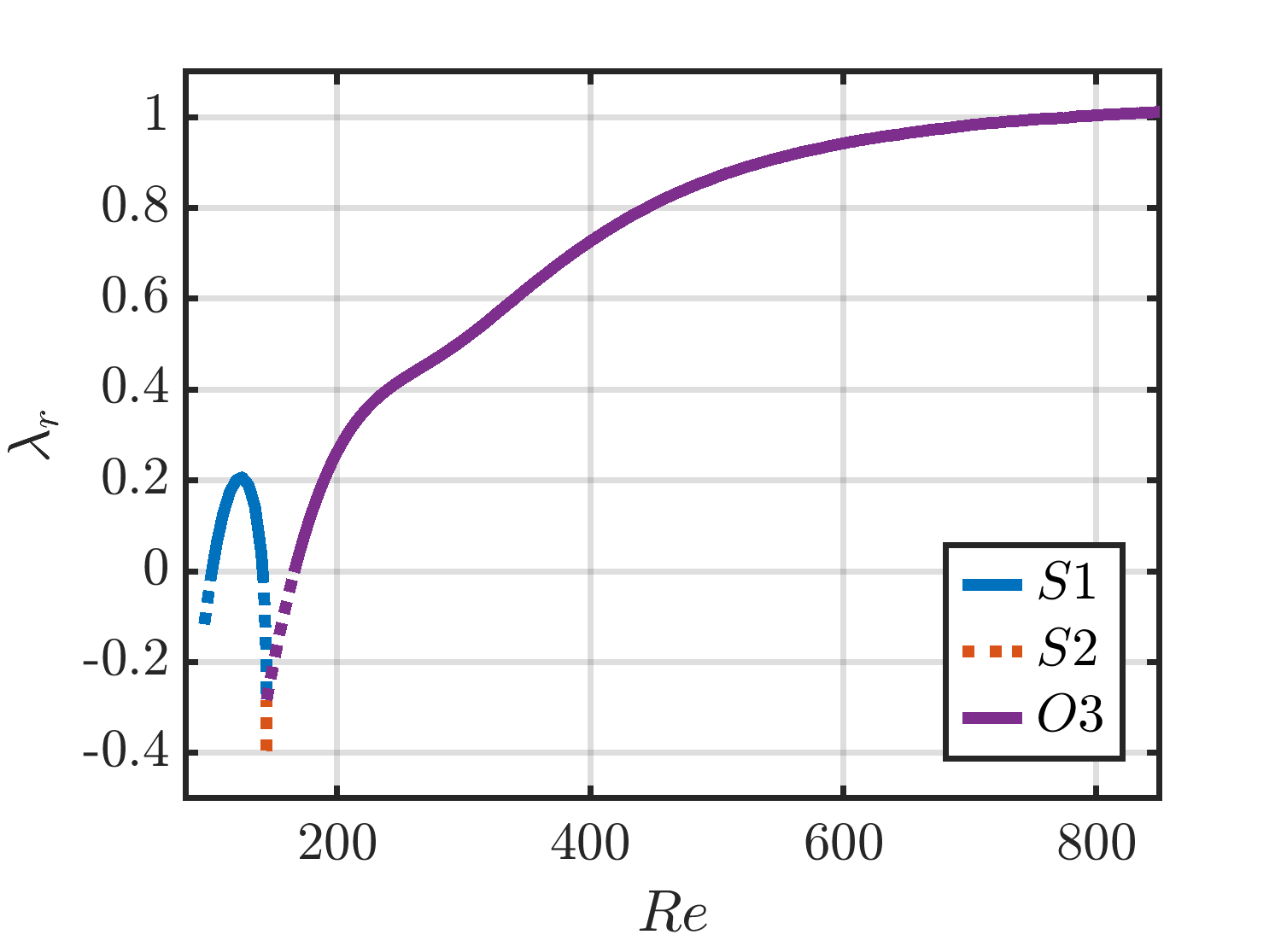}}
\subfloat[]{\includegraphics[scale = .8]{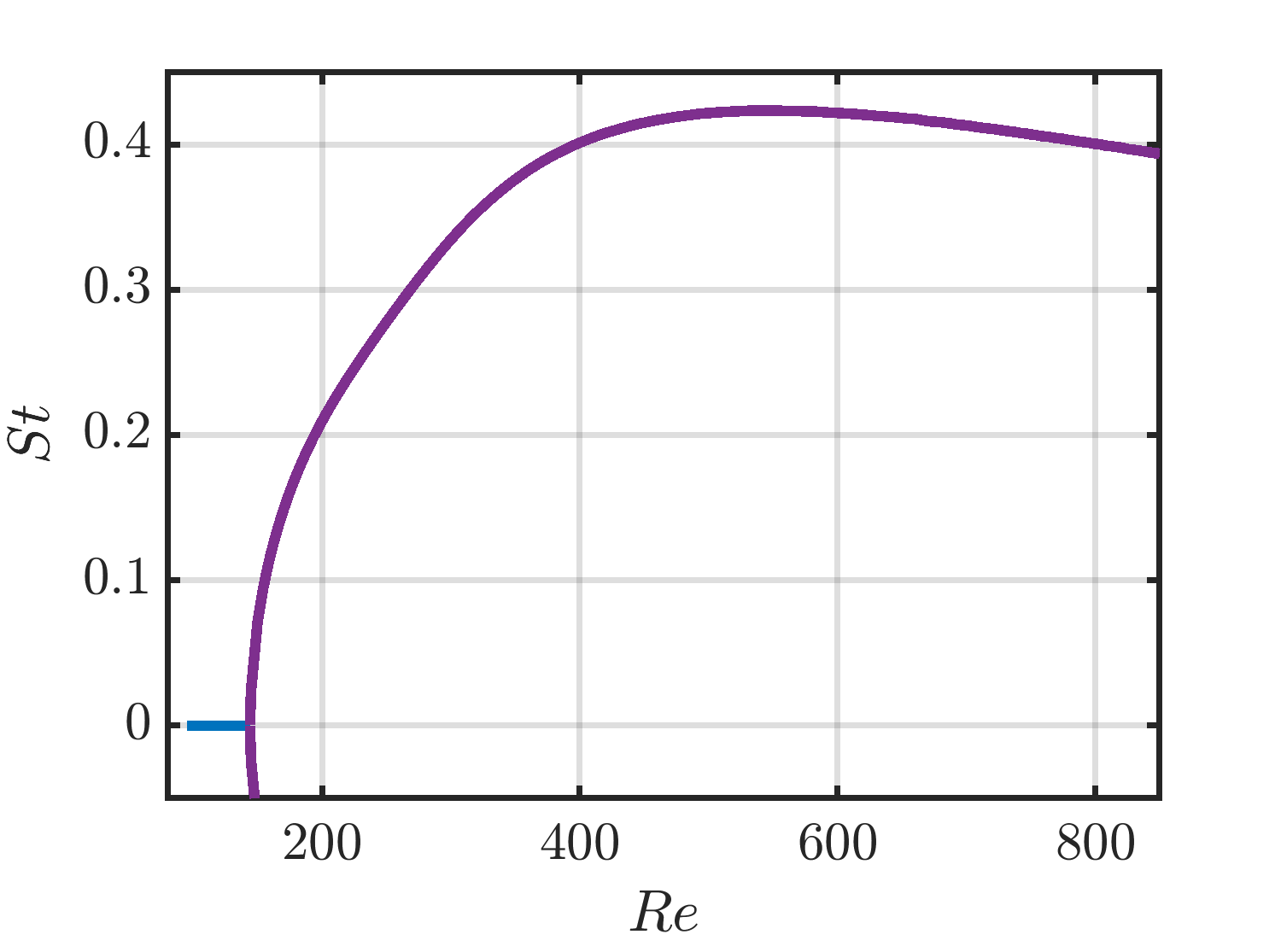}}
\caption{Amplification rate and Strouhal number for the three modes found for $L/d=2$ and $a/A=0.81$.}
\label{fig:NeutralCurves_asurA_0.81}
\end{figure}

Consider now an even more strongly confined flow with a section ratio $a/A = 0.81$ or a length ratio $d/D = 0.9$.
Despite the confinement parameters has slightly changed from the previous case with as results approximately the same characteristics
and axisymmetric base flow patterns (fig.  \ref{fig:Baseflow_asurA_0.81}), the 
the bifurcation scenario revealed by LSA approach seems different. 
Figure \ref{fig:NeutralCurves_asurA_0.81} displays the computed amplification rate and Strouhal number of the first three modes
as function of the Reynolds number. 
Again,  two non-oscillating eigenmodes $S1,S2$ and an oscillating mode $O3$ are found.  
Here, the $S1$ mode become unstable between the Reynolds numbers  $Re_c=101.06$ and $Re=141.15$, the amplification rate plot 
keeps its inverted parabola shape. 
The $S2$ mode is observed as a stable mode up to $Re \approx 150$ where a collision with the $S1$ mode gives rise to a pair of 
complex eigenvalues corresponding to the $O3$ mode.
The latter, first arises as a stable mode, and subsequently becomes destabilized through a Hopf bifurcation at $Re=166.20$. 
Then, the $O3$ remains the predominant mode over the range of parameter studied. There is a stable pocket between the appearance of the $S1$ and $O3$ mode where the $S1$ and $S2$ branches collide, as they are both stable, forming the $O3$ branch.
Note that the dimensionless frequency of the $O3$ mode has twice the value of the previous case, $St=0.4227$ at $Re=550$. 
It is not really surprising since the maximum axial velocity in the jet 
(as predicted by the annular Couette-Poiseuille solution given in appenxix A)  is about twice the value of the previous case.

%

Considering the differences between the present case and the previous one, one can postulate the existence of an intermediate value of the confinement ratio where the collision of the $S1$ and $S2$ modes and the destabilization of the $O3$ mode will occur simultaneously.
This situation, characterised by the existence of two simultaneously neutral modes with zero eigenvalues, corresponds to a codimension-two bifurcation of Takens-Bogdanov type. This point will be confirmed in the parametric study of Sec. \ref{sec:parametric}.

\begin{figure}
\centering
\begin{tabular}{c l}
\multicolumn{2}{c}{$S1$, $Re=140$}\\
\includegraphics[scale = .8,trim=0 106 0 120 ,clip]{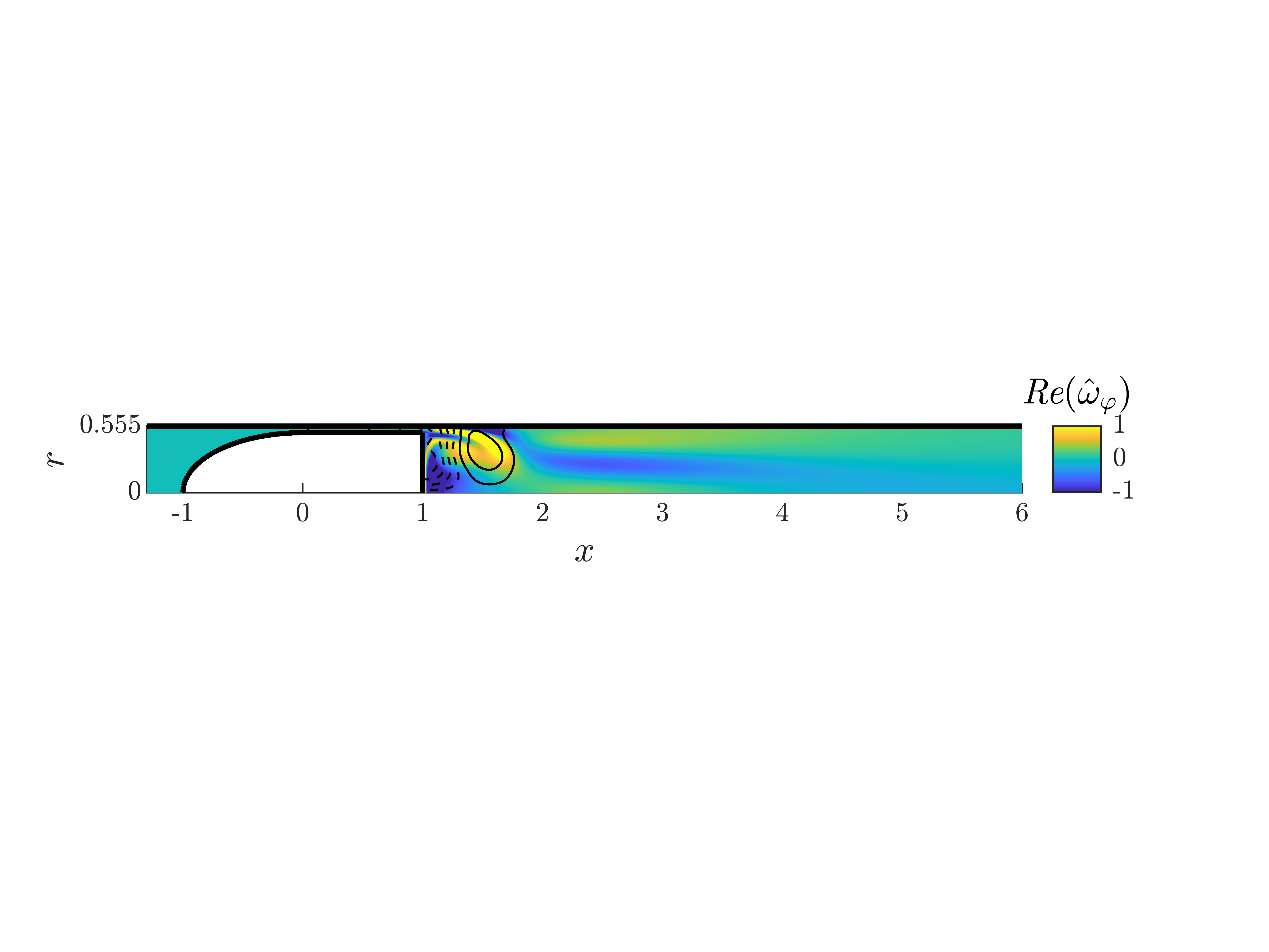} &
    \hspace{-4.em}\includegraphics[scale = .8,trim=10 5 0 0 ,clip]{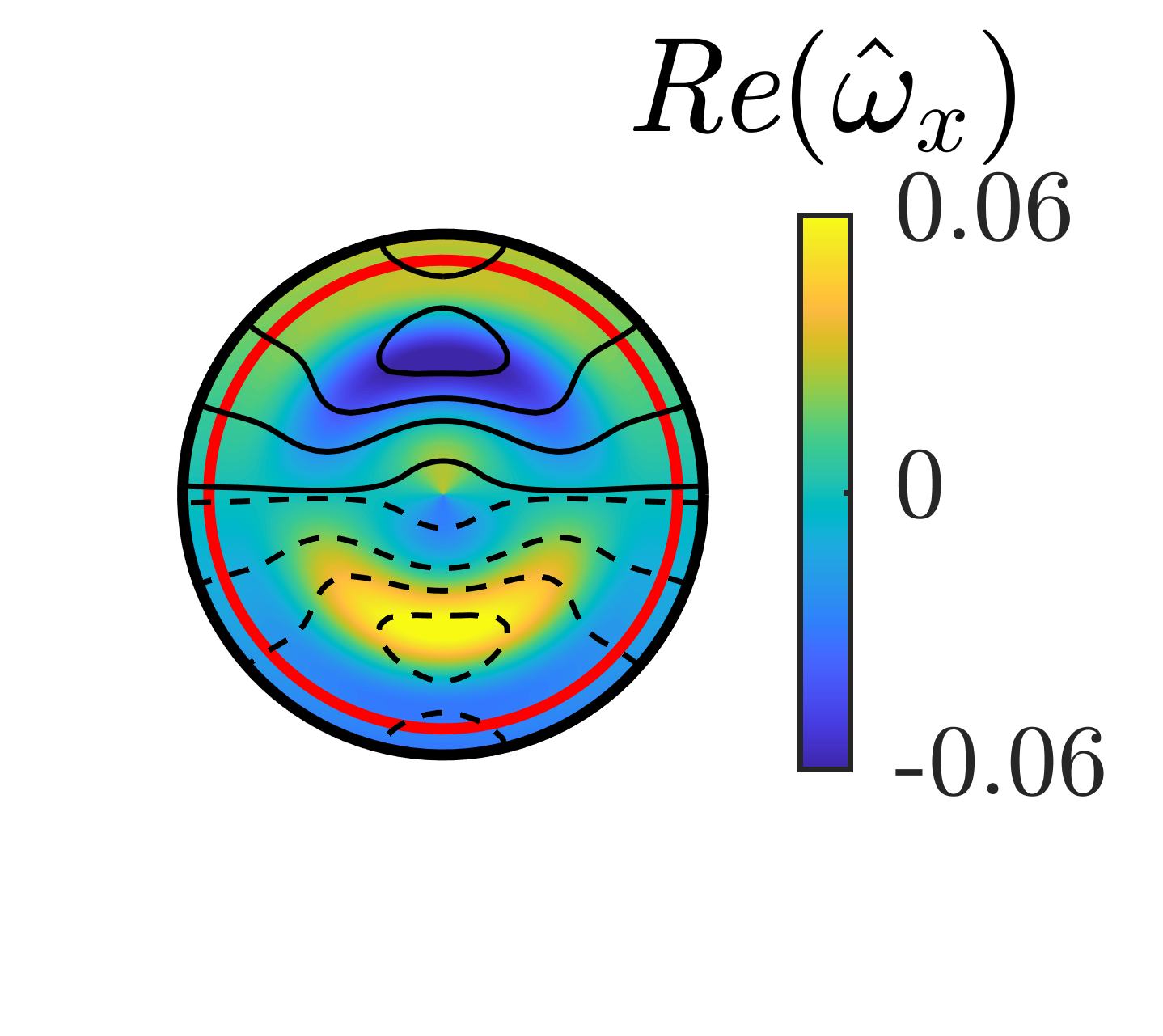}  \\
\multicolumn{2}{c}{$O3$, $Re=200$}\\
\includegraphics[scale = .8,trim=0 106 0 120 ,clip]{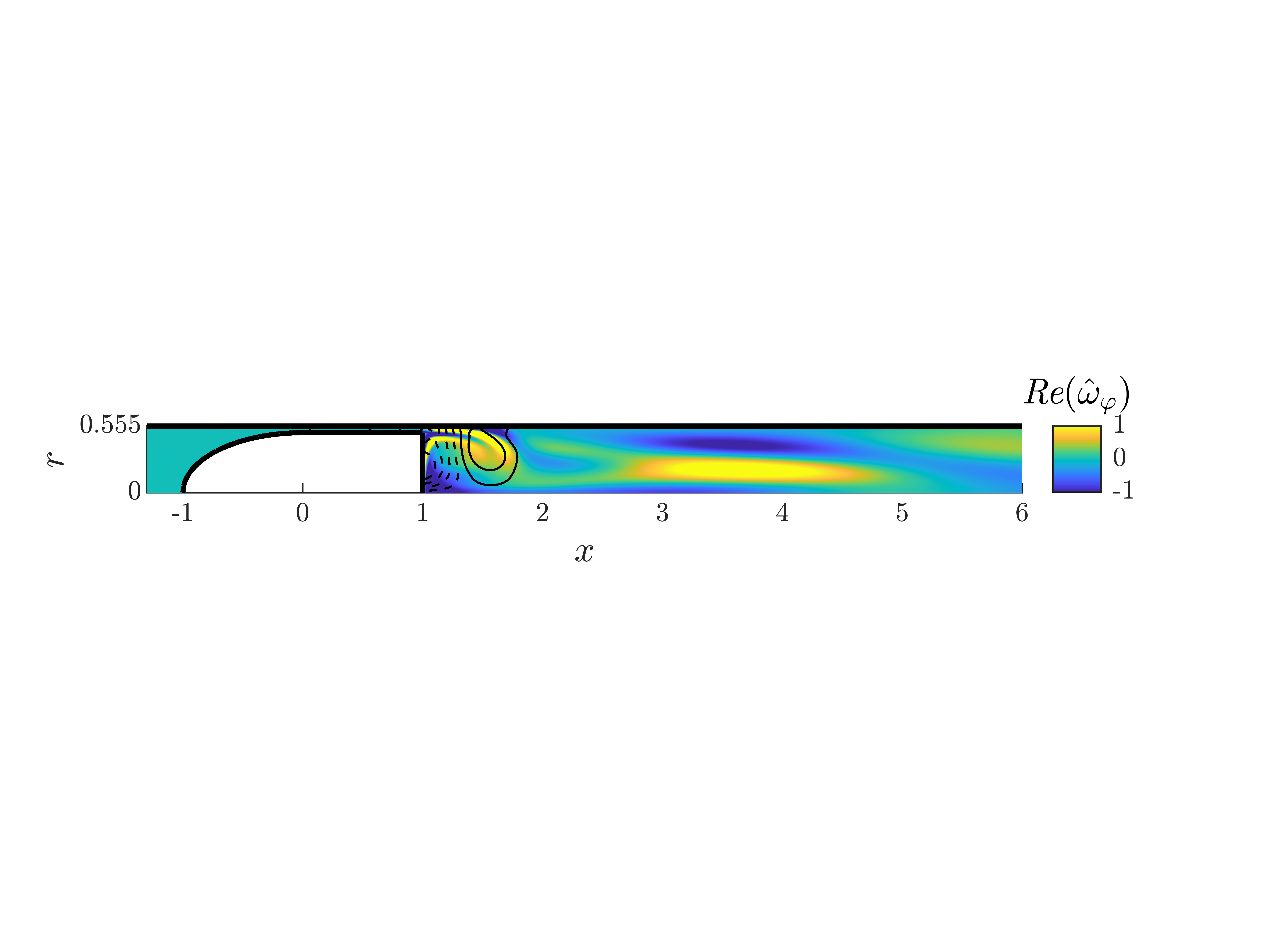} &
\hspace{-4.em}\includegraphics[scale = .8,trim=10 5 0 0 ,clip]{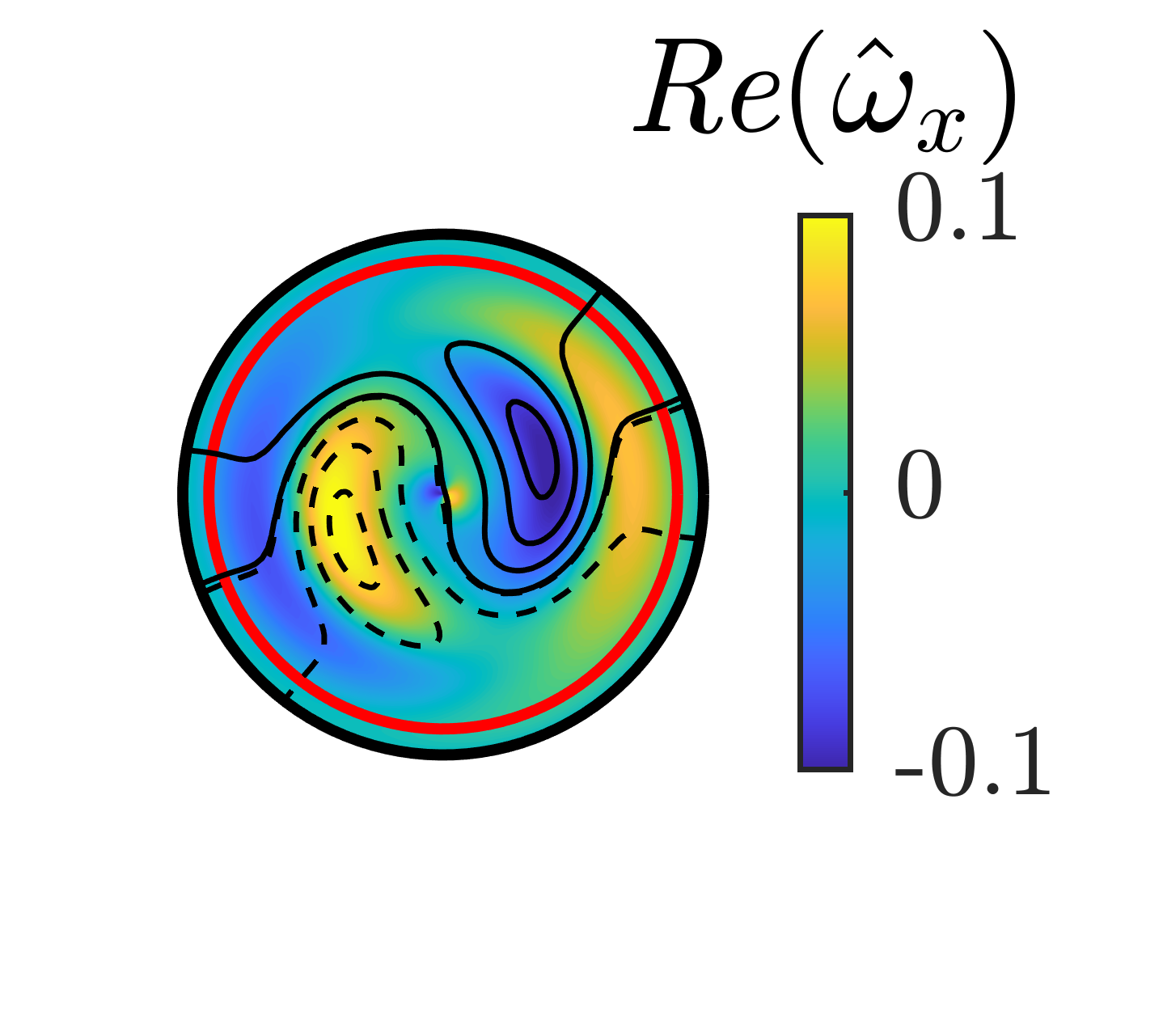}  \\
\end{tabular}
\caption{ Eigenmodes found for $a/A=0.81$, real parts of the vorticity with iso-levels of pressure. Slices are given for $x=2$
}
    \label{fig:modes_asurA_0.81_slice}
\end{figure}

To end up with characterisation of the $a/A=0.81$ case, figure \ref{fig:modes_asurA_0.81_slice} 
reveals the structure of the unstable modes $S1$ and $O3$. Observations made in the previous sub-section, for $a/A=0.75$, apply here. We can add that the influence of the confinement is noticeable in the $S1$ mode as the maximum of velocity of the base flow is higher compared to the previous case. The $O3$ mode also possess patch of alternated sign of azimuthal vorticity exhibiting higher extrema than the previous case, for the same reason cited just above. 
 





\subsection{Cartography of $m=\pm1$ modes in the $a/A$ - $Re$ plane  for $L/d=2$.}
\label{sec:parametric}

\begin{figure}
\includegraphics[scale = .8,trim=0 92 0 65,clip ]{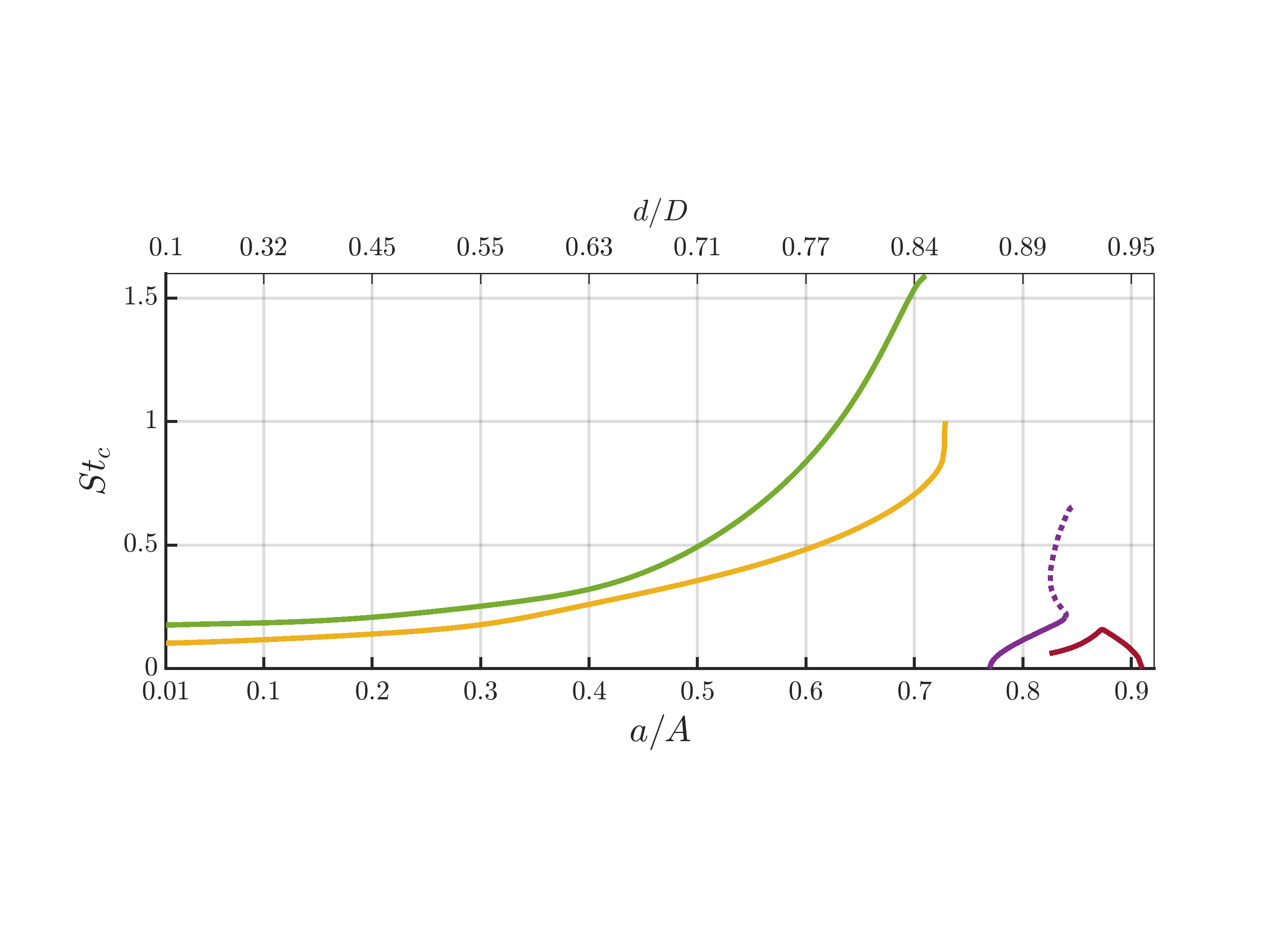}
\vspace{0.3em}
\includegraphics[scale = .8,trim=0 0 0 19,clip ]{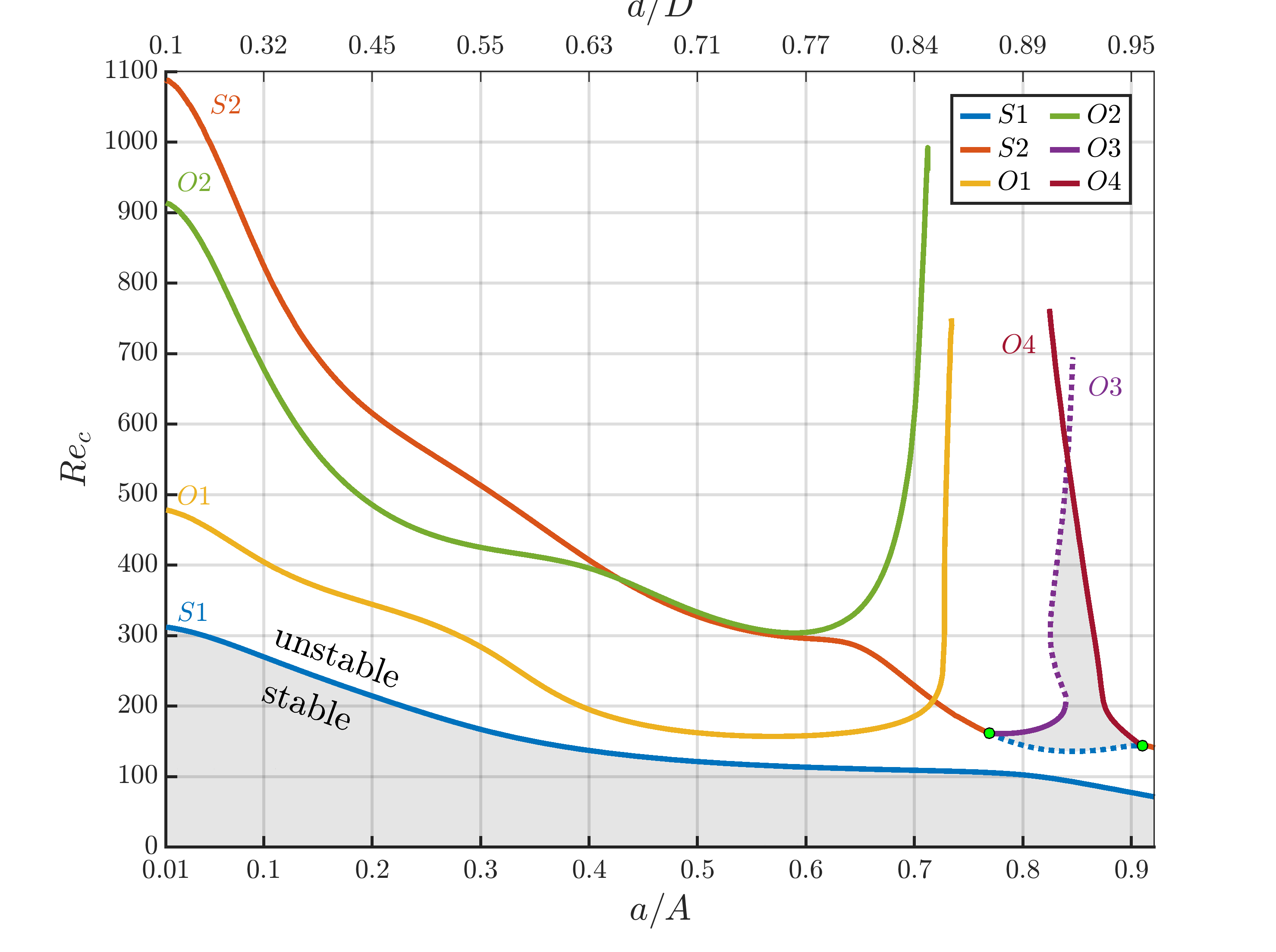}
\caption{Critical Reynolds number and critical Strouhal number as a function of the confinement ratio. The body length is kept constant, $L/d=2$.}
\label{fig:NeutralCurves_m1}
\end{figure}

A  first  exploration of the stability picture has been carried out for some selected values of $a/A$.
The study is now extended continuously to a larger range of the confinement parameter with  $a/A\in\left[0.01\, , 0.92\right]$.  The azimuthal wavenumber of the perturbation  and the length-to-diameter
aspect ratio  are kept respectively to $m=\mp1$  and  to $L/d=2$.

The neutral stability curve for any mode is the location of zero amplification perturbations. For each value of the ratio $a/A$ are obtained a critical Strouhal number $St_c$ and a critical Reynolds number
which are the limit of the instability for any mode. The neutral curves are displayed in 
Figure \ref{fig:NeutralCurves_m1}  for the three six modes of interest with a confinement ratio of
$\Delta (a/A)=0.001$ or a Reynolds number increment of $\Delta Re=1$.  A strategy has been developed
to ensure a continuous and accurate curve. 
A first sweep of the $(Re,a/A)$ plane have been initially performed to save computational time.  
Then a thorough computations are conducted following unstable branches.
For each confinement value,  $Re$ is increased  in order to find lower and upper bounds of it critical value $Re_c$, and a linear interpolation is completed to get a more accurate value 
such as $\lambda_r(Re_c)=0$.

The first important result observed from this figure is about  the destabilisation of the
axisymmetric base flow. It is always caused by the same mode, $S1$, for all area ratio $a/A$ in the range $0.01$ to $0.92$. The loss of axial symmetry always occurs through a stationary bifurcation.

The deep stability analysis has also revealed the existence, for the secondary modes, of two different regimes and a transition zone.
First, in the weakly confined regime, up to  $a/A < 0.7$, the secondary dominant mode is the $O1$ mode, and higher modes ($S2,O2$) arise in a much larger range of Reynolds number which makes their physical relevance unlikely. 
In addition the sequence of instabilities with a non-oscillating $S1$ mode followed by an oscillating $O1$ mode  is thus the same as observed for other blunt bodies in 
freestream flow \citep{natarajan1993instability,meliga2009unsteadiness,fabre2008bifurcations,auguste2010bifurcations}. 
The confinement is also found to be destabilizing for both these modes, as the critical Reynolds thresholds decrease as the confinement ratio $a/A$ grows.
Large confinement also increases the frequency of the oscillating $O1$ mode.
It  has ever been  explained by the fact that the mean velocity of the annular jet formed past the body increases with the confinement (for a given flow rate, decreasing section increases velocity).


Transition regime is found in the interval $0.7 < a/A < 0.76$. The threshold of the $S2$ mode first decreases after $a/A \approx 0.6$ to approach that of the $O1$ mode.
The latter is then strongly and abruptly stabilized, and it is not found anymore for $a/A > 0.72$.
In the range $0.72 < a/A < 0.76$ the stationary $S2$ mode is become the dominant secondary mode.

The strongly confined regime occurs for $a/A$ greater than $0.76$. 
At this regime, the oscillating modes $O1$ and $O2$ are not longer present and new ones ($O3$, $O4$) appear with low or moderate dimensionless frequency.
In figure \ref{fig:NeutralCurves_m1} let follow the mode evolution along a vertical line at $a/A$ close to $0.86$ and let us consider increasing Reynolds number.
It can be seen that the $S1$ initially stable becomes unstable on a short range of $Re$,
then it is unstable in a larger range of $Re$, and finally the flow instability
is generated by the appearance of the mode $O3$ and $O4$. In a short area, 
colored in grey in the figure, it is found a pocket of stability.
It can be noticed that the neutral curve of the $O3$ mode also  displays 
two turning points close to $Re_c\approx 200$, so in a narrow range around $a/A = 0.84$, the destabilization / restabilization sequence occurs twice as $Re$ is raised.
The complexity of the stability diagram for very strong confinement is a translation of the real physics complexity in this region with fast annular wall jet, separated flows
and vortical interactions. 


As already discussed, the emergence of the stable pocket is expected to be associated to a codimension-two bifurcation of Takens-Bogdanov type, where both $S1$ and $S2$ modes are simultaneously neutral. This statement  is confirmed in figure  \ref{fig:NeutralCurves_m1}, as indicated by the green point with coordinates $(a/A, Re_c)_{TB}^{\mathrm{O3}}= (0.769,161.57)$ from which the $O3$ neutral curve emerges.
Note that a second Takens-Bogdanov point is observed at coordinates 
$(a/A, Re_c)_{TB}^{\mathrm{O4}}= (0.91,143.96)$.
The latter bounds the stable pocked on the other side and is associated to the emergence of the $O4$ mode. As indicated in the upper plot, the critical Strouhal number 
of $O3$ and $O4$ modes is zero at the codimension-two points, as expected for a Takens-Bogdannov bifurcation. The Strouhal numbers of these modes raise
as one moves away from these points.

    
    

\subsection{Effect of the  $L/d$ aspect ratio}

\label{sec:LsurD}
\begin{figure}
	\centering
   	\includegraphics[scale = .8,trim= 0 22 0 0 ,clip]{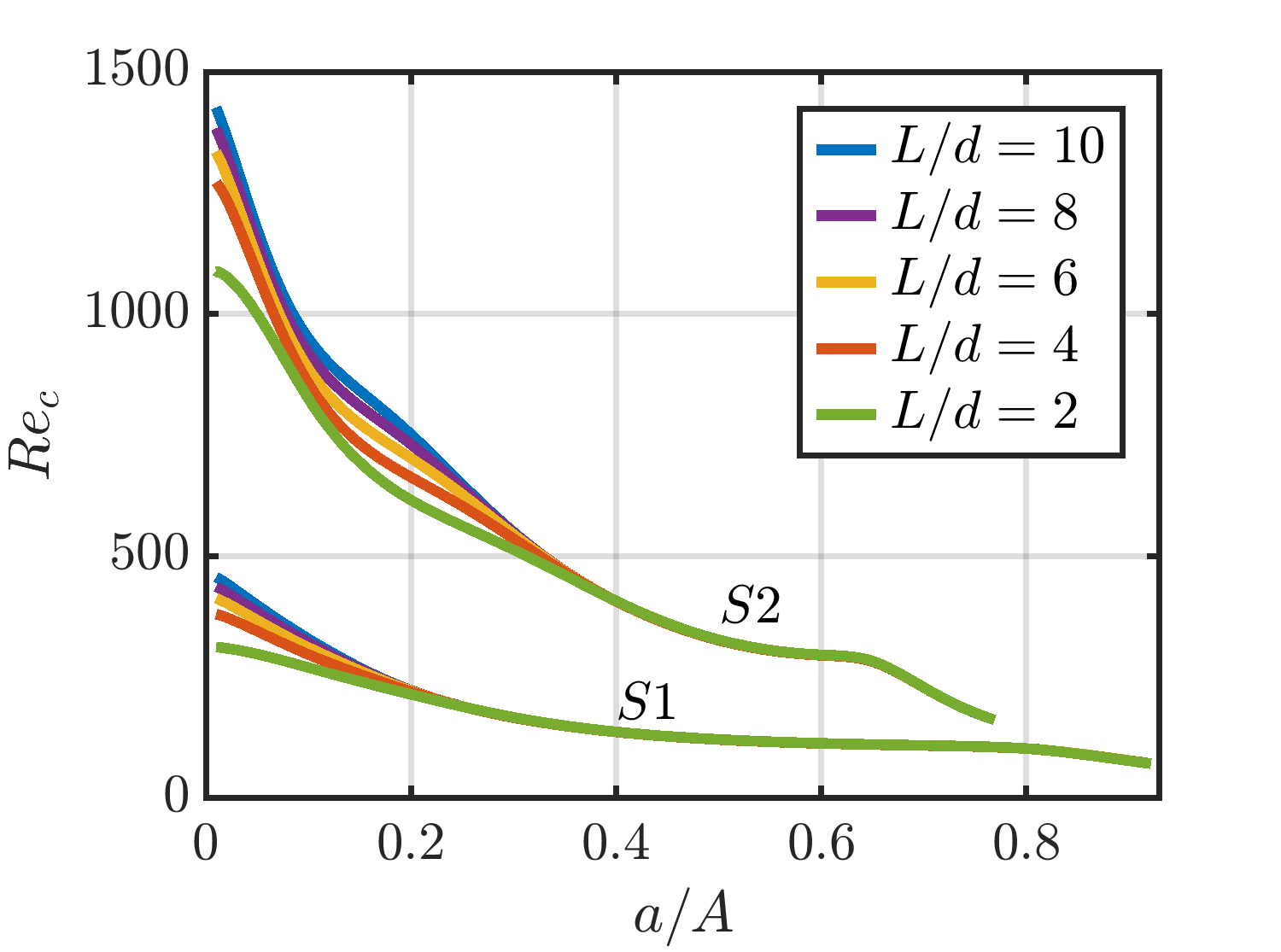}
    \phantom{\includegraphics[scale = .8,trim= 0 22 0 0 ,clip]{grfx_LSA/NeutralCurves_Re-c_dsurD_BluntBondy_S1-2_L.png}}
    
    \includegraphics[scale = .8,trim= 0 22 0 5 ,clip]{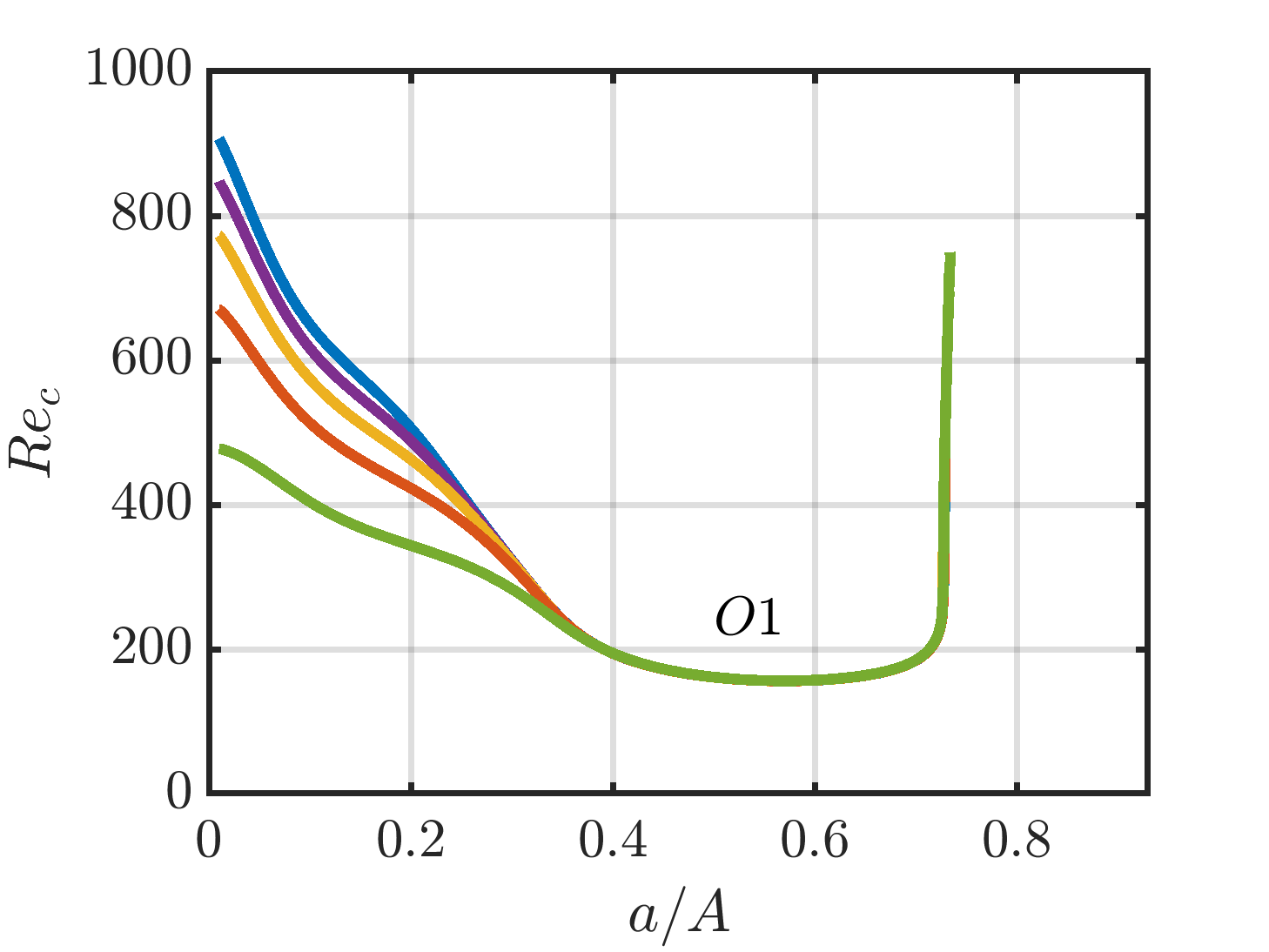}
    \includegraphics[scale = .8,trim= 0 22 0 5 ,clip]{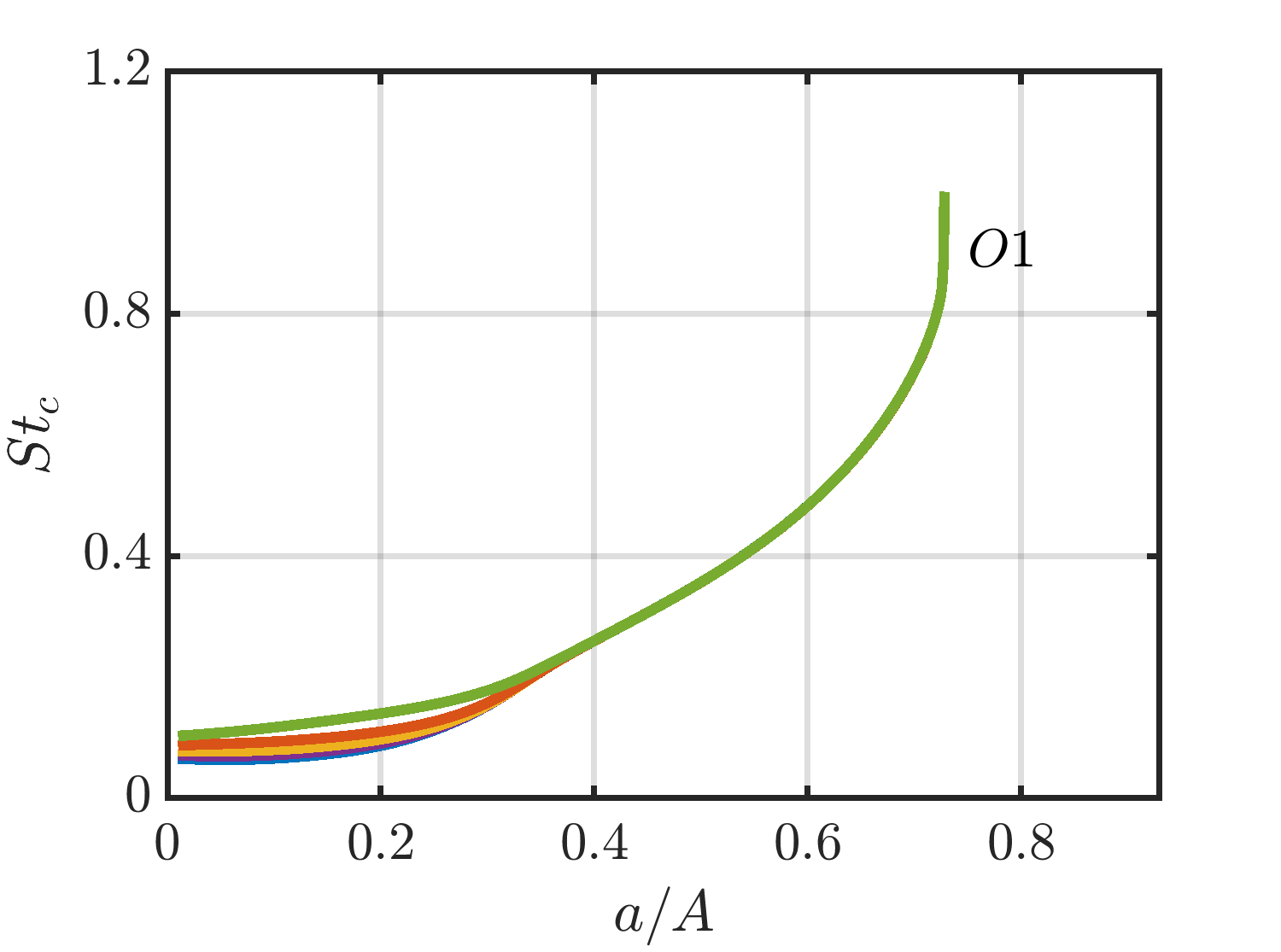}
    
    \includegraphics[scale = .8,trim= 0 0 0 5 ,clip]{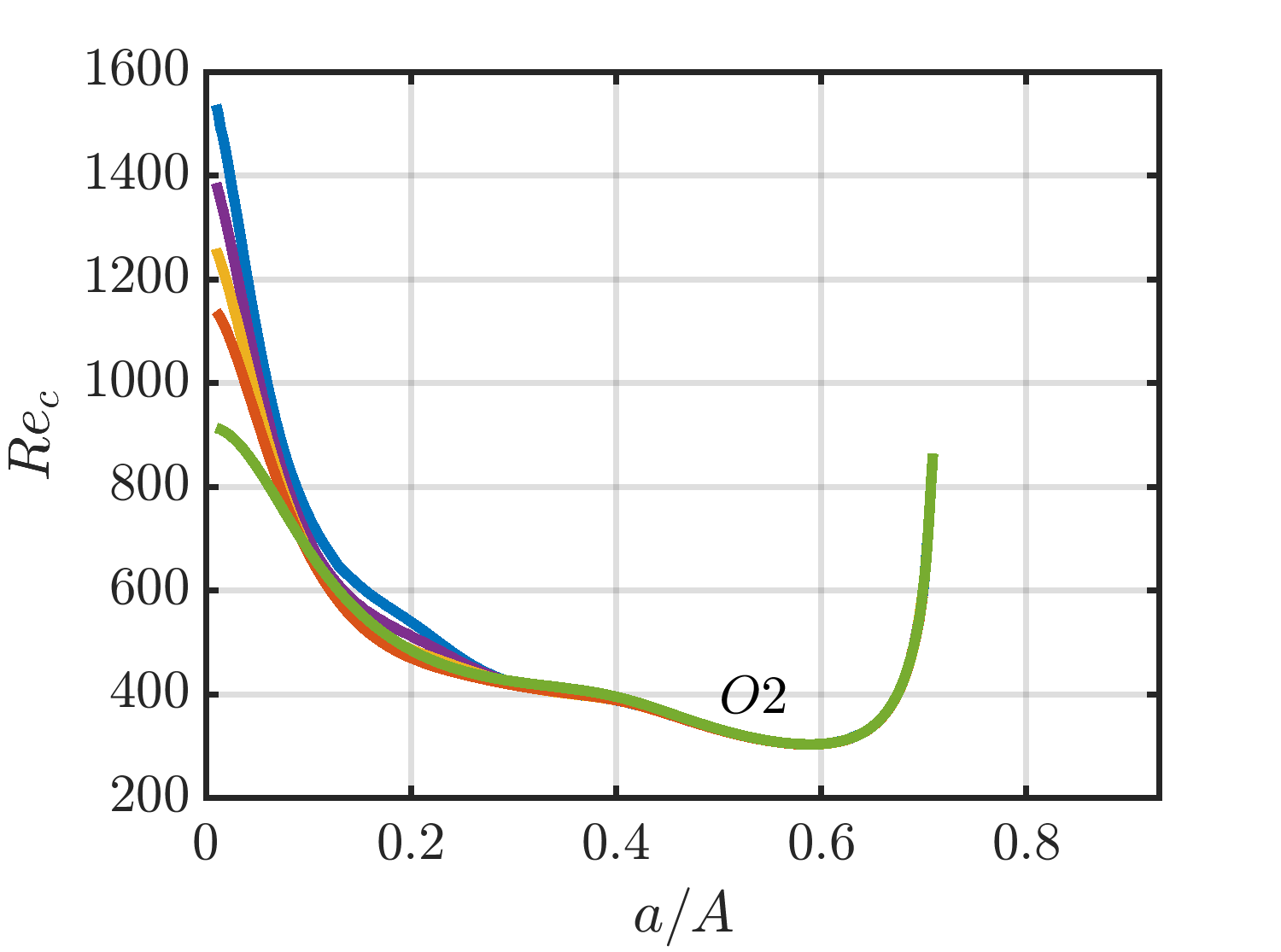}
    \includegraphics[scale = .8,trim= 0 0 0 5 ,clip]{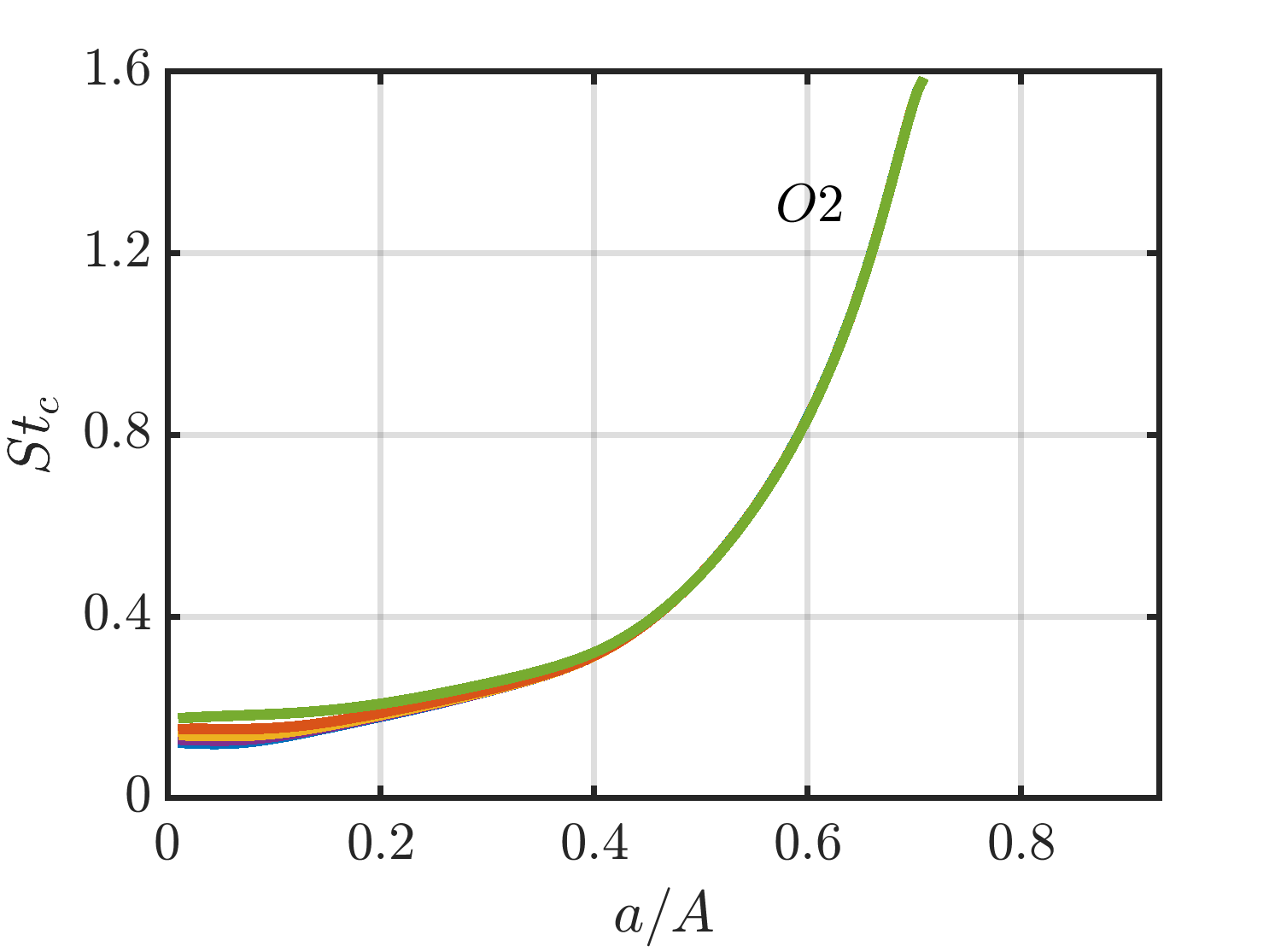}

	\caption{Neutral curves of the stationary modes for different body lengths.}
    \label{FIG:LengthInfluenceSmodes}
\end{figure}

\begin{figure}
	\centering
   	\includegraphics[scale = .8,trim= 0 0 0 0 ,clip]{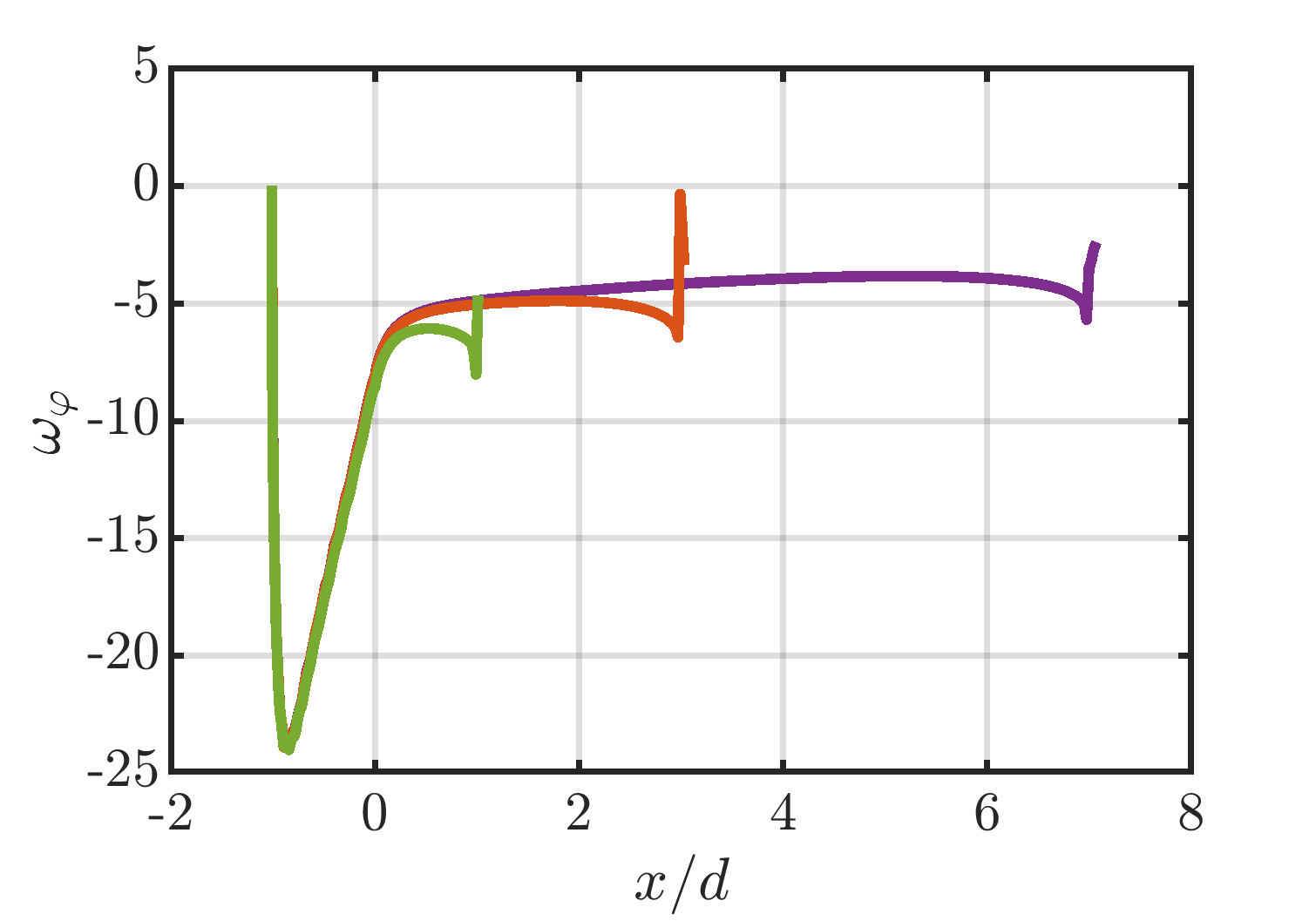}
   	\includegraphics[scale = .8,trim= 0 0 0 0 ,clip]{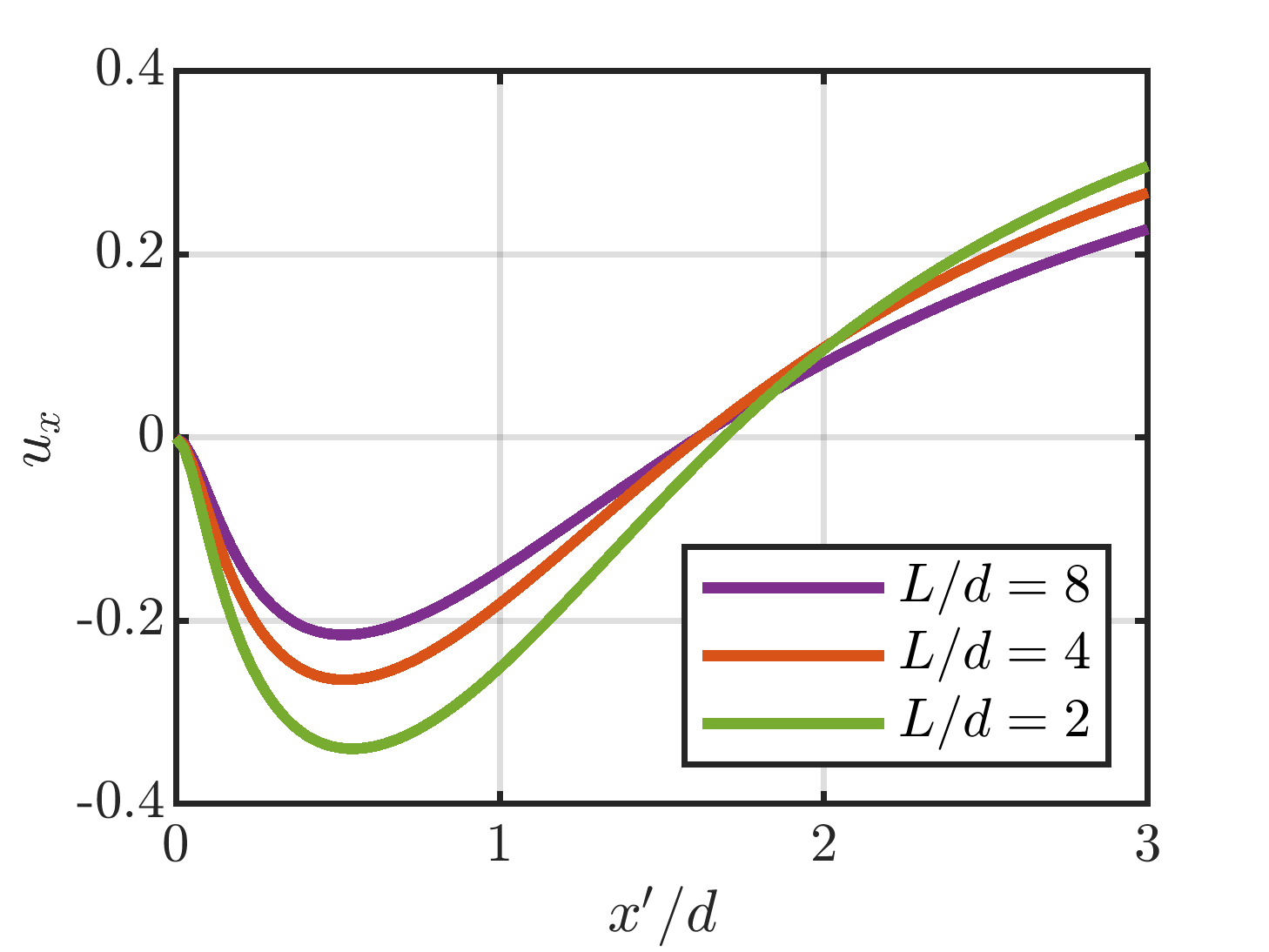}
   	
    \caption{Left, azimuthal vorticity of the base flow at the blunt body surface for $Re=330$. Right, axial velocity downstream the blunt body, the frame has been shifted in order to set the rear of the bodies at the same location. Only the base flow for $L/d=2$ has an unstable mode $S1$.}

    \label{FIG:LengthInfluenceBF}
\end{figure}

The effect of the length-to-diameter ratio $L/d$  of the blunt body is now investigated  keeping again the the restriction  to $m=\pm 1$ modes. 
This geometrical parameter is found to modify the stability properties only in the weakly confined regime at $a/A<0.7$ identified above.
Consequently, only the neutral curves of the modes  $S1$, $S2$, $O1$ and $O2$ relevant to this regime are tracked.
Figure \ref{FIG:LengthInfluenceSmodes} shows the neutral curves of these modes for different  values of $L/d=\left\{ 4\,6,\,8,\,10 \right\}$.
They are compared to the results of the reference case with $L/d=2$ presented in the previous paragraph (in green in figure  \ref{FIG:LengthInfluenceSmodes}).
For low confinement, $a/A<0.4$, the increase of the body length 
 stabilizes the flow  as pointed out by \cite{brucker2001spatio} in his experiments.
 He suggests a larger boundary thickness caused by a longer body is responsible for this stabilizing effect.
 To verify this argument, Figure \ref{FIG:LengthInfluenceBF} (left plot) shows the vorticity at the blunt body surface for different body lengths. On the ellipsoidal nose ($x<0$), the plots are superposed indicating the generation of the  the same amount of vorticity.
 Then, on the cylindrical surface of the blunt body ($x<0$), the vorticity reaches a higher value for short objects. Indeed, a streamline along the body and its recirculation zone is more curved for short objects, accumulating therefore more vorticity fueling the separared flow
 in the rear. 
 In conclusion, for shorter objects, the recirculation zone have stronger reverse velocities (see figure \ref{FIG:LengthInfluenceBF}, right plot), promoting wake instabilities at lower Reynolds number compared to the case of longer objects. We can also note that even if the vorticity intensities are quite different, their sizes do not differ much. 

Back to figure \ref{FIG:LengthInfluenceSmodes}, as the area ratio increases, all curves tend to collapse into one, either the $Re_c$ or the $St_c$. It means, for $a/A>0.4$, that the body length does have any influence on the onset of the four investigated instability modes. 
This is consistent with the fact that, as verified in Figure \ref{fig:CouettePoiseuille}, once a certain confinement is reached and whatever the length of the body, 
the velocity profiles is the same and correspond to the annular Couette-Poiseuille solution recalled in appendix A.


\subsection{Higher azimuthal wavenumber modes} 

To complete the parametric study, we now consider eigenmodes with azimuthal wavenumbers other than $\pm 1$. No axisymmetric ($m=0$) unstable mode was found, but numerous unstable modes with $|m|>1$ were detected. 
Most of them occur in ranges of Reynolds number far above the primary threshold of $m=\pm1$ modes so they are not likely to be observed in any real flow. Only two modes were detected with critical Reynolds number in the same range as $m=\pm1$ modes. Both of them  are non-oscillating, with respectively $m=2,\,3$ azimuthal wavenumbers, and will be referred to as $S_{m=2}$ and $S_{m=3}$. These modes arise in strongly confined regime $a/A>0.6$ where length of the body has negligible effect. In this section 
 we keep the body aspect ratio  $L/d=2$ but conclusions given in  this paragraph actually hold for all values of $L/d$.

\begin{figure}
\centering
\begin{tabular}{ c c}
\multicolumn{2}{c}{$S_{m=2}$}
\\
\includegraphics[scale = .6,trim= 0 105 30 100 
,clip]{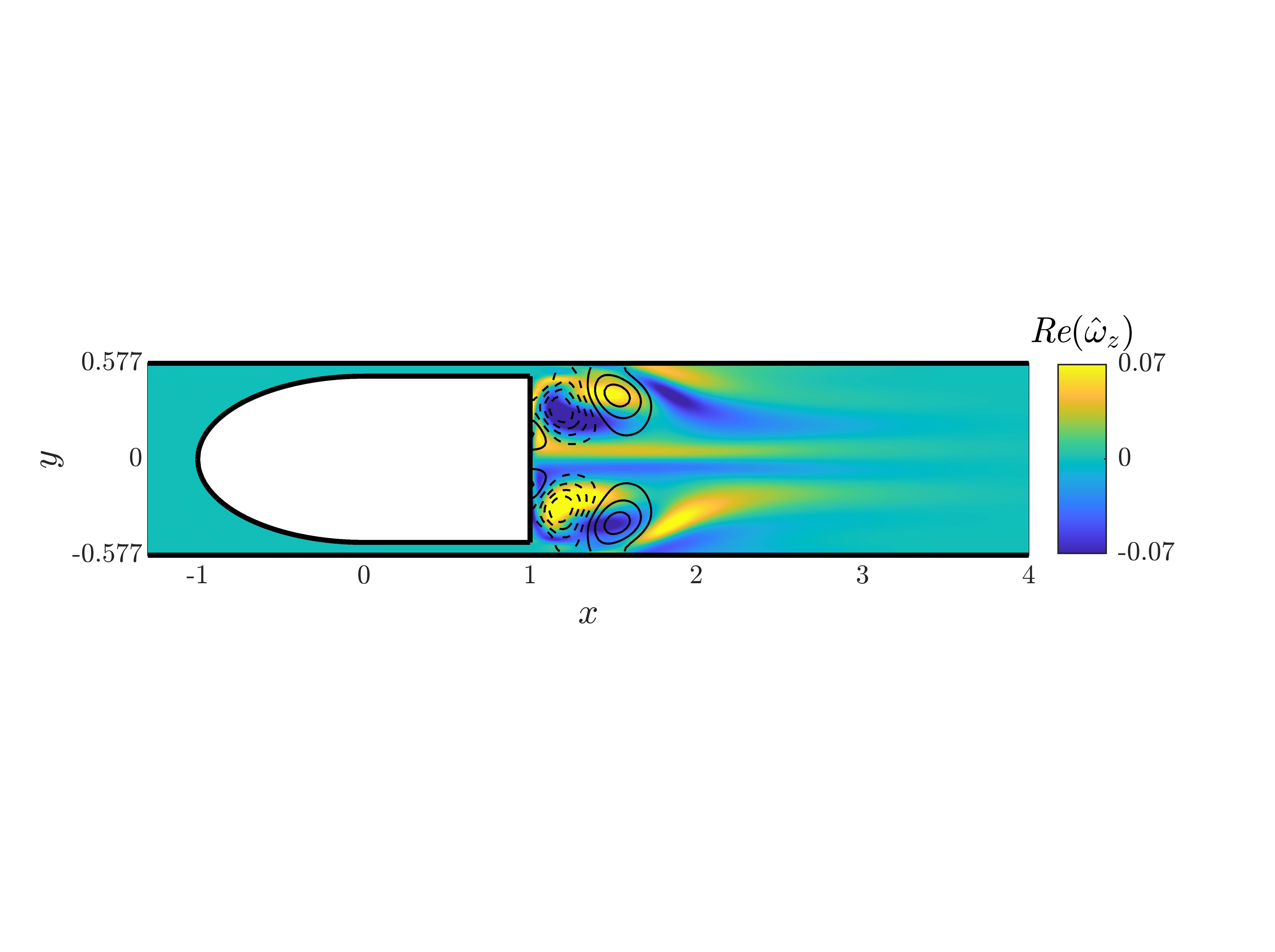}
&
\includegraphics[scale = .6,trim= 5 5 0 0 ,clip]{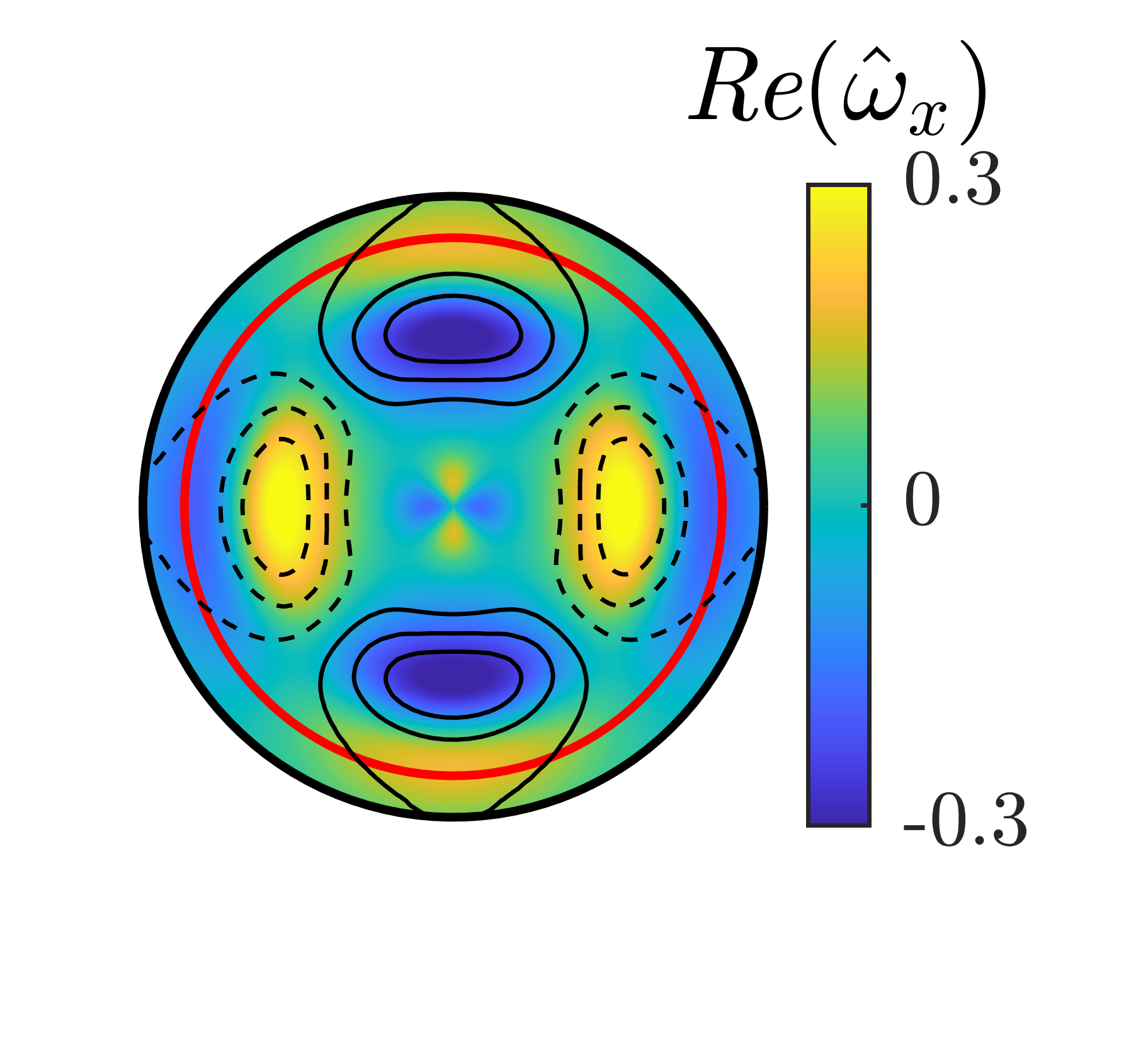}
\\
\multicolumn{2}{c}{$S_{m=3}$}
\\
\includegraphics[scale = .6,trim= 0 105 30 100
,clip]{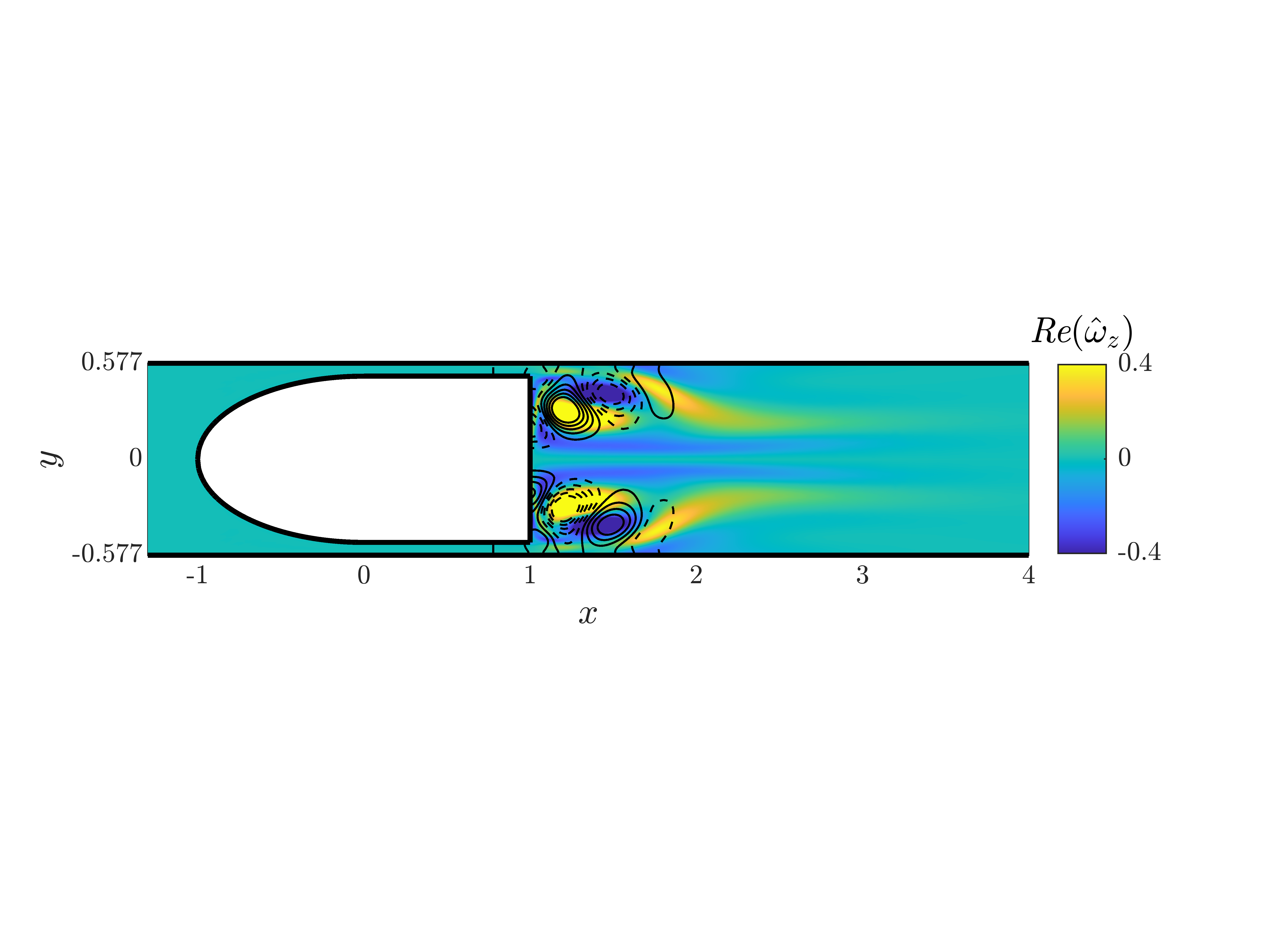}
&
\includegraphics[scale = .6,trim= 5 5 0 0 ,clip]{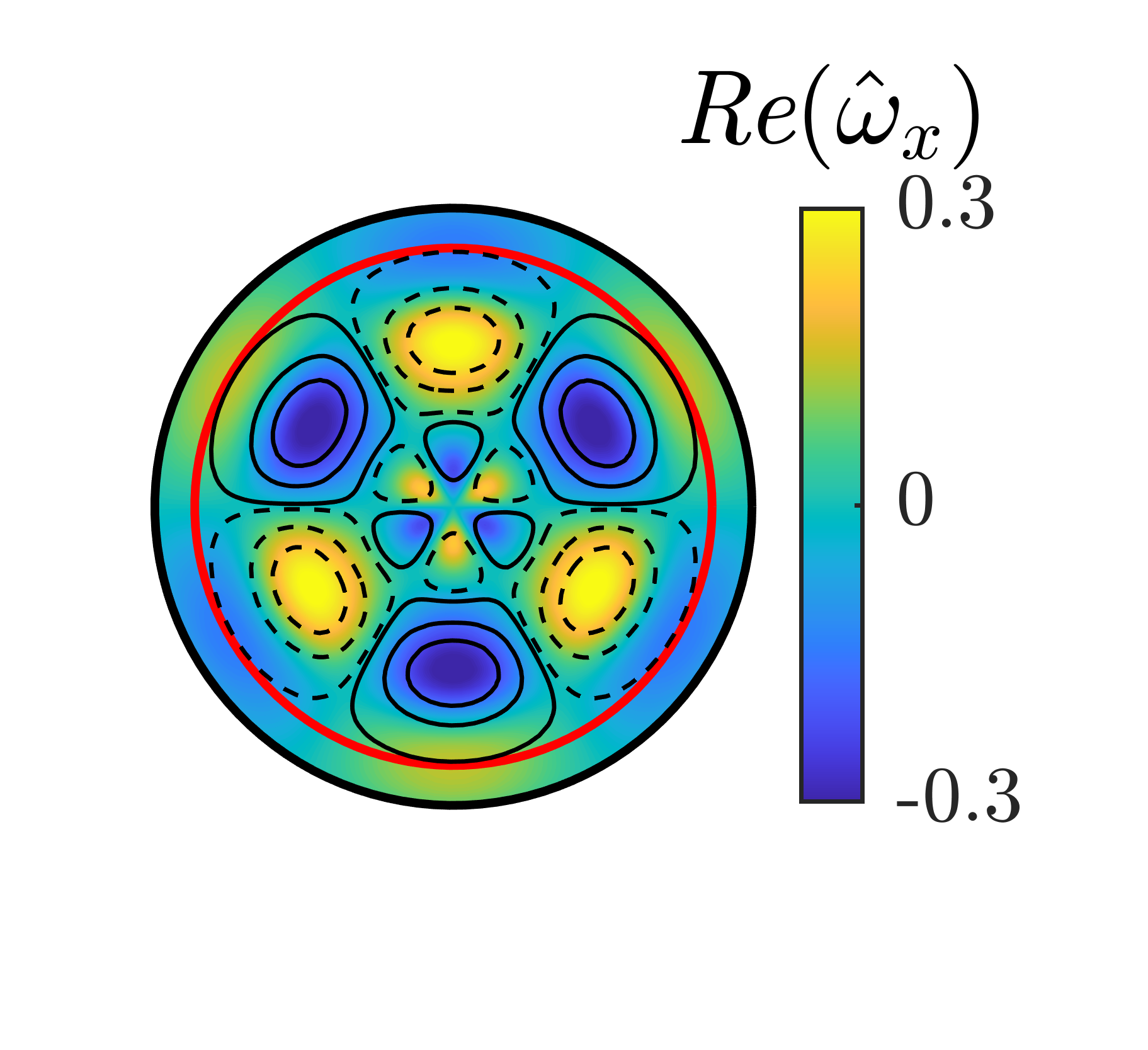}
\end{tabular}
\caption{Unstable non-oscillating modes at $Re=200$ and $a/A=0.75$. Left : $z$-component of the vorticity and iso-level of pressure. Right : slice in the transverse plane at $x=2$, streamwise component of the vorticity and iso-levels of pressure, the red circle represents the projection of the base of the blunt body in this plane. \label{fig:modes_m2-3} 
}
\end{figure}

Figure \ref{fig:modes_m2-3} illustrates the structure of these new eigenmodes.
Their geometry is best understood by looking at the views in a transverse $x$-plane (plots in the right column). Mode $O2$ is characterized by the existence of two orthogonal symmetry planes and displays four main structures of axial vorticity of alternated signs,  
while mode $O3$ has three planes of symmetry and six main vorticity structures. Secondary vorticity structures of opposed signs are also visible near the symmetry axis. The views in a vertical plane (left column) give a complementary picture. One can notice that compared to $m=\pm 1$ eigenmodes the present ones are more localized in the close wake and do not extend in the far wake.



\begin{figure}
\centering
\includegraphics[scale = .8]{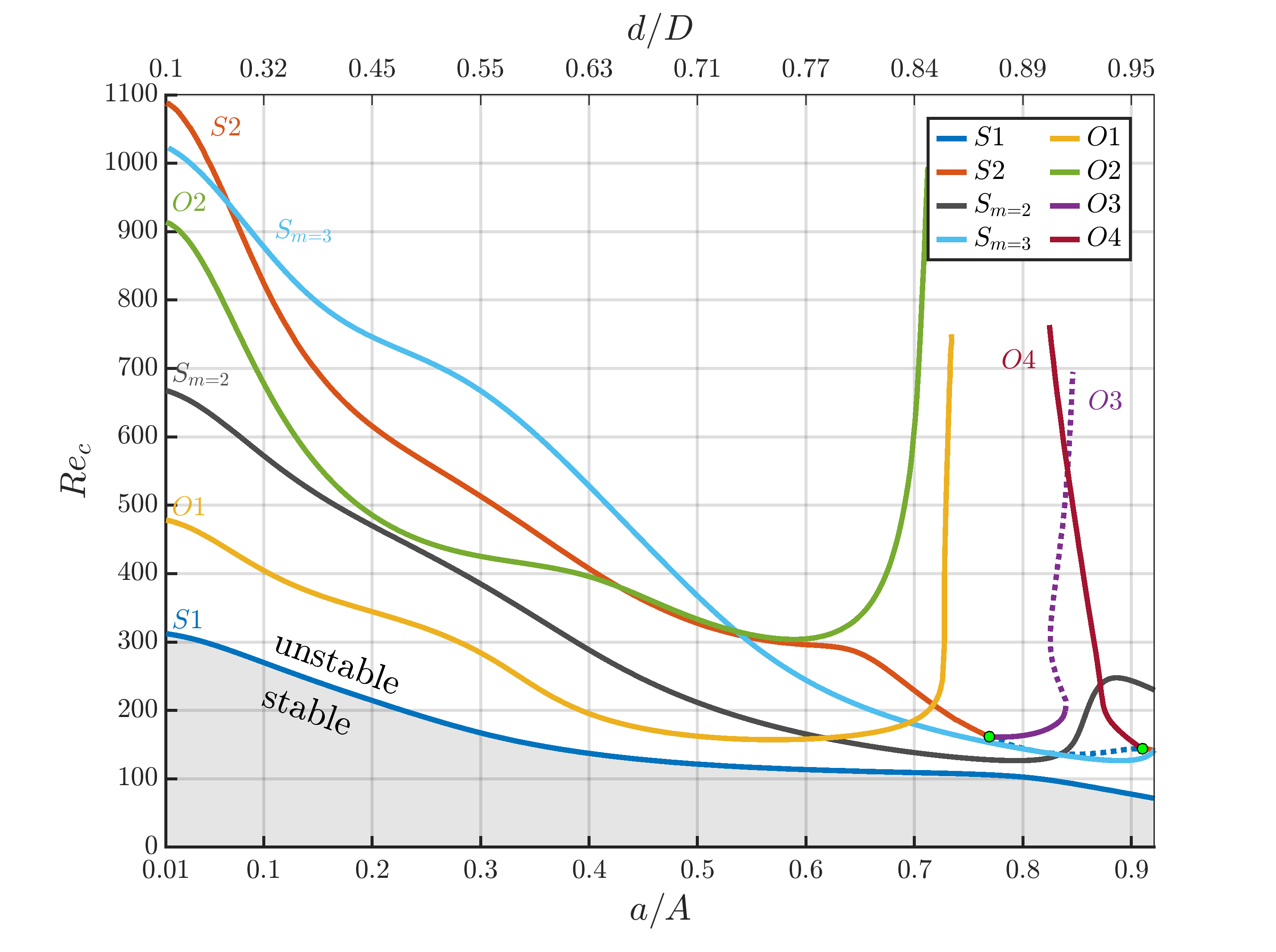}
\caption{Critical Reynolds number as a function of the confinement ratio for all modes considered. The body length is kept constant, $L/d=2$.}
\label{FIG:NeutralCurves_m1-2-3}
\end{figure}

The neutral curves in the $(Re_c,a/A)$ plane of the modes computed for $m= \{ 1,\,2,\, 3 \}$ are plotted in figure  \ref{FIG:NeutralCurves_m1-2-3} which completes the results of figure \ref{fig:NeutralCurves_m1} with additional azimuthal wave numbers. 
The Strouhal number are not displayed because the newly considered modes are stationary, $St_c=0$. 
For a low confinement, the stationnary modes $S_{m=2}$ and $S_{m=3}$  have critical Reynolds number much higher than the first unstable mode and it is rather unlikely to observe them in real experiments. 
However, as the confinement increases, their critical Reynolds number $Re_c$ decreases, 
and they alternatively become the second and third mode to be unstable for $a/A>0.7$. Interestingly, in this strongly confined regime, these two modes become unstable for 
 Reynolds number values very close to those corresponding to restabilization of the $S1$ mode.
 Hence, in such ranges they are the only unstable modes to exist.
 So non-axisymmetric flow characterised by azimuthal wavenumber $m=2$ or $3$ (or a superposition of both) are expected to be observed  without the presence of any $m=1$ component
  in experiments or simulations.
 Such structures is characterised by the absence of lift forces exerted on the body,
 as it justified for instance in \citep{tchoufag2014global}.
 Only $m=\pm 1$ modes can contribute to a lift force.



\section{{Exploration of nonlinear dynamics using DNS}}

In the previous section an exhaustive  mapping of the linear stability 
characteristics of the flow with respect to the aspect ratios and the Reynolds number has been performed.
In this section we now explore nonlinear dynamics using 
Direct Numerical Simulations (DNS). The aim is both to confirm the LSA predictions regarding the primary instability threshold 
 and to explore the nonlinear dynamics arising away from this threshold.


\subsection{Dynamical regimes detected by DNS and comparison with LSA}

\begin{figure}
\includegraphics[scale=0.9]{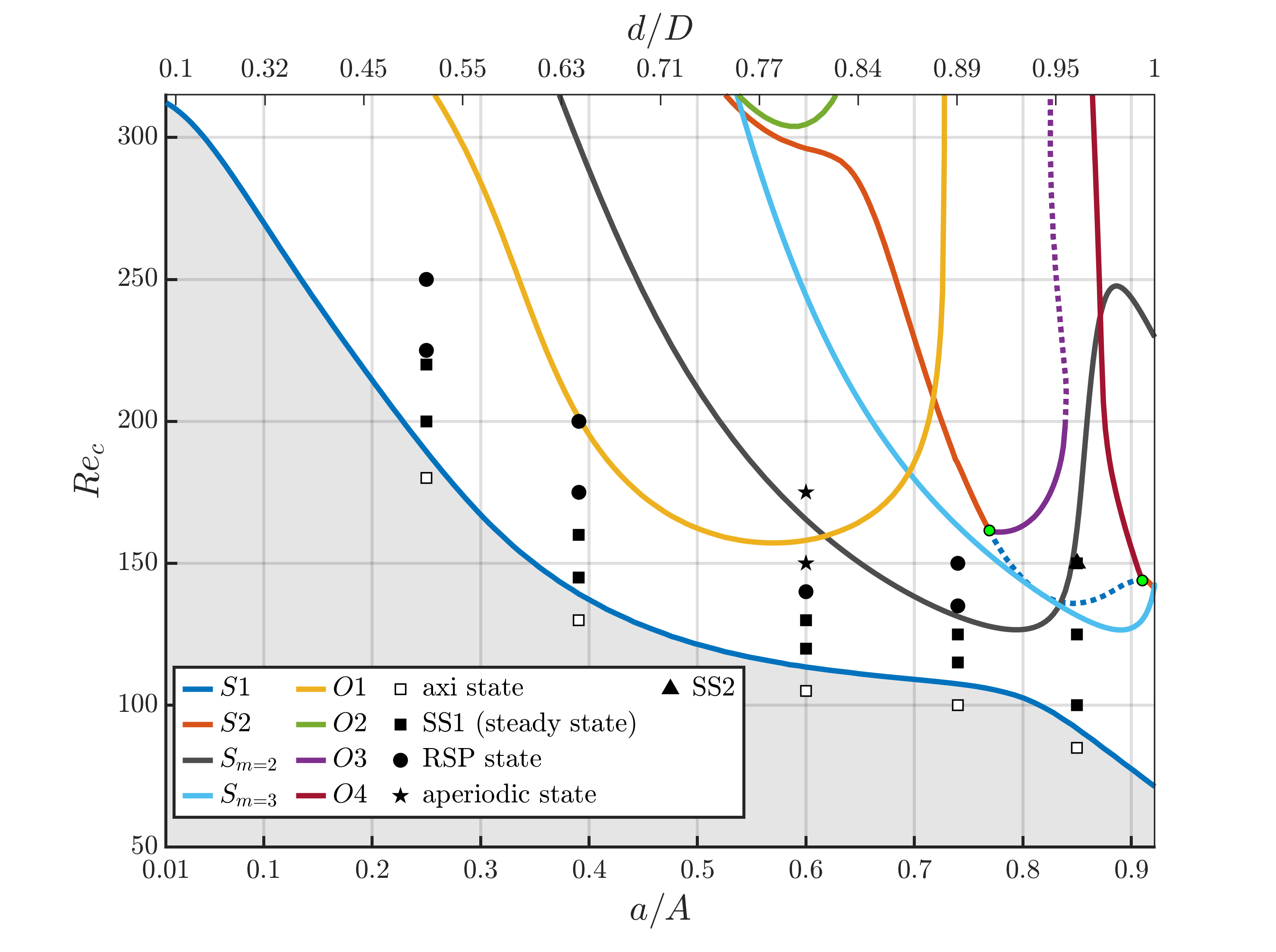}
\caption{Neutral curves computed previously using LSA and DNS results represented by the symbols.\label{fig:NeutralCurves_LSA_vs_DNS}}
\end{figure}

In the numerical exploration we selected five values of the confinement ratio $a/A$ covering the different regimes indicated by LSA, and ranges of $Re$ from slightly below the primary threshold found by LSA to about twice this value.
The conducted simulation runs are given by their coordinates in the parametric plane  
 $(Re-a/A)$  in figure \ref{fig:NeutralCurves_LSA_vs_DNS}.  

Five general kinds of solutions are observed and are displayed using different symbols. 
The first kind (white squares) is an axially symmetric state corresponding to a stable configuration with zero lift $\mathscr{L}$ i.e. the lift coefficient
$C_\ell=\frac{\mathscr{L}}{1/2\rho U_\infty \pi r^2}$ is measured lower than $10^{-4}$. The second (black squares) is a 3D steady state characterised by a constant lift and a symmetry plane This state is noted $SS1$ as its structure is a strong indication of the direct effect of a steady  $|m=\pm 1$ eigenmode.
 The third (black circles) is a reflexion-symmetry preserving (RSP) state.  This mode is defined by an oscillatory lift around a non-zero mean value, the wake still displaying a planar symmetry. Aperiodic behaviors (black stars) have also been observed. Finally, the fourth kind of solution (black triangle, noted $SS_3$) is a steady state with a structure characterized by an $m=3$ component.

The LSA predictions  are reproduced in figure \ref{fig:NeutralCurves_LSA_vs_DNS} to allow a comparison with DNS results.
We observe an  excellent agreement between the transition from the axisymmetric state and the steady, non-axisymmetric state $SS1$ revealed by DNS and the marginal stability curve $Re_{c,S1}(a/A)$ indicating destabilization of the $S1$ modes. This fully confirms that the nonlinear state  $SS1$  is effectively directly due to a supercritical nonlinear saturation of the $S1$ mode.

%

On the other hand, in the computed cases, the observed secondary bifurcations (leading either to a periodic RSP state or to an aperiodic state), does not directly match with any secondary bifurcation curve revealed by LSA.
This is not really surprising, since the secondary bifurcation occurs along the bifurcated 
steady state mode ($SS$) which differs from the axisymmetric solution used as a base-flow for the linear stability analysis. 
However, the nature of the secondary modes revealed by LSA may still be relevant to fully explain the nonlinear dynamics, as it will be demonstrated by a deeper exploration of  few cases in the next section.


\subsection{Towards nonlinear behavior, low confinement flow at $a/A=0.39$}

\begin{figure}
\begin{center}

(a)\quad{\includegraphics[scale = .3, trim= 0 190 600 50, clip]{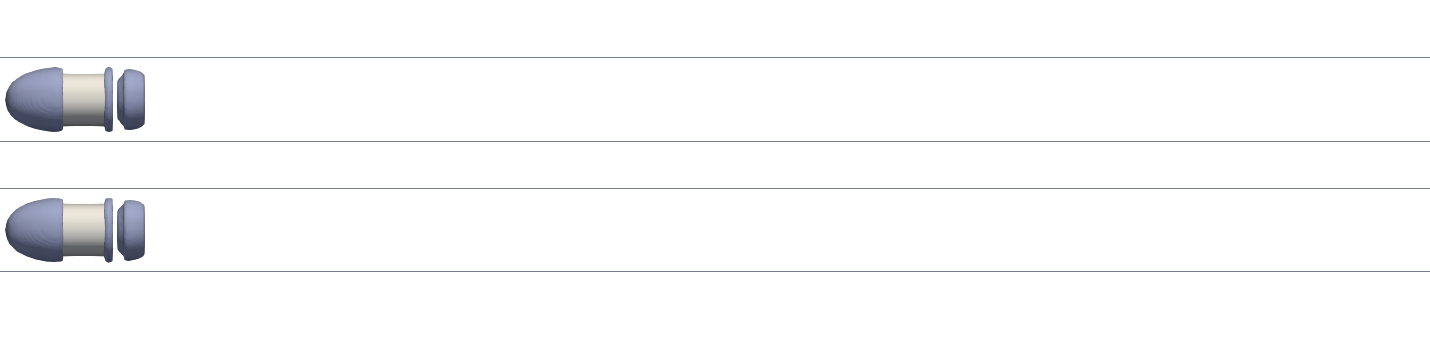}}

\phantom{(a)\quad}{\includegraphics[scale = .3, trim= 0 30 600 187, clip]{grfx_DNS/blunt_htt_aRatio0,39_Re130_Qcriterion.png}}

(b)\quad{\includegraphics[scale = .3, trim= 0 190 600 50, clip]{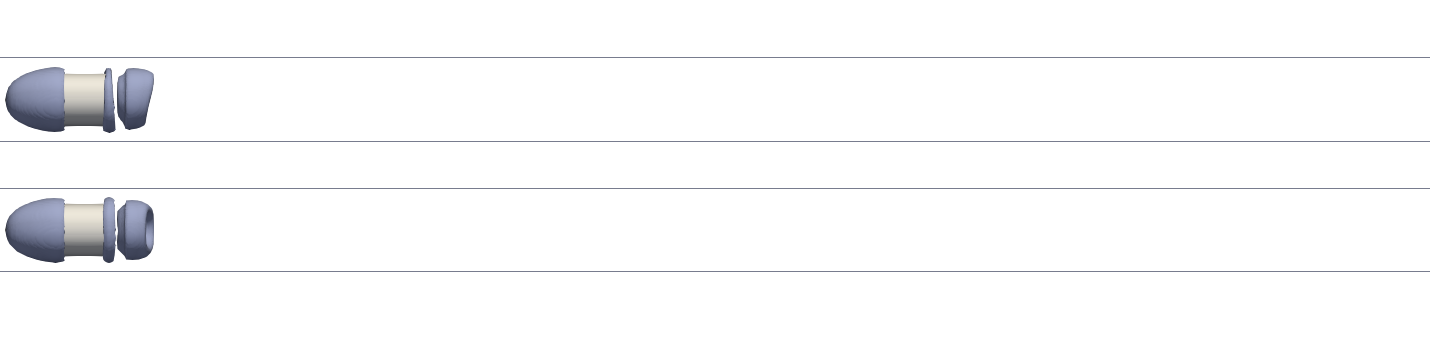}}

\phantom{(b)\quad}{\includegraphics[scale = .3, trim= 0 30 600 187, clip]{grfx_DNS/blunt_htt_aRatio0,39_Re145_Qcriterion.png}}


(c)\quad{\includegraphics[scale = .3, trim= 0 190 600 50, clip]{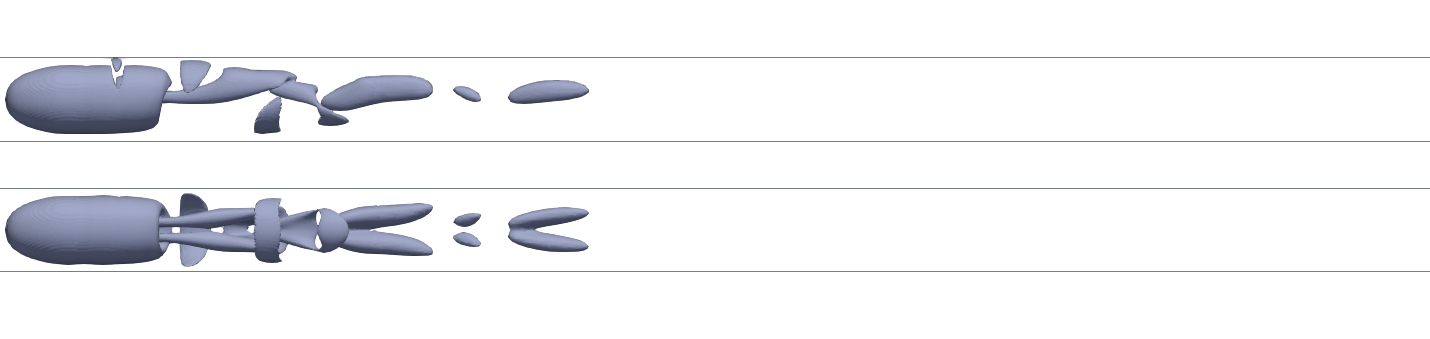}}

\phantom{(c)\quad}{\includegraphics[scale = .3, trim= 0 50 600 187, clip]{grfx_DNS/blunt_htt_aRatio0,39_Re175_Qcriterion.png}}

(d)\quad{\includegraphics[scale = .3, trim= 0 190 600 50, clip]{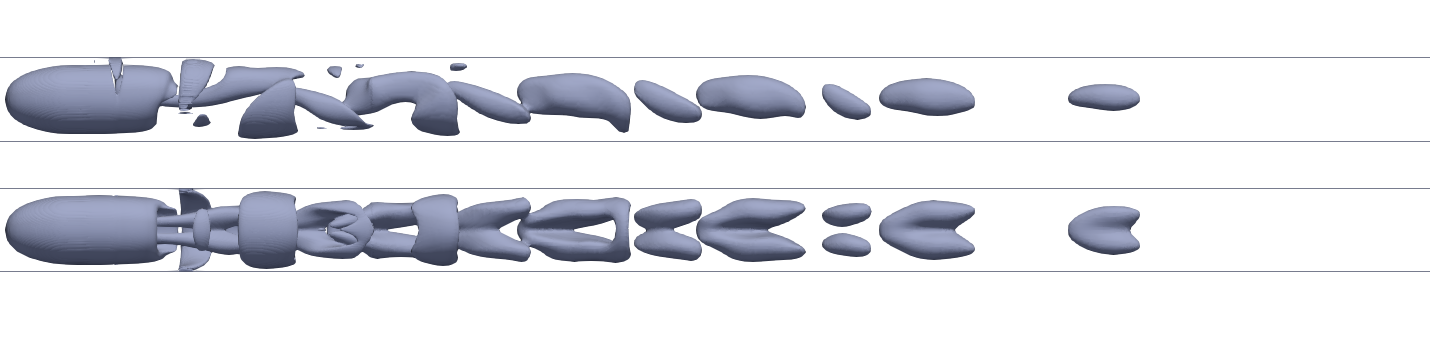}}

\phantom{(d)\quad}{\includegraphics[scale = .3, trim= 0 50 600 187, clip]{grfx_DNS/blunt_htt_aRatio0,39_Re200_Qcriterion.png}}

\end{center}
\caption{Iso-contour of Q-criterion for $a/A=0.39$, two perpendicular views are represented for each case. From top to bottom, $Re=130,\; 145,\; 175,\, 200$. }
\label{fig:Qcrit_BR0.39}
\end{figure}

The temporal evolution of the wakes for different Reynolds number and for a fixed area ratio
are now analysed with the DNS.
Fig. \ref{fig:Qcrit_BR0.39} displays the Q-criterion for the three numbers $Re=130,\; 145, 175, \; 200$, with two plots for each case corresponding to view in two orthogonal directions.


For $Re=130$ (fig. \ref{fig:Qcrit_BR0.39}$a$), the flow corresponds to the axially symmetric state, in accordance with LSA prediction. The wake is axially symmetric and the view is identical in both orthogonal directions.
The structure behind the blunt body is stationary and it consists in a toroidal recirculation bubble. 
For $Re = 160$ (fig. \ref{fig:Qcrit_BR0.39}$b$),
 is observed the $SS1$ steady, non-axisymmetric state. The breaking of axisymmetry results in a tilting of the  toroidal structure attached to the body, the latter expanding in one direction and retracting in the opposite one.

The next states for $Re=175$ and $Re=200$ (fig. \ref{fig:Qcrit_BR0.39}$c-d$) correspond to the reflectional symmetry preserving (RSP) oscillatory state. The toroidal recirculation gets destabilized and hairpin vortices are periodically advected in the streamwise direction. 
The interaction between those vortices and the wall is visible through wall shaped vortices which merge with the hairpin vortices as they move downstream. 

Up to here, the sequence of bifurcations and the structure of the observed states are identical to the unconfined case \citep{bury2012transitions,bohorquez2011stability}. The main difference is that due to confinement the bifurcations arise at much lower values of Re (for instance, \cite{bohorquez2011stability} report the first bifurcation for $Re = 319$ and the second for $Re =413$).



\begin{figure}
\begin{center}
\subfloat[]{\includegraphics[scale = 0.3]{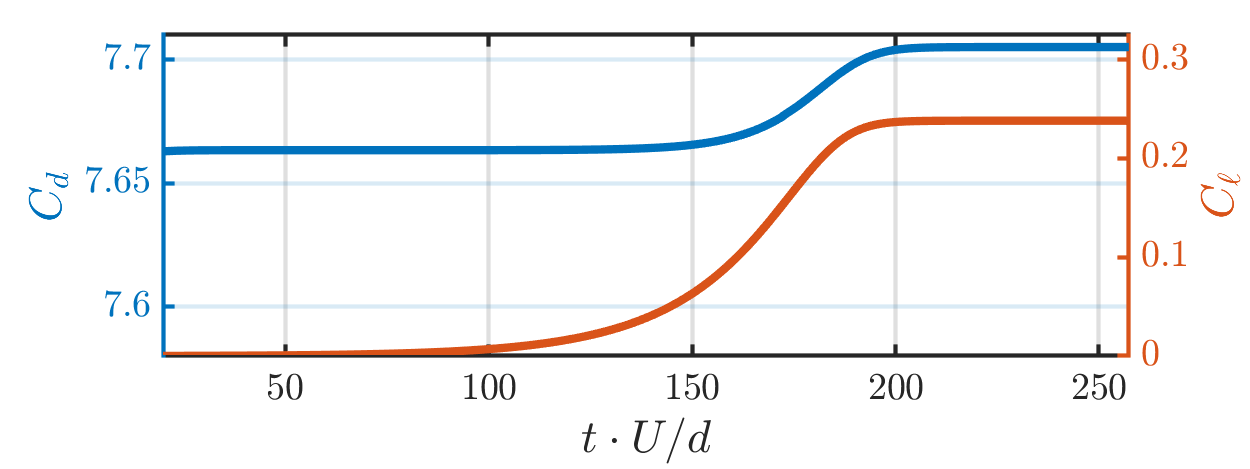}}\quad 
\subfloat[]{\includegraphics[scale = 0.3]{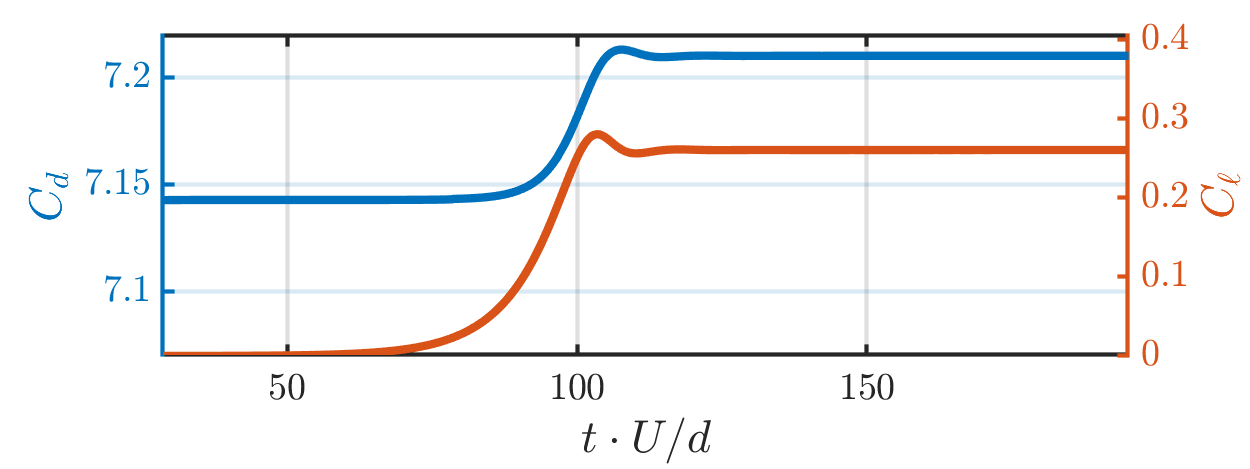}}\\
\subfloat[]{\includegraphics[scale = 0.3]{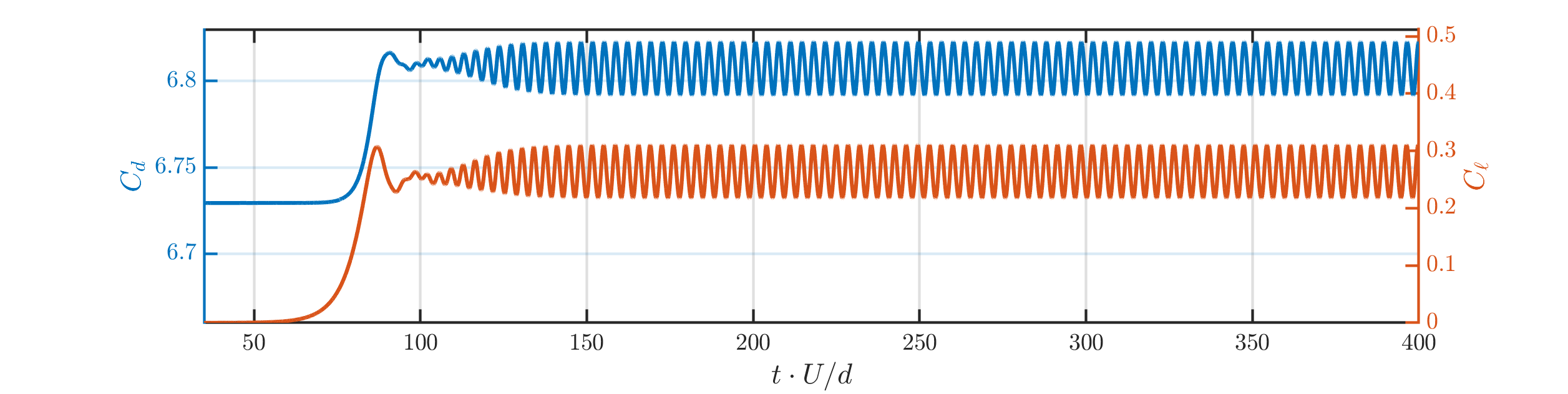}}\\
\subfloat[]{\includegraphics[scale = 0.3]{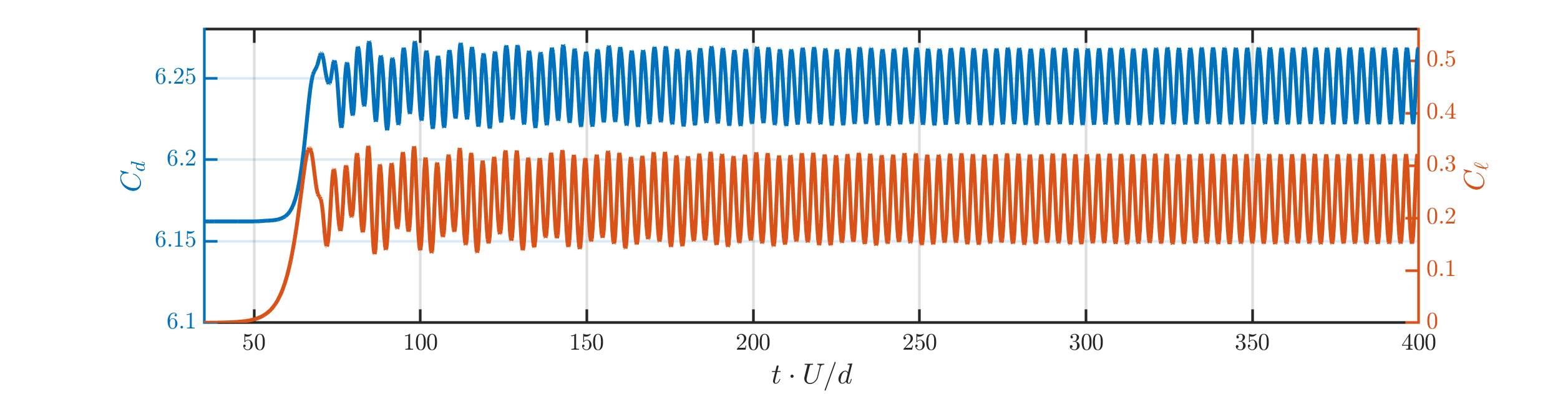}}\\
\end{center}
\caption{Lift and drag coefficients from DNS simulations ($L/d=2$ and $a/A=0.39$) versus dimensionless time, for (a) $Re=145$, (b)  $Re=160$, (c) $Re=175$ and (d) $Re=200$ }
\label{fig:DNS_lift-drag_BR0.39}
\end{figure}

Figure \ref{fig:DNS_lift-drag_BR0.39} displays the time histories of the lift and drag coefficients (noted $C_\ell$ and $C_d$) characterizing forces exerted on the body  calculated from DNS, again for $a/A=0.39$.

For $Re = 130$ (case not displayed), the lift converges to a zero value. For all other cases, after a short transient (not shown), the simulations first seem to converge towards an steady state with zero lift, approximately in the range $t \in [30-50]$. The later evolution shows however that this state is not stable, and a phase of linear instability characterized by exponential growth of both coefficients is seen. In this linear phase the observed behaviour of the lift coefficient corresponds to a purely exponential growth with non-oscillating behavior ($\approx e^{\sigma t}$ with real $\sigma$). This is a clear signature of the emergence of the non-oscillating mode $S1$ which is effectively the only unstable one detected by LSA for the values of $Re$ considered.




For  $Re=145$ (Fig. \ref{fig:DNS_lift-drag_BR0.39}$a$) and $Re = 160$
(Fig. \ref{fig:DNS_lift-drag_BR0.39}$b$), the subsequent nonlinear evolution is a saturation towards the $SS1$ steady state with non-zero lift. 
%
On the other hand, for  $Re=175$ (Fig. \ref{fig:DNS_lift-drag_BR0.39}$c$),  this steady state  seems to be transiently approached by the solution, but then a second phase of linear instability, this time with oscillating behaviour ($\approx e^{\sigma t}$ with complex $\sigma$) is observed. This trend is the signature of the existence of a non-oscillating mode related to the $O1$ mode. For $Re = 175$, the saturated state ultimately observed is the periodic, RSP state characterized by a lift force oscillating around a non-zero mean value.
For $Re=200$ (Fig. \ref{fig:DNS_lift-drag_BR0.39}$d$), the initial behaviour and ultimate state are similar, but transient towards the RSP state displays a low-frequency modulation which is eventually damped.

\begin{figure}
\begin{center}
\hspace{-2 cm}
\subfloat[]{\includegraphics[scale = 0.3,trim = 0 0 0 20]{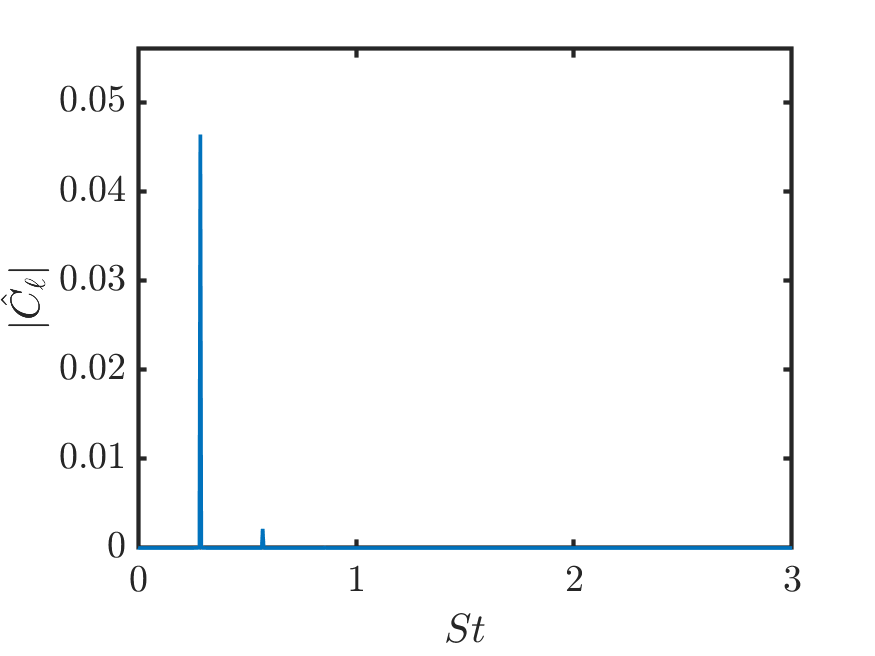}}
\subfloat[]{\includegraphics[scale = 0.3,trim = 0 0 0 20]{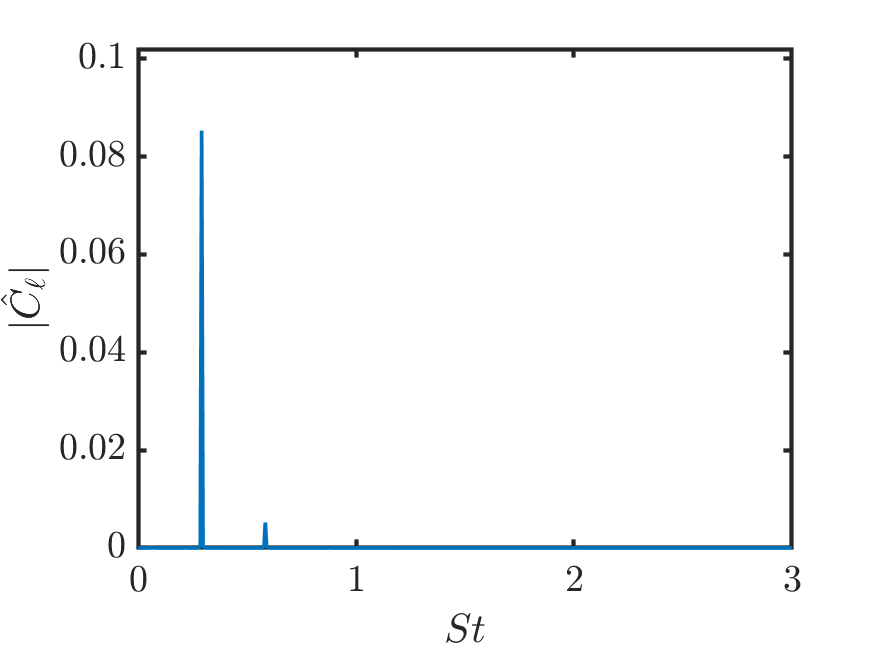}}
\subfloat[]{\begin{minipage}{0.15\textwidth}
\vspace{-3.5 cm}
\setstretch{1.}
    \begin{tabular}{c|c c}
        type    &     $Re$       & $St$      \\
        \hline \hline
        DNS (a)    &     $175$      & $0.28602$ \\
        DNS (b)    &     $200$      & $0.29196$ \\
        LSA     &     $201.2$    & $0.25159$ \\ \hline
    \end{tabular}\vspace{3em}
\setstretch{1.}
\end{minipage}}

\caption{Frequency spectra for DNS case $L/d=2$, $a/A=0.39$, $Re=175$ (a) $Re=200$ (b) and table of $St$ values of the higher peaks (DNS) with corresponding $St_c$ for LSA (c).}
\label{fig:pdf_BR0.39}
\end{center}
\end{figure}

The dimensionless frequency spectra of the two oscillating cases
are presented on figure \ref{fig:pdf_BR0.39}. 
Both spectra give similar results as the $Re$ are close.
They are performed from the $C_\ell$ signal of figure \ref{fig:DNS_lift-drag_BR0.39} and the transitional behaviors have been excluded. It has been verified  that the sample is large enough and does not influence the spectra. 
The two peaks can be interpreted as a fundamental frequency mode and its first harmonic. 
The amplitude of the first harmonic is much lower than the fundamental and therefore it is not visible to the naked eye on the signal which is very close to a pure sinusoid. As shown in the Table of Fig. \ref{fig:pdf_BR0.39}$c$, 
the influence of the Reynolds number on the Strouhal number $St$
is weak with this $Re$ range, the DNS  give a $2\%$ variation between $Re=175$ and $200$.
This behavior is substantiated by the quasi-constant $St$ values found using the LSA
for low confinement (see figure \ref{fig:asurA0.1_lambda}b). 
The $St$ values given by the DNS and LSA approaches are comparable even if a 
$16\%$ relative difference is measured between the DNS ($Re=200$) and LSA  at the threshold ($Re=201.2$). 
The discrepancy between those results can be explained by the fact that the $O1$ mode is obtained using an axially symmetric base flow whereas this base flow is no longer present in the DNS for $Re\geq175$, the RSP state oscillates around steady state which is non-axisymmetric. 


\subsection{Towards nonlinear behaviors, moderately confined cases ($a/A=0.6$ and $0.74$) }

\begin{figure}
\begin{center}
(a)\quad{\includegraphics[scale = 0.3, trim= 0 205 1120 50, clip]{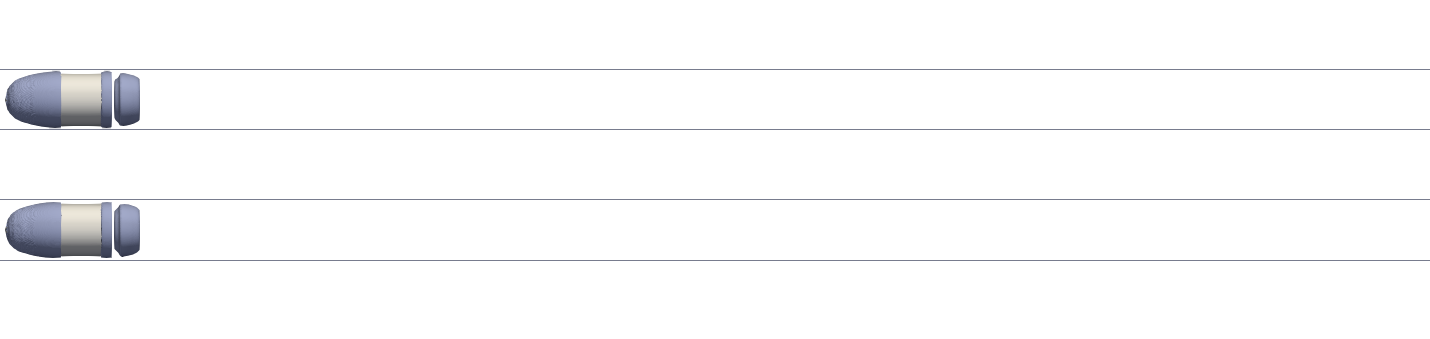}\quad
\phantom{
\includegraphics[scale = 0.3, trim= 0 205 1120 50, clip]{grfx_DNS/blunt_htt_aRatio0,74_Re100_Qcriterion.png}\quad
\includegraphics[scale = 0.3, trim= 0 205 1120 50, clip]{grfx_DNS/blunt_htt_aRatio0,74_Re100_Qcriterion.png}}
}

\phantom{(a)}\quad{\includegraphics[scale = .3, trim= 0 60 1120 195, clip]{grfx_DNS/blunt_htt_aRatio0,74_Re100_Qcriterion.png}\quad
\phantom{\includegraphics[scale = .3, trim= 0 60 1120 195, clip]{grfx_DNS/blunt_htt_aRatio0,74_Re100_Qcriterion.png}\quad
\includegraphics[scale = .3, trim= 0 60 1120 195, clip]{grfx_DNS/blunt_htt_aRatio0,74_Re100_Qcriterion.png}
}
}

(b)\quad{\includegraphics[scale = .3, trim= 0 205 1120 50, clip]{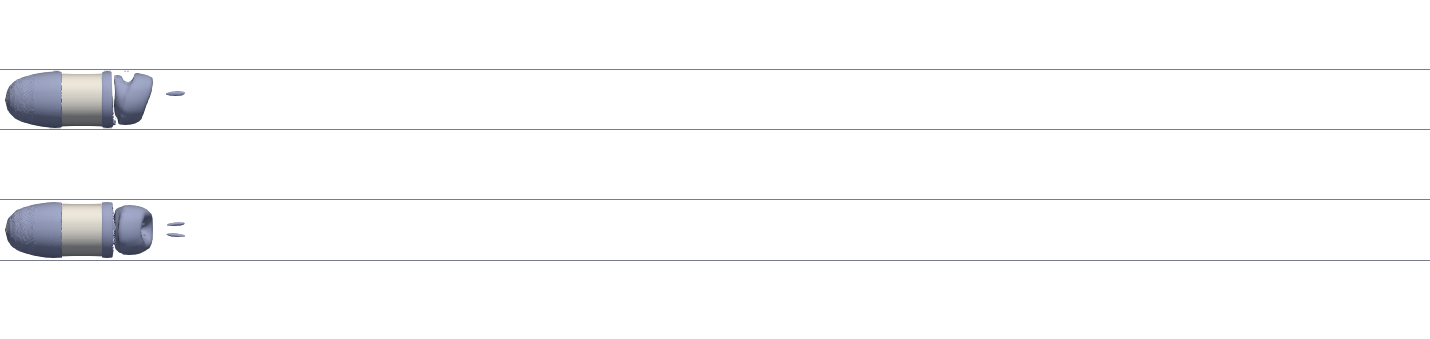}\quad
\phantom{
\includegraphics[scale = .3, trim= 0 205 1120 50, clip]{grfx_DNS/blunt_htt_aRatio0,74_Re100_Qcriterion.png}\quad
\includegraphics[scale = .3, trim= 0 205 1120 50, clip]{grfx_DNS/blunt_htt_aRatio0,74_Re100_Qcriterion.png}}
}

\phantom{(b)}\quad{\includegraphics[scale = .3, trim= 0 60 1120 195, clip]{grfx_DNS/blunt_htt_aRatio0,74_Re115_Qcriterion.png}
\quad
\phantom{\includegraphics[scale = .3, trim= 0 60 1120 195, clip]{grfx_DNS/blunt_htt_aRatio0,74_Re100_Qcriterion.png}\quad
\includegraphics[scale = .3, trim= 0 60 1120 195, clip]{grfx_DNS/blunt_htt_aRatio0,74_Re100_Qcriterion.png}
}
}

(c)\quad{\includegraphics[scale = .3, trim= 0 205 1120 50, clip]{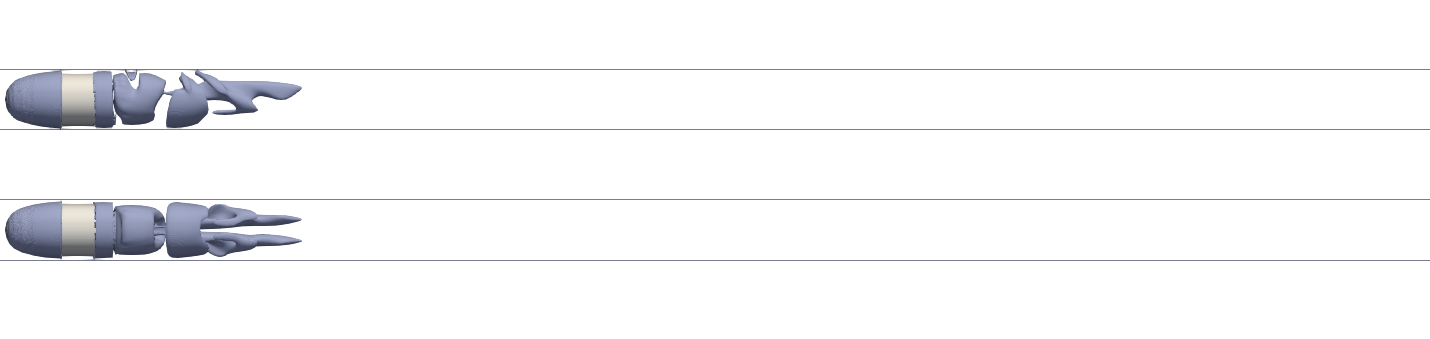}\quad
\includegraphics[scale = .3, trim= 0 205 1120 50, clip]{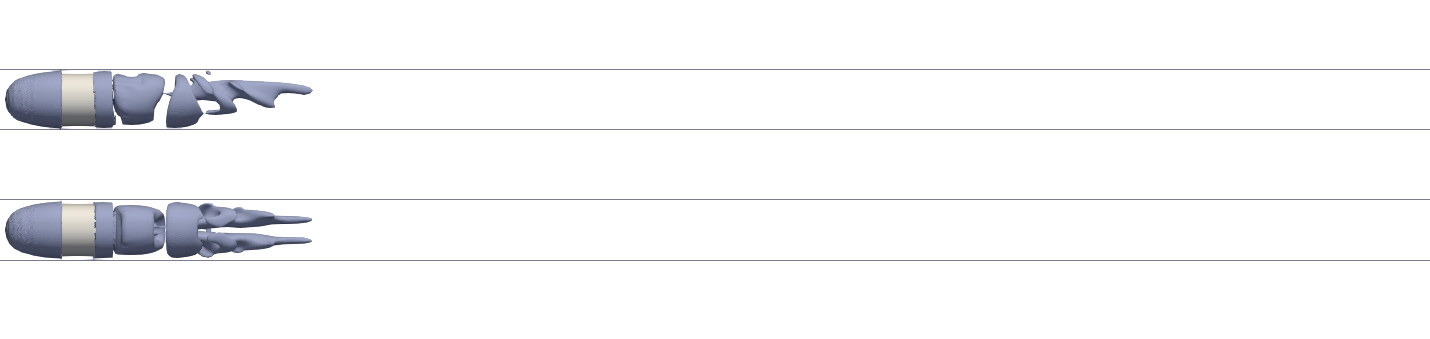}\quad
\includegraphics[scale = .3, trim= 0 205 1120 50, clip]{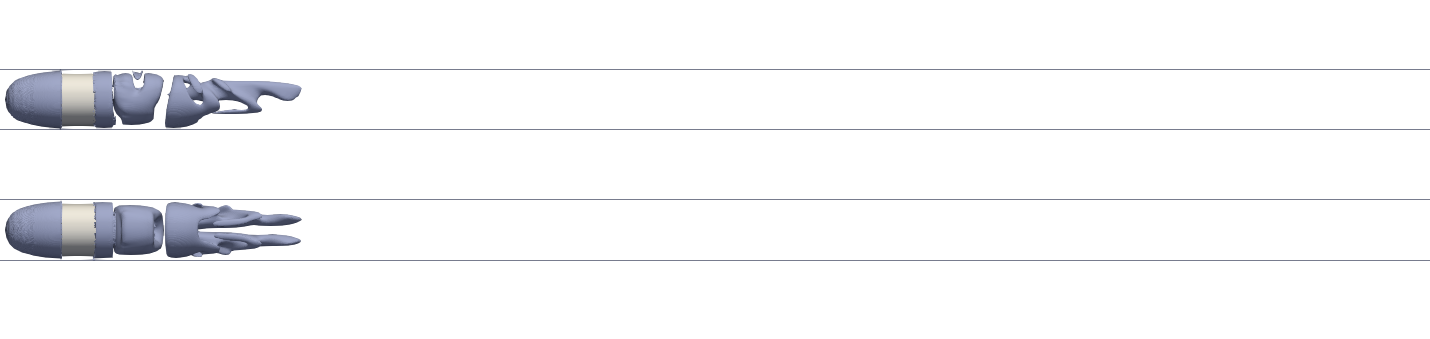}
}

\phantom{(c)}\quad{\includegraphics[scale = .3, trim= 0 60 1120 195, clip]{grfx_DNS/aRatio0,74_Re150.0003.png}\quad
\includegraphics[scale = .3, trim= 0 60 1120 195, clip]{grfx_DNS/aRatio0,74_Re150.0016.png}\quad
\includegraphics[scale = .3, trim= 0 60 1120 195, clip]{grfx_DNS/aRatio0,74_Re150.0029.png}
}

(d)\quad\includegraphics[scale = .3, trim= 0 205 1120 50, clip]{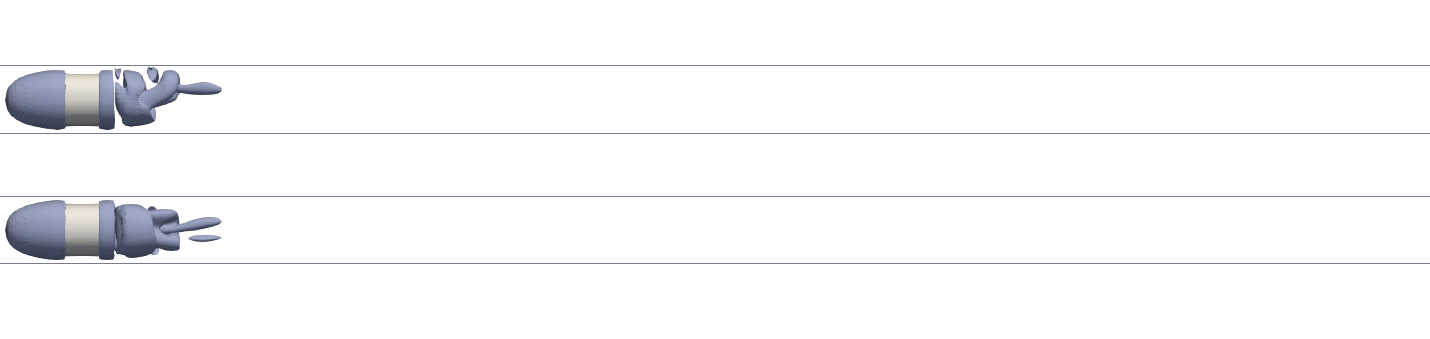}
\quad\includegraphics[scale = .3, trim= 0 205 1120 50, clip]{grfx_DNS/blunt_htt_aRatio0,6_Re150_Qcriterion.png}
\quad{\includegraphics[scale = .3, trim= 0 205 1120 50, clip]{grfx_DNS/blunt_htt_aRatio0,6_Re150_Qcriterion.png}
}

\phantom{(d)}\quad{\includegraphics[scale = .3, trim= 0 60 1120 195, clip]{grfx_DNS/blunt_htt_aRatio0,6_Re150_Qcriterion.png}\quad
\includegraphics[scale = .3, trim= 0 60 1120 195, clip]{grfx_DNS/blunt_htt_aRatio0,6_Re150_Qcriterion.png}
\quad
\includegraphics[scale = .3, trim= 0 60 1120 195, clip]{grfx_DNS/blunt_htt_aRatio0,6_Re150_Qcriterion.png}
}

\end{center}
\caption{Iso-contour of Q-criterion for moderately confined cases; two perpendicular views are represented for each case. $(a)$ $a/A =0.74$, $Re=100$,\; 
$(b)$ $a/A =0.74$, $Re=115$,\;
$(c)$ $a/A =0.74$, $Re=160$,\;
$(d)$ $a/A =0.6$, $Re=150$.
For the two last cases, 
instantaneous representations for three different instants are displayed.   }
\label{fig:Qcrit_BR0.74}
\end{figure}

Consider, now, the flow structures revealed by DNS in the range of moderately confined cases. The beginning of the bifurcation sequence is the same as described in the previous paragraph. With an initially symmetric state, followed by a steady, non-axisymmetric state. Fig. \ref{fig:Qcrit_BR0.74} $(a-b)$ displays these two states observed respectively for $a/A = 0.74$;$Re = 100$ and $a/A = 0.74$;$Re = 115$. Similar structures are obtained for $a/A =0.6$ and same values of $Re$ and are not displayed.

\begin{figure}
\begin{center}

\subfloat[]{\includegraphics[scale = .3]{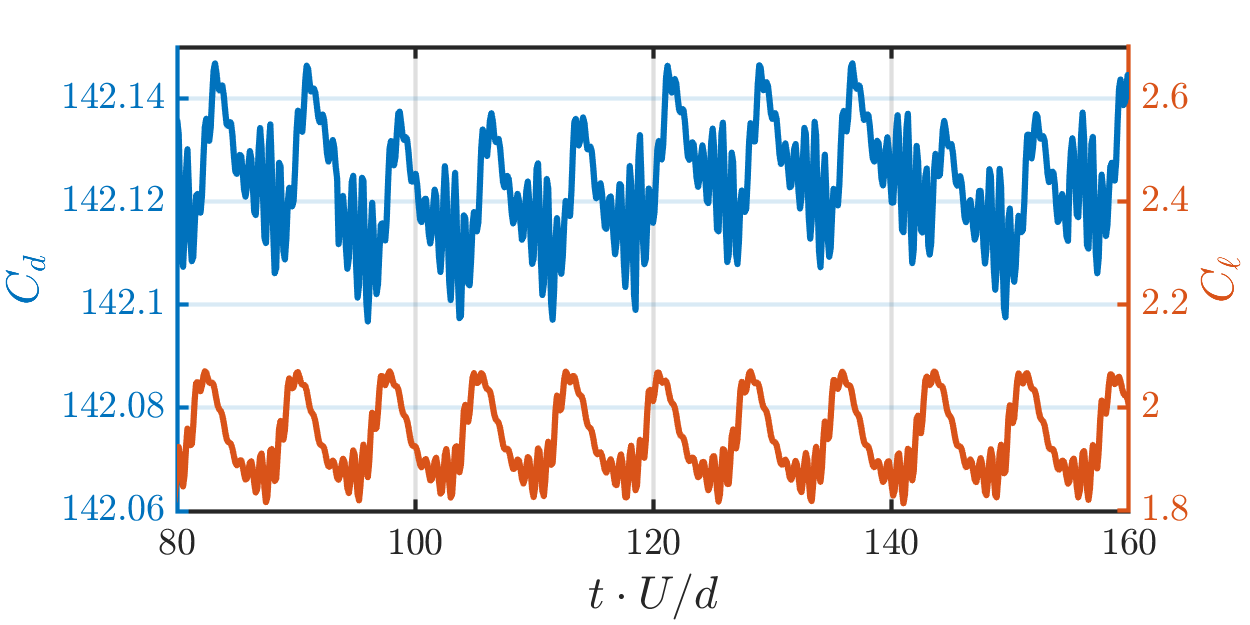}}
\quad
\subfloat[]{\includegraphics[scale = .3]{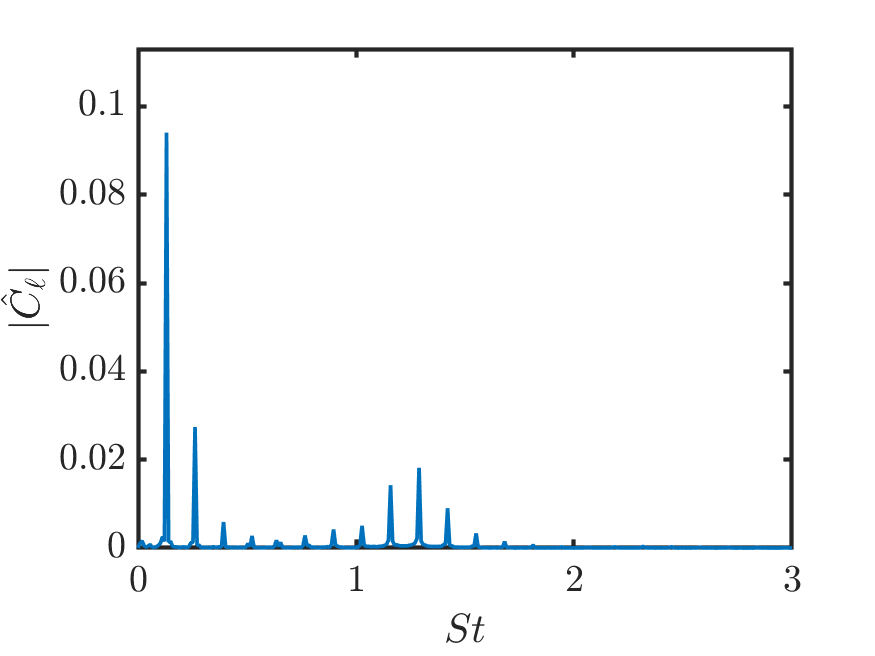}}
\caption{Characterization of time-dependent forces exerted on body for $Re=150$, $a/A=0.74$ : $(a)$ time-histories of Lift and drag coefficient, and $(b)$ 
Frequency spectrum of lift coefficient.
}
\label{fig:DNS_lift-drag_0.74}
\end{center}
\end{figure}

When raising the Reynolds number to $Re = 150$ in this range of moderately confined cases, nonlinearities lead to richer dynamics compared to the previous cases. Consider, first, the flow obtained for $a/A=0.74$
(Fig. \ref{fig:Qcrit_BR0.74}$c$). Although the 
flow symmetries still indicate the  (RSP) oscillatory state, the flow has a more complex structure than previously observed. 
 Two main oscillating regions can be seen: the first one is the upper part of toroidal recirculation, close to the body, where a separated structure periodically appears. The second one is formed by a more distant structure,  a $45^\circ$-inclined protrusion which is advected downstream. 
Sticking to the case $a/A=0.74$, $Re = 150$,
figure \ref{fig:DNS_lift-drag_0.74}$(a)$ displays the time-histories of the lift and drag coefficients. The lift force reveal a modulated where both a low-frequency component (with dimensionless period of order 7.5) and a high-frequency component (with a period about 10 times shorter) can be discerned. The drag coefficient displays similar patterns but the amplitude of oscillations are extremely small (less than $0,5\%$ of the average value).

Although the time-series may suggest a quasi-periodic behaviour, inspection of the Fourier-transform of the lift force (fig. \ref{fig:DNS_lift-drag_0.74}$b$) indicate that the behaviour is actually strictly periodic, as revealed by the existence of a fundamental frequency, $St_1=0.1308$ along with its harmonics. The spectrum also shows that apart form the fundamental, a high-frequeny content is centered around the harmonic number 10, corresponding to $St_{10}=1.308$. This matches with the high-frequency component detected in the time-series with a period about 10 times shorter compared to the low-frequency component. 

Trying to relate these dynamics to the LSA results is a bit puzzling, since in this range of $a/A$ no unsteady modes were detected: going back to figure \ref{fig:NeutralCurves_m1} shows that no unsteady modes exist for $a/A=0.74$ since the $O1$ and $O2$ are only detected for $a/A < 0.73$ and the low-frequency $O3$ mode only arises for $a/A>0.75$. However, again, the LSA results obtained considering the axisymmmetric base flow are only indicative here since the bifurcations arise from the steady non-axisymmetric state. 
$St_1=0.1308$
The order of magnitude of the Strouhal number $St_1$ characterizing the low-frequency oscillation is in the same range as the $O3$ mode which exists for $St \approx 0.1-0.2$, suggesting that the $O3$ mode actually play a role in the nonlinear solution given by the DNS.

\begin{figure}
\begin{center}
\subfloat[]{\includegraphics[scale = .3]{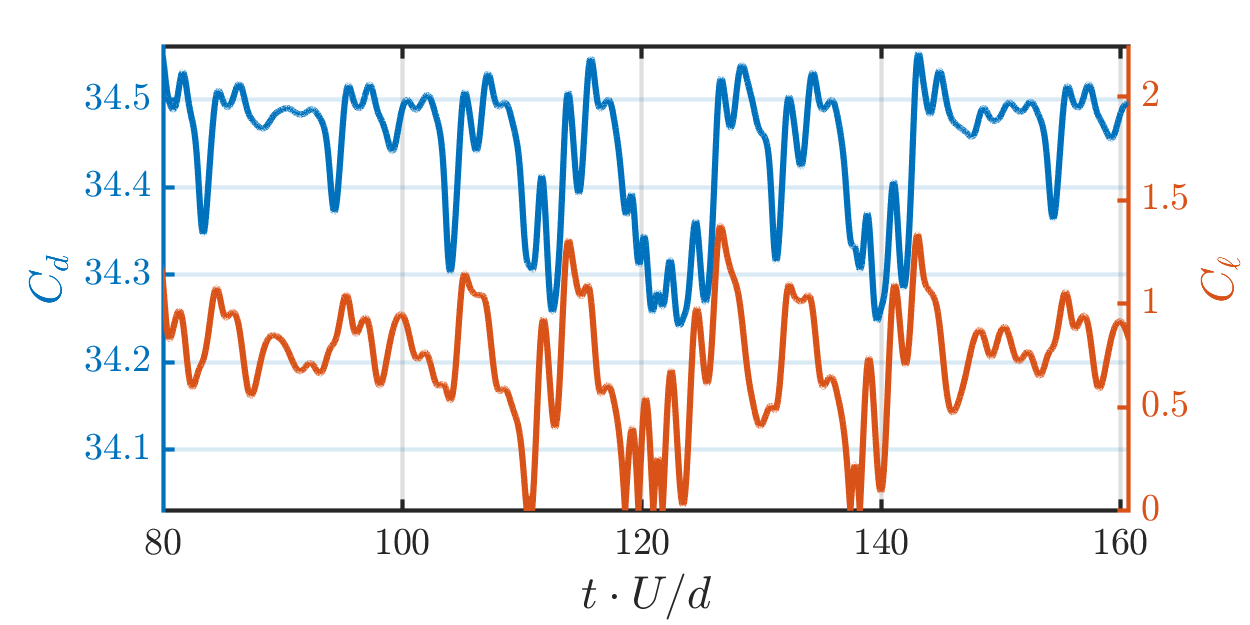}}
\quad
\subfloat[]{\includegraphics[scale = .3]{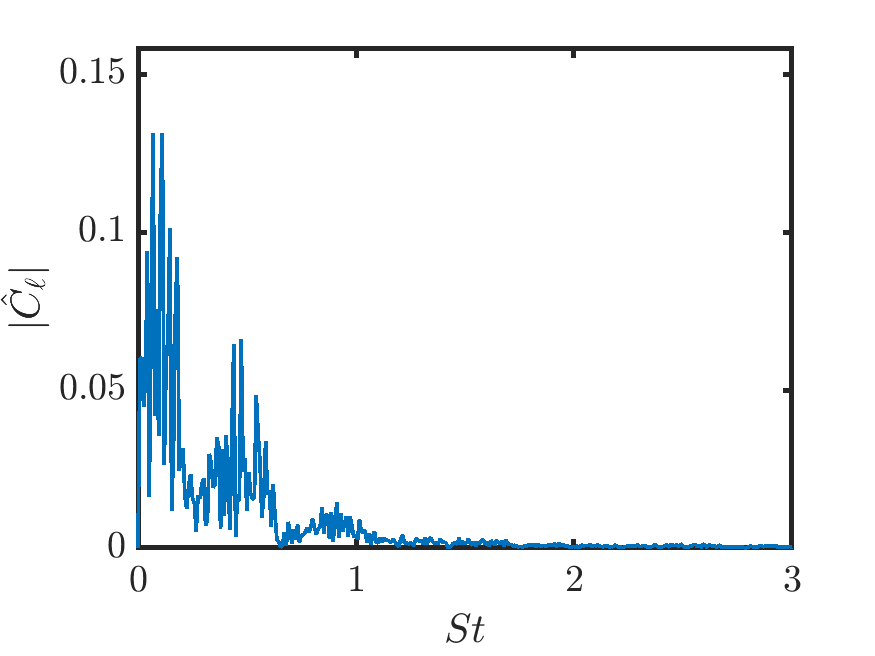}}
\caption{Characterization of time-dependent forces exerted on body for $Re=150$, $a/A=0.6$ : $(a)$ time-histories of Lift and drag coefficient, and $(b)$ 
Frequency spectrum of lift coefficient.}
\label{fig:DNS_lift-drag_0.6}
\end{center}
\end{figure}

 Consider, now, the flow obtained for $a/A=0.6$ and $Re = 150$ (Fig. \ref{fig:Qcrit_BR0.74} $d$). This time, the snapshots reveal that the planar symmetry is lost. Vortical structure of hairpin-like shape are shed, but their orientations and shapes are less ordered as in the previous cases. The time-series of the exerted forces (fig. \ref{fig:DNS_lift-drag_0.6}$a$) show that periodicity is clearly lost and indicate a chaotic behaviour. This is confirmed by examining the Fourrier transform of the lift force (fig. \ref{fig:DNS_lift-drag_0.6}$b$) which reveals a broadband spectrum.

\subsection{Towards nonlinear behaviors, high confinement flow at $a/A=0.85$}

 \begin{figure}
 \begin{center}
  \label{fig:SS3} 
 \subfloat[]{\includegraphics[scale = 0.6, trim= 50 500 250 195, clip]{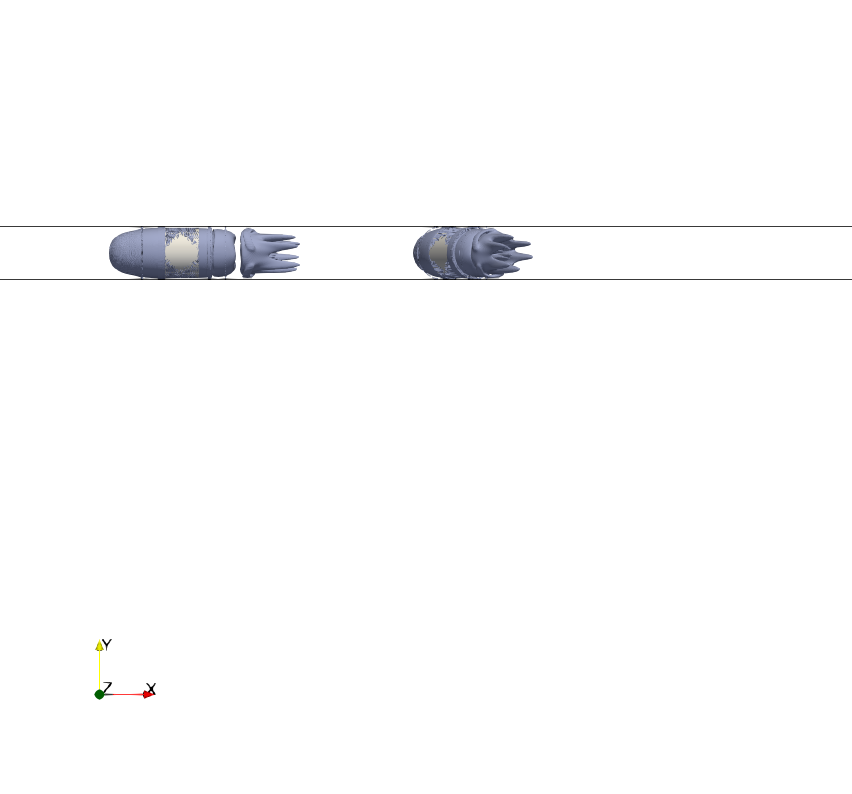}}
 \\
 
 \subfloat[]{\includegraphics[scale = 0.3, trim= 150 0 150 90, clip]{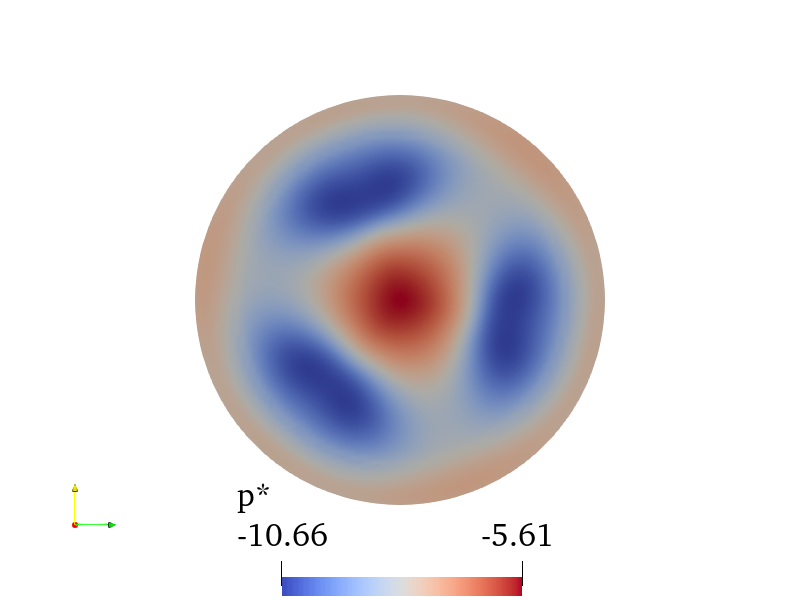}}
  \subfloat[]{\includegraphics[scale = 0.3, trim= 150 0  150 90, clip]{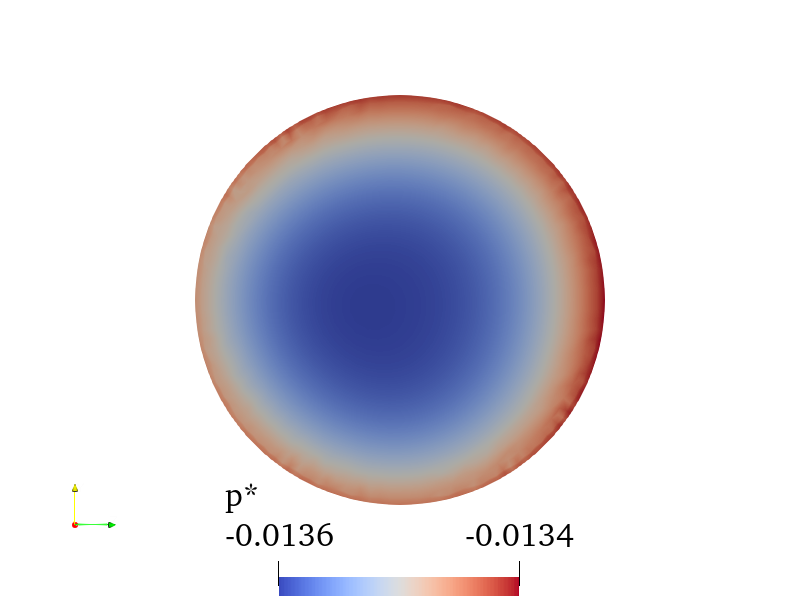}}
 \caption{Illustration of the wake for  $a/A=0.85$ and $Re=150$. Iso-contour of Q-criterion (a), side and rear diagonal views. Slices of the dimensionless pressure $p^*=p/(\rho U^2)$ in the wake of the body, at $x/D=1.2$ (b) and  at $x/D=9$ (c).   }
 \end{center}
 \end{figure}

 \begin{figure}
 \begin{center}
 \label{figlast}
 \includegraphics[scale = .3, trim= 0 0 0 0, clip]{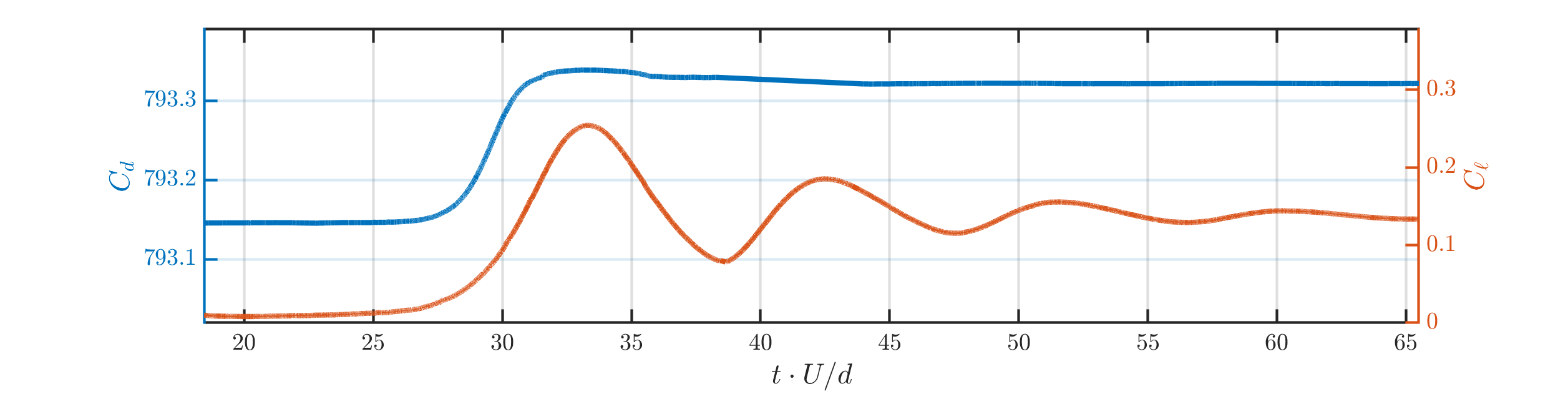}
 \caption{Lift and drag coefficient for DNS case $L/d=2$, $Re=150$ and $a/A=0.85$ versus dimensionless time.}
 \end{center}
 \end{figure}
 

To end up the exploration of nonlinear dynamics, consider now a highly confined case with $a/A=0.81$. The first bifurcation again leads to the steady, non-axisymetric state and is well explained by the onset of mode $S1$. On the other hand, when raising the Reynolds number, the next bifurcation does not lead to time-dependent vortex-shedding. Instead, the flow remains stationnary, but acquires a structure characterized by the shedding of six vortical structures instead of only two as in the $SS1$ state, as shown in figure \textcolor{blue}{25}$(a)$ for $Re=150$. This structure is a strong indication of the presence of an eigenmode with azimutal wavenumber $m=3$, and is fully in line with the LSA results which indeed predict the existence of the steady mode $S3$ in the same range of parameters. Plotting the pressure in a transverse slice just behing the body in Fig. \textcolor{blue}{25}$(b)$ indeed indicates a symmetry of order 3 (i.e. 3 symmetry planes).  However this symmetry is not perfect. Indeed, plotting the pressure in a slice  located farther downstream in Fig. \textcolor{blue}{25}$(c)$ reveals rather a symmetry of order one, visible in this plot by the fact that the region of largest pressure (blue levels) is slightly displaced towards the left. 
The presence of a $m=1$ component in the flow also manifests by the existence of a non-zero lift force, as indicated by the time-series in figure \textcolor{blue}{26}. 
This suggest that the observed flow structure actually results from the presence of both $S1$ and $S3$ modes.

%
\section{Conclusion}

In this study, the stability of the wake induced by a bullet-shaped blunt body moving
at constant velocity in moderate and strong confinement conditions has been investigated
by the mean of two different numerical approaches.
The first one is  the global linear stability analysis  and it has been performed
on a pretty exhaustive set of parameters (geometrical aspect ratios and Reynolds numbers),
more especially the $(a/A,Re)$ plane have been widely explored. 
One of the main conclusions arising  from this first study is that the first destabilisation of the axially symmetric is always associated to stationary mode with azimutal wave number $m=1$. 
In the low-confinement regime ($a/A<0.6$), one observes a sequence similar to the one observed in the unconfined case, characterized by the successive emergence of two  two stationary ($S1$ and $S2$) modes  and two oscillatory  modes ($O1$ and $O2$), all with azimuthal wavenumber $m=1$. Increasing the confinement results in a decrease of the associated critical Reynolds numbers and increase of the frequencies of the unsteady mode. The length of the body also influences the results and tends to delay the instability.
On the other hand, in the highly confined regime ($a/A > 0.75$), although the primary mode remains the $S1$ mode, the next ones to emerge are steady modes associated to wavenumbers $m=2$ and $m=3$. This range is also associated to a restabilization of most $m=1$ modes: the oscillating modes $O1$ and $O2$ completely disappear, and the primary stationary mode $S1$ restabilises, and new unsteady modes called $O3$ and $O3$ characterized by very low frequencies emerge. 
Interestingly, between these two latter events, there exists a range of Reynolds number where all eigenmodes with $m=1$ are stable and only unstable modes with $m=2,3$ exist. 
 Interestingly, in this high-confinement regime, the results become independent upon the length of the body. This is explained by the fact that a parallel flow of Couette-Poiseuille type establishes within the annular gap between the body and the wall.

The second part of this paper is a numerical exploration of the nonlinear dynamics. For this, direct numerical simulations are performed for various points of the $(a/A, Re)$ plane in order to confront the linear stability findings with numerical experiments.
The results of the DNS agrees well with the LSA close to the first instability threshold
as expected. For low confinement, the bifurcation scenario remains the same as the one observed for bullet-shaped blunt bodies.
First the loss of axial symmetry occurs through a stationary bifurcation implying a non-zero lift, and then an oscillatory behavior is exhibited  via the reflexion-symmetry preserving
(RSP) state. As the confinement raises, the  scenario is no longer valid
and other states emerge due to the wall presence.
For instance, aperiodic behavior can be observed for intermediary confinement, $a/A=0.6$.
The nonlinearity effects increase with the confinement as it is illustrated for 
 the RSP state when the section ratio is $a/A=0.74$. This state differs greatly from the one found for a low confinement: the wake oscillation gather a large number of harmonics of the same frequency as it has been demonstrated on spectra of the lift and drag coefficients. 


This study has been a first attempt to qualify the flow stability of such a configuration
where the flow is assumed laminar and incompressible. Two routes should be investigated, 
following the paper approaches, 
the effect of the gas compressibility and the stability of the flow in a mean turbulent flow.

\section*{Acknowledgments}

The authors acknowledge the support of Région occitanie under the project "HTT Analyse Aéro - Readynov Aero" number 18012298 under the program FEDER-FSE Midi-Pyrénées and Garonne 2014-2020.
Thanks to ISAE and to HTT teams for the valuable discussions.
Computations have been conducted in the CALMIP center, grant no. P20037.


\appendix
\section{Annular Couette-Poiseuille flow}

 \label{appendixA}

For strongly confined cases, the flow may be approximated by a parallel flow ${\bf u} = u_x(r) {\bf e}_x$.
With the assumptions the  Navier-Stokes equations written in the body frame  can be reduced to  :
\begin{equation}
\frac{1}{r} \frac{\partial}{\partial r}\left(r \frac{\partial u_x}{\partial r} \right) =  \mu \frac{d p}{d x}   
\end{equation}
where the axial pressure gradient can be shown to be constant.
The volume flow rate  is given from the product of front section of diameter $D$ and of the 
body velocity i.e the external wall velocity  $u_w$ in the body frame.
Let us use for convenience  the aspect ratio $\xi = d/D$, which is lower than 1.
The reference length and reference velocity  are respectively set to $d/2$
and $u_w$.
The nondimensional analytical solution (referred now as $u_x$) is easily found by integration, with the help of  the no slip-velocity on walls  ($u_x(1)=0$, $u_x(1/\xi) = 1$) and of the conservation of the volume flow rate $q_v = \xi^{-2}~u_w~\pi d^2/4$.
The solution reads
\begin{equation}
u_x(r) = \frac{\xi^2 (\eta^2-1) - (1 + \xi^2) \log \eta}
{(1 + \xi^2) \log \xi + 1 - \xi^2}, 
\qquad \eta = \frac{2 r} {d}, \qquad 1 \leqslant \eta \leqslant 1/\xi
\label{eq:anpo}
\end{equation}


In addition the nondimensional pressure gradient and the Reynolds number  $Re_{d/2}$ are related to $\xi$ by :
$$\displaystyle
\ \frac{\xi^2}
{(1 + \xi^2) \log \xi + 1 - \xi^2} = \frac{1}{4~Re_{d/2}} \frac{d p}{d x}
$$
Obviously we can see that $Re_{d/2} = Re / 2$.


\section{Mesh convergence for DNS simulations}
\label{appendixB}

\begin{table}
\setstretch{1.3}
\centering
\begin{tabular}{ c | c |  c | c|  c  c  c  c  c  }
 \makecell{Mesh\\ref.}  & \makecell{Cells\\ $(\times 10^6 )$}& $n_c/D$ & $n_{BL}$ & $\overline{C_d}$ & $\sigma_{C_d}$ & $\overline{C_\ell}$ & $\sigma_{C_\ell}$ & $St$ \\ \hline \hline
A & $0.28$ & 17 & 5 & $\quad6.8043\quad$ & $\quad0.013311\quad$ & $\quad0.26079\quad$ & $\quad0.034502\quad$ & $\quad0.28439\quad$ \\
 & & & & $0.46\%$ &  $30\%$ &  $2.3\%$ &  $5.6\%$ &  $0.77\%$  \\ 
B & $1.29$ & 35 & 3 & $\quad6.8075\quad$ & $\quad0.010950\quad$ & $\quad0.26237\quad$ & $\quad0.032857\quad$ & $\quad0.28603\quad$ \\
 & & & & $0.42\%$ &  $6.8\%$ &  $1.7\%$ &  $2.0\%$ &  $0.20\%$  \\
C & $1.55$ & 35 & 5 & $\quad6.8292\quad$ & $\quad0.011845\quad$ & $\quad0.26390\quad$ & $\quad0.034830\quad$ & $\quad0.28711\quad$\\
 & & & & $0.01\%$ &  $15\%$ &  $1.2\%$ &  $7.0\%$ &  $0.18\%$ 
 \\ 
D & $3.13$ & 40 & 5 & $\quad6.8381\quad$ & $\quad0.011418\quad$ & $\quad0.26505\quad$ & $\quad0.034094\quad$ & $\quad0.28813\quad$ \\
 & & & & $0.03\%$ &  $11\%$ &  $0.73\%$ &  $5.8\%$ &  $0.54\%$ 
 \\
E & $8.46$ & 70 & 3 & $\quad6.8270\quad$ & $\quad0.009906\quad$ & $\quad0.26628\quad$ & $\quad0.031337\quad$ & $\quad0.28617\quad$ \\
 & & & & $0.13\%$ &  $3.2\%$ &  $0.27\%$ &  $2.6\%$ &  $0.14\%$ 
 \\ 
F & $9.50$ & 70 & 5 & $\quad6.8381\quad$ & $\quad0.010255\quad$ & $\quad0.26700\quad$ & $\quad0.032210\quad$ & $\quad0.28659\quad$ \\
 & & & & ref. &  ref. &  ref. &  ref. &  ref. 
\\ \hline
\end{tabular}\caption{Comparison of global quantities for different meshes ($Re=175$, $a/A=0.39$, $d/D=0.625$, $L/d=2$).  \label{table:convergence_openFoam} }
\end{table}

Mesh convergences ha been verified to trust the OpenFoam simulations. 
The outputs parameters of the convergence analysis are some global quantities as the time average and variance
of respectively the drag coefficient ($\overline{C_d}$, $\sigma_{C_d}$) and the lift coefficient 
($\overline{C_\ell}$, $\sigma_{C_\ell}$). The last output quantity is a Strouhal number evaluated in the body wake.
Three main parameters can qualify the mesh quality and they are described in the following.

The first parameter is the number of cells per blunt body diameter $n_{c/D}$. In this study it  chosen in $\left\{17,\; 35,\; 40,\; 70\right\}$.  The refinement levels of the boundary layer developed along the body, referred as $r_{BL}$
chosen in $\left\{3,\; 5\right\}$ is the second parameter.  The last parameter if the resulting number 
of cells (automatically generated).
During the mesh processing, the cells in contact with a solid wall are divided in layers
tangentially to this wall to ensure a better capture of the the boundary layer. The $r_{BL}$ parameter is
simply the number of layers defined by the user.

The lengths of the computational domain have been carefully chosen.
The extent of the each mesh in the streamwise direction is  $82 \times d$. 
The distance between the inlet and the body nose is set to $20 \times d$ and 
the distance between the body rear and the outlet is $60 \times d$. 
The convergence study has been performed for a Reynolds number of $Re=175$.
The parameters and the values of the output quantities are reported in table 
\ref{table:convergence_openFoam}. 
The mesh referred as F is the finest one and the results from simulations are considered as the reference.
Relative errors to $F$ mesh results are also added in the table.
It can be seen that the values of the drag coefficient $\overline{C_d}$ and of the Strouhal $St$ are converged for all meshes, even for the mesh A the coarsest one which could be assumed of bad quality.
The relative errors are lower than $0.46\%$ for $C_d$  coefficients and lower than $0.77$ for the nondimensional frequencies $St$. 
The lift coefficient $C_\ell$  can still be considered 
as converged with for all meshes with  a  maximal relative error up to  $2.3\%$ for mesh A.
Nevertheless, it seems to be more difficult to reach convergence for the variances of $C_d$ and $C_\ell$.
The explanation  can be found in the very low level of these variances  compare to the mean value of 
$C_d$ and $C_\ell$ coefficients. It indicates a very low amplitude of the fluctuations and that  always 
numerically requires some large dense meshes in such a case. 

\begin{figure}
    \centering
    \resizebox{\linewidth}{!}{
    \includegraphics[trim= 0 100 0 30, clip]{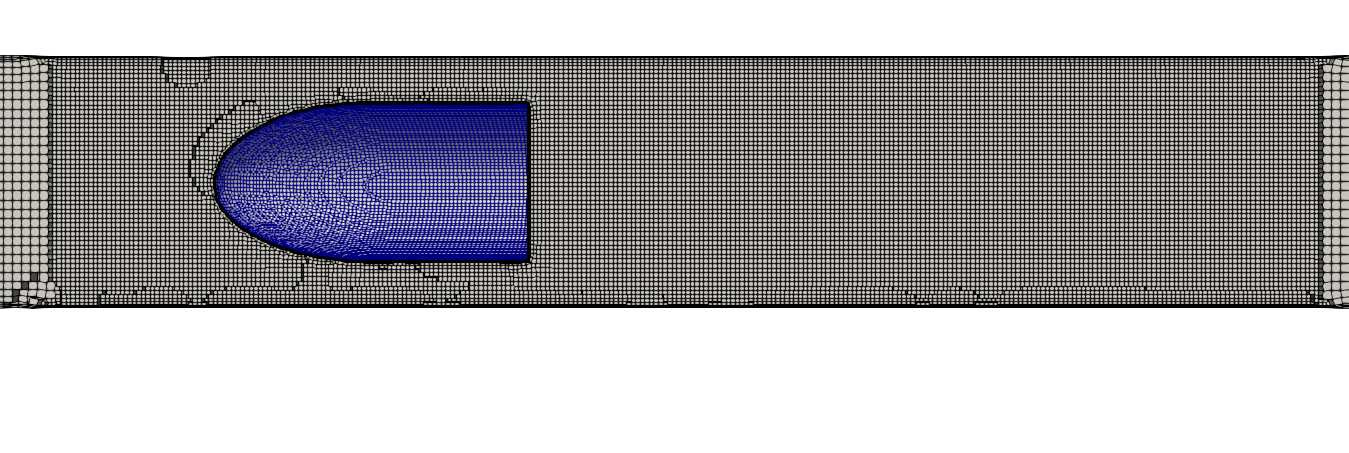} }
    \caption{Axial cut of the mesh B}
    \label{fig:mesh}
\end{figure}

Finally, mesh B (see figure \ref{fig:mesh}) has been selected for the main computations in this paper
because it is the best compromise between accuracy and precision. The variances are quite low respectively for
the $C_\ell$ and $C_d$ to $6.8\%$ and $2\%$
For other $Re$ than $175$,  the  number of cells in the boundary layer has been kept by 
using the scaling  given by laminar law relative to the boundary layer thickness and the Reynolds number $\delta_{BL} / d \propto  Re^{-1/2}$.



 \bibliographystyle{jfm}
\bibliography{Bibliography}

\end{document}